\documentclass[aps,showpacs]{revtex4}
\newcommand{\cd}{\makebox[0.08cm]{$\cdot$}}
\usepackage{epsf}
\usepackage{subfigure}
\begin{document}


\def\GreenYellow{} 
\def\Yellow{} 
\def\Goldenrod{} 
\def\Dandelion{} 
\def\Apricot{} 
\def\Peach{} 
\def\Melon{} 
\def\YellowOrange{} 
\def\Orange{} 
\def\BurntOrange{} 
\def\Bittersweet{} 
\def\RedOrange{} 
\def\Mahogany{} 
\def\Maroon{} 
\def\BrickRed{} 
\def\Red{} 
\def\OrangeRed{} 
\def\RubineRed{} 
\def\WildStrawberry{} 
\def\Salmon{} 
\def\CarnationPink{} 
\def\Magenta{} 
\def\VioletRed{} 
\def\Rhodamine{} 
\def\Mulberry{} 
\def\RedViolet{} 
\def\Fuchsia{} 
\def\Lavender{} 
\def\Thistle{} 
\def\Orchid{} 
\def\DarkOrchid{} 
\def\Purple{} 
\def\Plum{} 
\def\Violet{} 
\def\RoyalPurple{} 
\def\BlueViolet{} 
\def\Periwinkle{} 
\def\CadetBlue{} 
\def\CornflowerBlue{} 
\def\MidnightBlue{} 
\def\NavyBlue{} 
\def\RoyalBlue{} 
\def\Blue{} 
\def\Cerulean{} 
\def\Cyan{} 
\def\ProcessBlue{} 
\def\SkyBlue{} 
\def\Turquoise{} 
\def\TealBlue{} 
\def\Aquamarine{} 
\def\BlueGreen{} 
\def\Emerald{} 
\def\JungleGreen{} 
\def\SeaGreen{} 
\def\Green{} 
\def\ForestGreen{} 
\def\PineGreen{} 
\def\LimeGreen{} 
\def\YellowGreen{} 
\def\SpringGreen{} 
\def\OliveGreen{} 
\def\RawSienna {} 
\def\Sepia{} 
\def\Brown{} 
\def\Tan{} 
\def\Gray{} 
\def\Black{} 
\def\White{} 

\title{Two-fermion relativistic bound states in Light-Front Dynamics}

\author{M. Mangin-Brinet\thanks{Now at
D.P.N.C Universit\'e de Gen\`eve, 24 Quai Ansermet,
CH-1211 Geneva 4, Switzerland, e-mail: Mariane.Mangin-Brinet@cern.ch},
J. Carbonell\thanks{e-mail: carbonel@isn.in2p3.fr}}
\affiliation{Institut des Sciences Nucl\'{e}aires, 53, Av. des Martyrs, 38026 Grenoble, France}
\author{V.A. Karmanov\thanks{e-mail: karmanov@sci.lebedev.ru, karmanov@isn.in2p3.fr}}
\affiliation{Lebedev Physical Institute, Leninsky Pr. 53, 119991 Moscow, Russia}%

\date{\today}

\begin{abstract}
{In the Light-Front Dynamics, the wave function equations and
their numerical solutions, for two fermion bound systems, are
presented. Analytical expressions for the ladder one-boson
exchange interaction kernels corresponding to scalar,
pseudoscalar, pseudovector and vector exchanges are given.
Different couplings are analyzed separately and each of them is
found to exhibit special features. The results are compared with
the non relativistic solutions.}
\end{abstract}


\maketitle

\section{Introduction}

The  two-fermion system covers a huge number of applications
in atomic (e$^+$e$^-$), nuclear (NN, \={N}N) and subnuclear (q\={q}) physics.
The interest in using a relativistic
description for such systems  appeared in the early days of quantum mechanics
\cite{KGE,Dirac_PRSL_A117_28} and has constantly been pursued since by many authors.  This
interest has recently found a new \'elan due to the measurements performed at
Jefferson Laboratory \cite{JLAB_ed_98,JLAB_AQ2_A_99,JLAB_AQ2_C_99,JLAB_t20_00}
where simple nuclear systems have been -- and are being -- probed at momentum
transfers much larger than their constituent masses.
This experimental activity  motivated a consequent number of works on relativistic dynamics.
Extensive reviews on the past and recent deuteron results can be found in \cite{GVO,GG}.

Most of the approaches developed for describing relativistic two-body systems
are based on Bethe-Salpeter (BS) equation
\cite{BS_51,FT_PRD21_80,ZT_PRC22_80,VFT_PRC33_86,RT_PRC45_92,PA_PRC54_96,BBMST_02}
or three-dimensional reductions of it
\cite{LT_63,BBS_66,Gross,Gross_et_al,Wallace,HT_PRC42_90}.

An alternative approach is provided by the Light-Front Dynamics
(LFD). In its standard version,  following Dirac's classification
of relativistic theories \cite{Dirac_49}, the state vector is
defined on the $\sigma=z+t$ surface. Wave functions - defined as
the Fock components of the state vector - are the formal objects
of this theory and are directly comparable to their non
relativistic counterparts. LFD has been developed and used by many
authors \cite{ LS_AP_78, CCKP_PRC37_88,COESTER_92,
Fuda_90,FZ_PRC_95, Ji_94,Burkart_96,
BKT_NPB158_79,LB_PRD52_95_2,SB_PRD58_98,Bakker, Hiller,KP_91,
BPP_PR_98,GPW_PRD45_92,TP_NP90_00,Tritmann_IJMP_01,
FP_PRD64_01,FPZ_PRD_02,
Miller_PPNP_00,MM_PRC60_99,CMP_PRC61_00,CM_02, GHPSW_PRD47_93,
SFCS_PRC61_00, FSMS_90,GKS_94, Weber_91,CGNSS_95_1} and represents
a promising approach to non perturbative Hamiltonian Quantum Field
Theory, specially when dealing with composite relativistic
systems. The interested reader can be aware of the last advances
and more complete references set in the proceedings of the last
conferences devoted to the subject
\cite{LCM_Heid_00,LCM_Trento_01}.

The explicitly covariant version of Light-Front Dynamics (ECLFD) was initiated by one of the authors  in a
series of papers \cite{VAK_76,VAK_80,VAK_81}.
The state vector is there defined on a space-time hyperplane whose
equation is given by $\omega\cdot x=\sigma$, where $\omega$
is a four-vector determining the orientation of the light-front plane
and satisfies $\omega^2=0$.
This choice is not only a mathematical
{\it delicatesse} but a way to carry everywhere in the theory the $\omega$-dependence
in an explicit way.
It has several advantages, all related to the fact that $\omega$ is a four vector
with well defined transformation properties.
This approach provides explicitly covariant expressions
for the on shell amplitudes,
a property  which is often hidden in the standard formulation,
recovered by fixing the value $\omega=(1,0,0,-1)$.
This value is however associated to a particular reference frame
and it is not valid in any other one.
The formalism and some of its first applications
to few-body systems has been reviewed in \cite{CDKM_PR_98}.

Approximate light-front solutions for the NN system \cite{CK_NPA581_95,CK_NPA589_95}
were found in a perturbative way over the Bonn model wave functions
\cite{Bonn} and successfully applied to calculate the deuteron electromagnetic
form factors \cite{CK_EPJA_99} measured at  Jefferson Lab.
Latter applications to heavier nuclei \cite{Antonov_02,Gaidarov_02} have shown
the pertinence of this approach in describing high momentum components of
the NN correlation functions.

These successes stimulated a series of works aiming at
developing some formal problems of the theory
and to obtain exact solutions in the ladder approximation for systems of increasing complexity.
Results concerning bound states of two scalar particles
can be found in \cite{MC_PLB_00,MCK_Heid_00,KCM_Taiw_01,MC_Evora_01}.
We present in this paper the formalism and numerical
solutions describing bound two fermion systems interacting via the usual --
scalar, pseudoscalar, vector and pseudovector -- one-boson exchange (OBE) kernels.
Results are limited to $J=0$ and $J=1$ states.
Our main interest in this work is to study the solutions
of the LFD equations as they are provided by the OBE ladder sum with special interest
in their stability, their comparison to the non relativistic limits
 and the construction of non-zero angular momentum states.
For this purpose, we have studied each coupling separately
and the only physical system considered is positronium.
The first conclusions concerning the Yukawa model
have been published in \cite{MCK_PRD_01,KMC_Prague_01,KCM_LCM_01,MCK_LCM_01}
and a more detailed  derivation of equations and kernels can be found in \cite{MMB_These_01}.
This series of works is also being extended to the two-body scattering solutions
and to three-particle systems. The case of three-bosons interacting
via zero range forces was considered in \cite{3bosons}.
In references \cite{CMK_Varna_02,KC_Rila_02} the ensemble of these results
is briefly reviewed.

It worth mentioning preceding works on two-fermion system using
the LFD approach. In \cite{GHPSW_PRD47_93}, the relativistic
bound-state problem in the light-front Yukawa model was
considered. In \cite{GPW_PRD45_92,TP_NP90_00}, positronium and
heavy quarkonia calculations in discretized light cone
quantization  were carried out. The formalism was used in
\cite{FZ_PRC_95} to build one boson exchange kernels and to
calculate nucleon-nucleon phase shifts as well as deuteron
properties. Recent application to meson spectra can be found in
\cite{FP_PRD64_01,FPZ_PRD_02}. LFD was also applied in
\cite{MM_PRC60_99,CM_02} to describe the NN system and nuclear
matter equation of state.

The paper is organized as follow.
In section \ref{wf} we establish the structure and main properties of the explicitly covariant
Light-Front wave functions, the two-body equation and the OBE kernels.
In section \ref{angl} the problem of angular momentum $J$ is discussed and
states with $J=0,1$ are constructed.
In section \ref{J0} we derive the coupled equations for the
wave function components of states with angular momentum $J=0$.
The corresponding equations for $J=1$ states are derived in section \ref{J1}.
The non-relativistic limit and perturbative calculations are  discussed in section \ref{NR}.
In sections \ref{Res_S}, \ref{Res_Ps} and \ref{Res_V} we present the results
of numerical calculations.
In order to disentangle their different behaviors each coupling is separately analyzed.
Section \ref{concl} contains a summary of the results and the concluding remarks.

\section{Wave function, equation and kernels}\label{wf}

\begin{figure}[htbp]
\begin{center}
\mbox{\epsfxsize=7.cm\epsffile{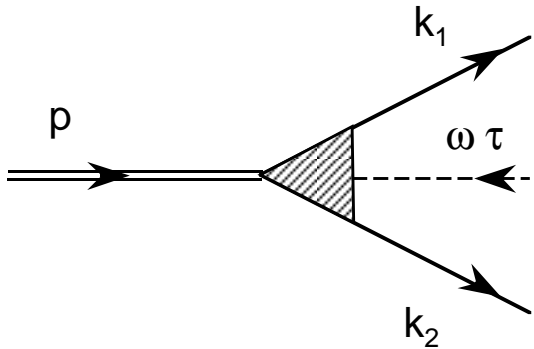}}
\caption{Graphical representation of the Light Front  two-body wave function. Dash line corresponds to the spurion (see text).\label{fwf}}
\end{center}
\end{figure}
Wave functions  we deal with are Fock components of the state vector
defined on the light-front plane $\omega\cd x=0$.
For a two-fermion system -- shown graphically in Fig. \ref{fwf} --  it reads:
\begin{equation}\label{wf1}
{\mit \Phi}_{\sigma_2\sigma_1}={\mit \Phi}_{\sigma_2\sigma_1}(k_1,k_2,p,\omega\tau)
\end{equation}
were $\sigma_i$ are the constituent angular momenta.
The general form of the wave function is obtained by constructing all possible
spin structures compatible with the quantum numbers of the state.
The four-vector $\omega$ enters in the wave function on the same
ground than the particles four-momenta,
giving rise to a number of structures larger than in non relativistic dynamics.
Each of them is mastered by a scalar function, denoted $f_i$ all through the paper,
which can be interpreted as a wave function component on the spin space.
The number N of such independent amplitudes simply follows from the
dimension of the spin matrix forming the two-fermion wave function with total momentum
$J$, i.e. $N={1\over2}(2J+1)(2\sigma_1+1)(2\sigma_2+1)$ with
a factor ${1\over2}$ to take into account the parity conservation.
In the case $\sigma_1=\sigma_2={1\over2}$, it gives $N$=2 amplitudes for $J$=0 states and $N$=6 for $J$=1.
These wave function components will be specified in the  coming sections.

Since the Fock-space component is, by construction,  the coefficient
of the state vector decomposition in the creation operators basis:
$a^{\dagger}_{\sigma_2}(\vec{k}_2)
a^{\dagger}_{\sigma_1}(\vec{k}_1)\vert 0\rangle$, the independent variables are
the three-dimensional vectors ($\vec{k}_1,\vec{k}_2$) and the particles energies
are expressed  through them.
Consequently all four-momenta are on corresponding mass
shells: $k_1^2=k_2^2=m^2$, $p^2=M^2$, $(\omega\tau)^2=0$ and satisfy the conservation law:
\begin{equation}\label{wf2}
k_1+k_2=p+\omega\tau.
\end{equation}
This equation  generalizes  the $(\perp,+)$-components
conservation  in the standard approach; the minus components are
not constrained. In the light-front coordinates with
$\omega=(1,0,0,-1)$,  the only non-zero component of $\omega$ is
$\omega_-=\omega_0-\omega_z=2$. The four-vector $\omega\tau$ just
incorporates the non-vanishing difference
$2\tau=k_{1-}+k_{2-}-p_{-}$. In this sense the ECLFD wave function
is off energy shell. Since the four-momentum $\omega\tau$ enters
in the wave function on equal ground with the particle momenta, we
associate it for convenience with a fictitious particle -- called
spurion -- showed in Fig. \ref{fwf} by a dash line. We would like
to emphasize however that the Fock space basis does not contain
for all that any additional and unphysical degree of freedom. By
spurion, we mean only the difference -- proportional to $\omega$
-- between non-conserved particle four-momenta in the
off-energy-shell states.

It is convenient to introduce other kinematical variables, constructed from the initial
four-momenta as follows:
\begin{eqnarray}\label{cdm}
\vec{k}&=& L^{-1}({\cal P})\vec{k}_1 = \vec{k}_1 -\frac{\vec{\cal P}}{\sqrt{{\cal P}^2}}
\left[k_{10}- \frac{\vec{k}_1\cd\vec{{\cal P}}}{\sqrt{{\cal P}^2}+{\cal P}_0}\right]\cr
\vec{n}&=& \frac{L^{-1}({\cal P})\vec{\omega}}{\mid L^{-1}({\cal P}) \vec{\omega}\mid}
\end{eqnarray}
where ${\cal P}=p+\omega\tau$, and $L^{-1}({\cal P})$ 
results from 
the Lorentz boost into the reference system where $\vec{\cal P}=0$.
In these variables the wave function (\ref{wf1}) is represented as:
\begin{equation}\label{wf3}
{\mit \Phi}_{\sigma_2\sigma_1}
={\mit \Phi}_{\sigma_2\sigma_1}(\vec{k},\vec{n}).
\end{equation}
Under rotations and  Lorentz transformations of four-momenta
$k_1,k_2,p,\omega\tau$, variables ($\vec{k},\vec{n}$) are only rotated,
so the three-dimensional parametrization (\ref{wf3}) is also explicitly covariant.
In practice, instead of the formal transformations (\ref{cdm}),
it is enough to consider the wave function and the equation in the
c.m. system where $\vec{\cal P}=\vec{k}_1+\vec{k}_2=0$ and set $\vec{k}_1=\vec{k}$, $\vec{k}_2=-\vec{k}$,
$\vec{\omega}=\vec{n}|\vec{\omega}|$.
Because of covariance, the result is the same as after transformation
(\ref{cdm}). Since $\vec{\omega}$ determines only the orientation of the
light-front plane, the modulus $|\vec{\omega}|$ disappears from the wave
functions and amplitudes. Note that in
the c.m. system,  the  momentum $\vec{p}$ is not zero: $\vec{p}=-\vec{\omega}\tau$.

The light-front graph techniques is a covariant
generalization of the old fashioned perturbation theory. The
latter was developed by Kadyshevsky \cite{kadysh} and adapted to
the explicitly covariant version  in \cite{VAK_76,CDKM_PR_98}.

\begin{figure}[hbtp]
\centerline{\epsfbox{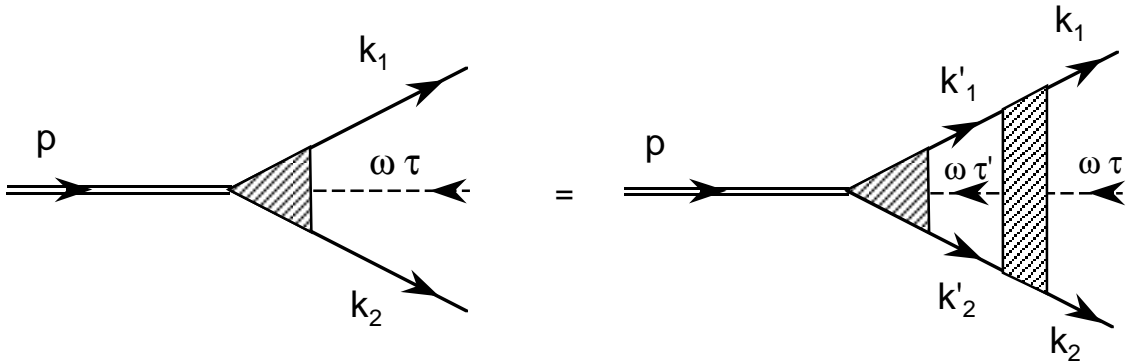}}
\caption{Equation for the two-body wave function.\label{feq}}
\end{figure}

The equation for the wave function is shown graphically in Fig.\ref{feq}.
It is the  projection on the two-body sector of the general mass equation $P^2\phi=M^2\phi$.
Its analytical form is obtained by applying the rules of the graph techniques
to the diagrams in Fig. \ref{feq}. In  variables (\ref{cdm}) this equations reads:
\begin{equation}\label{eq5d}
\left[4(\vec{k}\,^2+m^2)-M^2\right]{\mit \Phi}_{\sigma_2\sigma_1}
(\vec{k},\vec{n})=-\frac{m^2}{2\pi^3} \int \sum_{\sigma'_1\sigma'_2}
K_{\sigma_2\sigma_1}^{\sigma'_2\sigma'_1}(\vec{k},\vec{k}\,',\vec{n},M^2)
{\mit \Phi}_{\sigma'_2\sigma'_1}(\vec{k}\,',\vec{n})\frac{d^3k'}{\varepsilon_{k'}}\ .
\end{equation}
where $K_{\sigma_2\sigma_1}^{\sigma'_2\sigma'_1}(\vec{k},\vec{k}\,',\vec{n},M^2)$
is the interaction kernel.
We detail in what follows the LFD one-boson exchange  kernels corresponding to the
interaction Lagrangians:
\begin{itemize}
\item[({\it i})] Scalar (S):
\begin{equation}\label{4.1b}
{\cal L}^{int} = g_s\ \bar{\psi} \psi \phi^{(s)}
\end{equation}
\item[({\it ii})] Pseudoscalar  (PS):
\begin{equation}\label{4.1a}
{\cal L}^{int} = i\ g_{ps}\ \bar{\psi} \gamma_5 \psi\ \phi^{(ps)}
\end{equation}
\item[({\it iii})] Pseudovector (PV):
\begin{equation}\label{input1}
{\cal L}^{int} = -\frac{f_{pv}}{2m}\
\bar{\psi} \gamma^{\mu} \gamma_5\ \psi\ \partial_{\mu}\ \phi^{(ps)}
\end{equation}
\item[({\it iv})] Vector (V):
\begin{equation}\label{4.1c}
{\cal L}^{int}=\bar{\psi}[g_v\gamma^{\mu}\phi^{(v)}_{\mu}+{f_t\over4m}
\sigma^{\mu\nu}(\partial_{\mu}\phi^{(v)}_{\nu}-\partial_{\nu}\phi^{(v)}_{\mu})]\psi
\end{equation}
with
$$\sigma^{\alpha'\alpha}=\frac{i}{2}(\gamma^{\alpha'}\gamma^{\alpha}-\gamma^{\alpha}\gamma^{\alpha'}).$$
\end{itemize}
\begin{figure}[htbp]
\centerline{\epsfbox{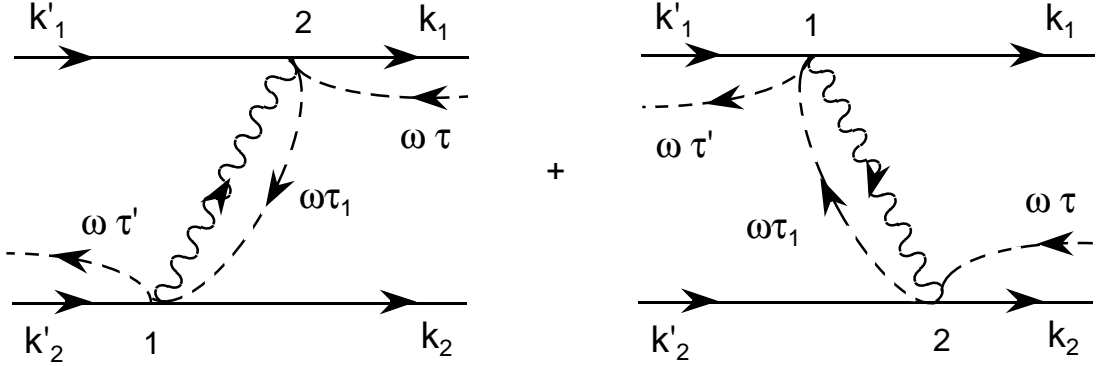}}
\caption{One boson exchange kernel.\label{fkern}}
\end{figure}
The LFD ladder kernels have two contributions
corresponding to the two time-ordered diagrams (in the light-front time)
shown in Fig. \ref{fkern}. For S, PS and PV couplings they have the structure:
\begin{eqnarray}\label{4.2}
&&K^{\sigma'_2\sigma'_1}_{\sigma_2\sigma_1}
(k_1,k_2,\omega\tau; k'_1,k'_2,\omega\tau')
=-\frac{1}{4m^2}
  \left[\bar{u}^{\sigma_2}(k_2)O_2 u^{\sigma'_2}(k'_2)\right]
\;\left[\bar{u}^{\sigma_1}(k_1)O_1 u^{\sigma'_1}(k'_1)\right]\qquad \qquad \\
&\times&
\left\{\frac{\theta\left(\omega\cd (k'_1 - k_1)\right)}{\mu^2-
(k'_1-k_1)^2 + 2\tau'\omega\cd (k'_1-k_1)} \right.
+\left. \frac{\theta\left(\omega\cd (k_1-k'_1)\right)} {\mu^2 -
(k_1-k'_1)^2 + 2\tau\omega\cd (k_1-k'_1)}\right\}.\nonumber
\end{eqnarray}
For scalar exchange $$O_1=O_2=g_s,$$ for pseudoscalar
$$O_1=O_2=i\gamma_5g_{ps}$$ and for pseudovector
$$O_1=\left\{\begin{array}{ll}
\left(1-\frac{\hat{\omega}\tau}{2m}\right)i\gamma_5f_{pv},& \mbox{if} \quad \omega\cd (k_1 - k'_1)>0\\
\left(1+\frac{\hat{\omega}\tau'}{2m}\right)i\gamma_5f_{pv},& \mbox{if} \quad \omega\cd (k_1 - k'_1)<0
\end{array}\right.$$
$$O_2=\left\{\begin{array}{ll}
\left(1+\frac{\hat{\omega}\tau'}{2m}\right)i\gamma_5f_{pv},
& \mbox{if} \quad \omega\cd (k_1 - k'_1)>0\\
\left(1-\frac{\hat{\omega}\tau}{2m}\right)i\gamma_5f_{pv},
& \mbox{if} \quad \omega\cd (k_1 - k'_1)<0
\end{array}\right.$$
with:
\[\tau= \frac{4\varepsilon_k^2-M^2}{2\omega\cd p},\quad \tau'=\frac{4\varepsilon_{k'}^2-M^2}{2\omega\cd p}. \]

For values $\tau,\tau'\neq 0$ the kernels are off energy shell. In
this case the pseudoscalar and pseudovector kernels differ from
each other but coincide on energy shell ($\tau=\tau'=0$).

We use the notation $\hat{\omega}=\omega_\mu\gamma^\mu$. Writing
the propagators in the center of mass variables, (\ref{4.2}) gets
the simpler form:
\begin{equation}\label{eq6}
K^{\sigma'_2\sigma'_1}_{\sigma_2\sigma_1}
=-\frac{1}{4m^2}\frac{1}{Q^2+\mu^2}
\left[\bar{u}_{\sigma_2}(k_2)O_2 u_{\sigma'_2}(k'_2)\right]\,
\left[\bar{u}_{\sigma_1}(k_1)O_1 u_{\sigma'_1}(k'_1)\right],
\end{equation}
with
\begin{eqnarray}\label{Q2}
Q^2&=& (\vec{k}-\vec{k}')^2
-(\vec{k}\cd\vec{n})(\vec{k}'\cd\vec{n})\frac{(\varepsilon_k -\varepsilon_{k'})^2}{\varepsilon_k\varepsilon_{k'}}
+\left(\varepsilon_{k}^2+\varepsilon_{k'}^2-\frac{1}{2}M^2\right)
\left|\frac{\vec{k}\cd\vec{n}}{\varepsilon_{k}}-\frac{\vec{k}'\cd\vec{n}}{\varepsilon_{k'}}\right|
\end{eqnarray}

The kernel for the vector coupling is given by a contraction of
terms similar to (\ref{eq6}) with the tensor structures
$L_{\alpha\beta}$. It reads
\begin{equation}\label{v3}
K^{\sigma'_2\sigma'_1}_{\sigma_2\sigma_1}
=-\frac{1}{4m^2}\frac{1}{\mu^2+Q^2} L_{\alpha\beta}
\left[\bar{u}(k_1)O_1^{\alpha}
u(k'_1)\right]\;\left[\bar{u}(k_2)O_2^{\beta} u(k'_2)\right],
\end{equation}
with
\begin{equation}\label{v4}
L_{\alpha\beta}=\left\{
\begin{array}{ll}
-g_{\alpha\beta}+
{1\over\mu^2}(k_1-k'_1-\omega\tau)_{\alpha}(k'_2-k_2-\omega\tau')_{\beta},\quad
&\mbox{if}
 \quad \omega\cd(k_1-k'_1)>0\\
-g_{\alpha\beta}+{1\over\mu^2}(k'_1-k_1-\omega\tau')_{\alpha}
(k_2-k'_2-\omega\tau)_{\beta},\quad &\mbox{if} \quad
\omega\cd(k_1-k'_1)<0
\end{array}
\right.
\end{equation}
and  vertex operators
\begin{equation}\label{v5}
O_1^{\alpha}=\left\{\begin{array}{ll}g_v\gamma^{\alpha}+
\frac{\displaystyle{f_t}}{\displaystyle{2m}}\sigma^{\alpha'\alpha}(-i)
(k_1-k'_1-\omega\tau)_{\alpha'},&\mbox{if}\quad \omega\cd(k_1-k'_1)>0\\
g_v\gamma^{\alpha}+\frac{\displaystyle{f_t}}{\displaystyle{2m}}\sigma^{\alpha'\alpha}(i)
(k'_1-k_1-\omega\tau')_{\alpha'},&\mbox{if}\quad
\omega\cd(k_1-k'_1)<0\end{array}\right.
\end{equation}
\begin{equation}\label{v6}
O_2^{\beta}=\left\{
\begin{array}{ll}
g_v\gamma^{\beta}+\frac{\displaystyle{f_t}}{\displaystyle{2m}}\sigma^{\beta'\beta}(i)
(k'_2-k_2-\omega\tau')_{\beta'},
&\mbox{if}\quad \omega\cd(k_1-k'_1)>0\\
g_v\gamma^{\beta}+\frac{\displaystyle{f_t}}{\displaystyle{2m}}\sigma^{\beta'\beta}(-i)
(k_2-k'_2-\omega\tau)_{\beta'},&\mbox{if}\quad
\omega\cd(k_1-k'_1)<0\end{array}\right.
\end{equation}

Hereafter we will not take into account the tensor coupling, that is we put
$f_t=0$ and $O_1^{\alpha}=O_2^{\alpha}=g_v\gamma^{\alpha}$. In this case, vector kernel (\ref{v3}) simplifies into:
\begin{eqnarray}\label{v3a}
K^{\sigma'_2\sigma'_1}_{\sigma_2\sigma_1}
&=&\frac{g_v^2}{4m^2}\frac{1}{\mu^2+Q^2}\left\{
\left[\bar{u}(k_1)\gamma^{\alpha}
u(k'_1)\right]\;\left[\bar{u}(k_2)\gamma_\alpha
u(k'_2)\right] - \frac{\tau\tau'}{\mu^2}\left[\bar{u}(k_1)\hat{\omega}
u(k'_1)\right]\;\left[\bar{u}(k_2)\hat{\omega} u(k'_2)\right]\right\}
\end{eqnarray}

In  the $\mu=0$ case, e.g. one-photon or one-gluon exchange
kernels, the $L_{\alpha\beta}$ expressions depend on the gauge.
Using the Feynman gauge, one has
$L_{\alpha\beta}=-g_{\alpha\beta}$, i.e. the $\mu$-dependent terms
on (\ref{v4}) and (\ref{v3a}) are simply dropped out.

It will be often necessary to regularize the LFD kernels by
means of vertex form factors. Unless explicit mention of the
contrary we will take the form used in the Bonn model \cite{Bonn},
i.e.
\begin{equation}\label{BFF}
F(Q^2) =\left( { \Lambda^2-\mu^2\over \Lambda^2+ Q^2}\right)^n
\end{equation}
where $\Lambda$ and $n$ are parameters whose values depend on the coupling.
Form factors appear in the kernels multiplying each of the vertex operators $O_i$.
In the non relativistic limit, $Q^2\approx(\vec{k}-\vec{k}')^2$ and $F$ is local in configuration space.
This locality is however broken from the very beginning in LFD due to the $\vec{n}$-dependent terms on $Q^2$.

\section{Angular momentum}\label{angl}

In LFD the construction of states with definite angular momentum is a delicate problem.
Working in the explicitly covariant version,
we have developed a method to overcome this difficulty. It will be explained in this section.
In contrast to the equal-time approach,
the LFD generators $J_{\rho\gamma}=J^0_{\rho\gamma}+J^{int}_{\rho\gamma}$
of four-dimensional rotations are not kinematical, but contain interaction
in $J^{int}_{\rho\gamma}$. The
interaction also enters in the angular momentum operator, i.e. the Pauli-Lubansky vector:
\begin{equation}\label{eqf1b}
S_{\mu}=\frac{1}{2}\varepsilon_{\mu\nu\rho\gamma}P^{\nu}J^{\rho\gamma}.
\end{equation}
Like the action of the Hamiltonian on the Schr\"odinger wave function is expressed through the time derivative
\[ H^{int}\, \phi= i\partial_t\phi \ ,\]
the action of $J^{int}_{\rho\gamma}$ on the LFD state vector is expressed
through derivatives with respect to the four-vector $\omega$ \cite{karm82}:
\begin{equation}\label{kt12}
J^{int}_{\mu\nu}\, \phi(\omega)= L_{\mu\nu}(\omega)\phi(\omega)\ ,
\end{equation}
where:
\begin{equation}\label{kt13}
L_{\mu\nu}(\omega) =i\left(\omega_{\mu}
\frac{\partial}{\partial\omega^{\nu}} -\omega_{\nu}
\frac{\partial}{\partial\omega^{\mu}}\right)\ .
\end{equation}

Equation (\ref{kt12}) is called {\it angular condition} and can also be written in the form:
\begin{equation}\label{kt13p}
S_{\mu}\, \phi(\omega) = W_{\mu}\, \phi(\omega)
\end{equation}
with
\begin{equation}\label{kt19}
W_{\mu}= \frac{1}{2}\epsilon_{\mu\nu\rho\gamma}P^{\nu}M^{\rho\gamma}
\end{equation}
and
\[M_{\mu\nu} =J^0_{\mu\nu} +L_{\mu\nu}(\omega).\]
$W_{\mu}$ is a kinematical Pauli-Lubansky vector.
As far as the angular condition is satisfied, the dynamical Pauli-Lubansky
vector $S_{\mu}$  can be replaced by the kinematical one $W_{\mu}$.
The great benefit of doing so is that the problem of constructing
angular momentum states with operator (\ref{kt19}) becomes purely kinematical.
In practice one rather prefers to start
constructing  states with definite angular momentum using $W_{\mu}$,
and then take into account the restriction imposed by the angular condition (\ref{kt12}).

\bigskip
It is worth noticing that without this condition, there is an ambiguity in defining
the state vector with given angular momentum. This can be seen by introducing the operator:
\begin{equation}\label{kt30}
\hat{A}^2 =\left(\frac{W\cd\omega}{P\cd\omega}\right)^2.
\end{equation}
It commutes with $P_{\mu}$ and $W_{\mu}$ and -- taking $A^2$ instead of $A$ -- with the parity operator.
The state vector is then characterized not only by its mass $M^2$, momentum $p$,
angular momentum $J$ -- defined by means of (\ref{kt19}) --
and parity $\pi$ but also by $a$, square root of the $A^2$ eigenvalue:
\begin{equation}\label{A2}
\hat{A}^2\ \phi^{(a)} =a^2\ \phi^{(a)}\ .
\end{equation}
For a total angular momentum  $J$ there are $J+1$ eigenstates $\phi^{(a)}$.
In principle one could imagine
any of these eigenstates to be an acceptable solution.
It turns out however that, except for $J=0$, none of these
eigenstates can satisfy the angular condition (\ref{kt13p}).
Indeed if $\phi(\omega)$ is an eigenstate of $A^2$,
the right hand side of (\ref{kt13p}) -- $W_{\mu}\phi(\omega)$ -- is still an eigenstate of $A^2$
whereas this is not possible in its left hand side  -- $S_{\mu}\phi(\omega)$ --
due to the non zero commutator $[S_{\mu},A^2]\neq0$.
What is then the state vector?

A solution of the angular condition --  the only
remaining equation to be fulfilled --
is therefore provided by a linear combination of different eigenstates $\phi^{(a)}$:
\begin{equation}\label{ELC}
\phi =\sum_{a=0}^{J} c_a \,\phi^{(a)} .
\end{equation}
The coefficients $c_a$ can in principle be determined by inserting
 (\ref{ELC}) into (\ref{kt12}) or (\ref{kt13p}).

\bigskip
We would like to emphasize this result,
which is, to our opinion, an important issue of Light-Front Dynamics.
It tells us that the state vector is necessarily a superposition of different $A^2$ eigenstates.
This conclusion does not depend on the approximation
resulting from  any eventual Fock-space truncation.

In an exact solution of the problem, i.e. with the generators
satisfying the Poincar\'e algebra, the eigenstates $\phi^{(a)}$
are degenerate in mass and the superposition  (\ref{ELC}) is
furthermore a solution of the mass equation (\ref{eq5d}). Indeed,
as already noticed, $S_{\mu}\phi^{(a)}$ is not an eigenstate of
$A^2$ but a superposition of different $A^2$ eigenstates. On the
other hand, the commutation relation  $[S_{\mu},P_{\nu}]=0$
implies $S_{\mu}\phi^{(a)}$ to have the same mass than
$\phi^{(a)}$. This is possible only if the masses of different
states $\phi^{(a)}$ are equal.

Due to the Fock-space truncation, or to some other kind of approximation,
the Poincar\'e algebra is violated.
The eigenstates $\phi^{(a)}$ are no longer degenerate
and the solution (\ref{ELC}), built with eigenstates of
different mass, cannot satisfy  equation (\ref{eq5d}).
However, while this equation
is an approximate one, the form (\ref{ELC}) for the state vector remains valid.
Each term in (\ref{eqf1a}) is an exact solution of the truncated mass equation
(\ref{eq5d}) with eigenvalue $M^2_a$.
Their superposition satisfies no any mass equation
but has the proper form  of the non-truncated hamiltonian problem.
The corresponding mass squared -- at the same level of approximation -- is given by:
\begin{equation}\label{M}
 M^2= \sum_{a=0}^{J} c_a^2 M_a^2
\end{equation}

The ensemble $(\phi,M)$  obtained that way, constitutes the solution of the problem
compatible with the degree of approximation considered.

\bigskip
This formalism is translated to $J=1$ states in the two-body sector as follows.
The interaction kernel $K(\vec{k}\,',\vec{k},\vec{n},M)$ depends on scalar
products of  vectors $\vec{k}\,',\vec{k},\vec{n}$ and also on scalar
products with Pauli matrices: $\vec{k}\cd \vec{\sigma}$, $\vec{k}\,'\cd \vec{\sigma}$,
$\vec{n}\cd \vec{\sigma}$.
Therefore the total angular momentum operator constructed as
\begin{equation}\label{ac3}
\vec{J} = -i [\vec{k}\times\partial_{\vec{k}}\,]
          -i [\vec{n}\times\partial_{\vec{n}}\,] + \vec{s}_1+\vec{s}_2\ ,
\end{equation}
commutes with the kernel ($\vec{s}_{1,2}$ are the fermion spin operators).
In the c.m. system this operator is proportional
to the  kinematical Pauli-Lubansky vector $\vec{W}$
given in (\ref{kt19}). The solutions of equation (\ref{eq5d}) correspond to
definite $J$ and $J_z$ eigenvalues  of the operators $\vec{J}^2,J_z$.

Since $A^2$ is applied to states with definite $p$,  it has the form:
\begin{equation}\label{ac6}
A^2=(\vec{n}\cd\vec{J})^2.
\end{equation}
$A^2$ commutes with the kernel $K$ since $\vec{J}$ commutes with
$K$ and $\vec{n}$ is a parameter. It commutes also with $\vec{J}$
since $A$ is a scalar. Thus, like in the case of a full state
vector (\ref{A2}), the truncated solutions in the two-body sector
are also labelled by $a$:
\begin{equation}\label{eqf1}
A^2\;\vec{\psi}^{(a)}(\vec{k},\vec{n})=
a^2\;\vec{\psi}^{(a)}(\vec{k},\vec{n}).
\end{equation}
and the two-body wave function is a superposition of $A^2$
eigenstates $\vec{\psi}^{(a)}$ with different $a$ values:
\begin{equation}\label{eqf1a}
\vec{\psi}(\vec{k},\vec{n})= c_0\;\vec{\psi}^{(0)}(\vec{k},\vec{n})+c_1\;\vec{\psi}^{(1)}(\vec{k},\vec{n}).
\end{equation}

The mass equations determining the eigenstates $\vec{\psi}^{(a)}$ with
different $a$ are decoupled; in particular, the  $J=0$ state  is determined by one single equation.
We would like to comment here that the decoupling  into subsystems takes
place in any formulation of LFD, both in the explicitly covariant and in the standard one.
However, in the latter approach it looks as a {\it matter of art}, whereas in ECLFD
this splitting has  transparent reasons. For example, in
\cite{GHPSW_PRD47_93} the four equations system for the
wave function components with angular momentum projection $m=0$ was split, by a proper
transformation, in two subsystems with two equations each. In ECLFD
this corresponds to the $a=0$ eigenstate of
$J=0$ and $J=1$ states, each of them having two components.

Because the truncation of the  Fock space, the states $\vec{\psi}^{(a)}$ are not degenerate.
Their splitting was effectively calculated in case of scalar particles in
\cite{CMP_PRC61_00,MCK_Heid_00,KCM_Taiw_01} for $J=1,2$ as a function of the coupling constant.
It has been shown in \cite{CMP_PRC61_00} that this splitting
indeed decreased when the interaction kernel incorporates larger number of
particles in the intermediate states.
However, the number of states taken into account in any practical calculation
will be always very limited.
The splitting, though decreased, will remain, specially for strongly bound systems like
$q\bar{q}$  mesons.
The problem of determining the state vector at a given level of approximation is thus not solved by this way.
These are some of the reasons why, as explained before, our approach to deal with this problem
 follows a different philosophy.
Despite the non degeneracy of $\vec{\psi}^{(a)}$, we search the
physical two-body wave function in the form (\ref{eqf1a}), the
same  as for the full state vector (\ref{ELC}). The corresponding
mass squared is given by:
\begin{equation}\label{M2}
 M^2= c_0^2 M_0^2+ c_1^2 M_1^2
\end{equation}
where $M^2_a$ is the mass associated with $\vec{\psi}^{(a)}$.
The $M^2$ value thus obtained is always between $M_0^2$ and $M_1^2$, where would be the exact solution.

To determine in practice coefficients $c_a$, we use a method
proposed in \cite{MCK_Heid_00,KCM_Taiw_01,MMB_These_01}, without explicitly solving (\ref{kt12}).
It is based on the fact that, when the momentum $k\to 0$, the interaction part in
(\ref{kt12}) is irrelevant and the angular condition reads simply $L_{\mu\nu}\phi=0$.
Thus, in this limit, $\vec{\psi}$ does not depend on the
light-front direction $\vec{n}$ anymore.
Such a requirement unambiguously determines the coefficients of the superposition.
The method was applied to a model with scalar particles \cite{KCM_Taiw_01}
and found to give very accurate results.
The procedure will be detailed in section \ref{J1}
and illustrated by numerical calculations in section \ref{Res_S}.

\section{$J=0$ states}\label{J0}

The $J=0^+$ two-fermion wave function can be written in the form \cite{CK_NPA589_95,CDKM_PR_98}:
\begin{equation}\label{Phi_J0}
{\mit \Phi}_{\sigma_2\sigma_1}(k_1,k_2,p,\omega\tau)=
\sqrt{m}\overline{u}_{\sigma_2}(k_2)
\;\phi\; U_c \overline{u}_{\sigma_1}(k_1),
\end{equation}
where \begin{equation}\label{eq74}
u_{\sigma}(k)= \sqrt{\varepsilon_{k}+m} \left(\begin{array}{c}1\\
\frac{\displaystyle\vec{\sigma}\cd\vec{k}}{\displaystyle(\varepsilon_{k}+m)}
\end{array}\right) w_{\sigma},
\end{equation}
is the Dirac spinor
normalized to $\bar{u}_{\sigma}u_{\sigma'}=2m\delta_{\sigma\sigma'}$,
$w_{\sigma}$ the Pauli spinor normalized to
$w^{\dagger}_{\sigma} w_{\sigma'}=\delta_{\sigma\sigma'}$
and $\varepsilon_{k}=\sqrt{\vec{k}\,^2+m^2}$.
$U_c=\gamma^2 \gamma^0$ is the charge conjugation matrix.
In its turn, $\phi$ is  written as a superposition of two independent spin structures $S_i$
\begin{eqnarray}\label{eq1_2}
S_1&=&\frac{1}{2\sqrt{2}\varepsilon_k}\gamma_5\\\nonumber
S_2&=&\frac{\varepsilon_k}{2\sqrt{2} mk\sin\theta}
\left(\frac{2m}{\omega\cd p}\hat\omega-\frac{m^2}{\varepsilon_k^2}\right)\gamma_5
\end{eqnarray}
whose coefficients $f_{i}$, scalar functions depending on variables $(k,\cos\theta=\vec{n}\cd\vec{k}/k)$,
are the wave function components in the spin-space:
\begin{equation}\label{eq1_1}
\phi=f_1S_1 +  f_2 S_2,
\end{equation}

The existence of one additional component with respect to the non-relativistic theory
is due to the $\hat{\omega}=\omega_\mu\gamma^\mu$ term.
The number of independent amplitudes determining the wave function
is however the same, whatever will be the LFD  version used.
We have shown in a preceding work \cite{MCK_PRD_01} that the $J^{\pi}=0^+$ state
we are considering is strictly equivalent in the standard approach  to the $(1+,2-)$ one
\cite{GHPSW_PRD47_93} which is described also by two components $\Phi^{1+},\Phi^{2-}$.

In the reference system where $\vec{k}_1+\vec{k}_2=0$ the
wave function (\ref{Phi_J0}) takes the form:
\begin{equation}\label{eq00}
{\mit \Phi}_{\sigma_2\sigma_1}=\sqrt{m} w_{\sigma_2}^{\dagger}\psi(\vec{k},\vec{n})w_{\sigma_1}^{\dagger}
\end{equation}
with
\begin{equation}\label{psi_J0}
\psi(\vec{k},\vec{n})=\frac{1}{\sqrt{2}}
\left(f_1+ \frac{i\vec{\sigma}\cd [\vec{k}\times\vec{n}]}{k\,\sin\theta}f_2\right)\sigma_y,
\end{equation}
The definition of the components themselves
is to some extent arbitrary, as are arbitrary the choices of structures (\ref{eq1_2}).
Our choice (\ref{eq1_2}) is justified by the clear separation
of $\vec{n}$-independent and dependent terms it induces in the wave
function (\ref{psi_J0}).

The normalization condition reads:
\begin{eqnarray}\label{na5}
\frac{1}{(2\pi)^3}\sum_{\sigma_2\sigma_1}\int \left|{\mit
\Phi}_{\sigma_2\sigma_1}\right|^2 {d^3k\over\varepsilon_k}  &=&
{m\over (2\pi)^3}\int Tr\{\bar{\phi}(\hat{k}_2+m)\phi(\hat{k}_1-m)\}
{d^3k\over\varepsilon_k}\nonumber\\
&=&{m\over (2\pi)^3}\int Tr\{\psi^{\dag}(\vec{k},\vec{n})\psi(\vec{k},\vec{n})\}{d^3k\over\varepsilon_k}
={m\over (2\pi)^3}\int (f_1^2+ f_2^2) {d^3k\over\varepsilon_k}=1,
\end{eqnarray}
where we denote $\bar{\phi}=\gamma_0 \phi^\dagger \gamma_0$.
The spin structures $S_{i}$ introduced in (\ref{eq1_2})
are  orthonormalized relative to the trace:
\begin{equation}\label{na5_1}
Tr\{\bar{S}_i(\hat{k}_2+m)S_j(\hat{k}_1-m)\}=\delta_{ij},
\end{equation}
where $\bar{S}_i=\gamma_0 S_i^\dagger \gamma_0$, that is:
\begin{eqnarray}\label{eq1_3}
\bar{S}_1&=&-\frac{1}{2\sqrt{2}\varepsilon_k}\gamma_5\cr
\bar{S}_2&=&-\frac{\varepsilon_k}{2\sqrt{2} mk\sin\theta}\gamma_5\left(\frac{2m\hat{\omega}}{\omega\cd p}-\frac{m^2}{\varepsilon_k^2}\right)
\end{eqnarray}

Substituting in  (\ref{eq5d}) the wave function (\ref{Phi_J0}),
multiplying it on left by $u(k_2)$, on right by $u(k_1)$
and using relation $\sum_{\sigma}u^{\sigma}(k)\bar{u}^{\sigma}(k) =\hat{k}+m$, we find:
\begin{eqnarray}\label{eq8d}
&&[4(\vec{k}\,^2+m^2)-M^2] \;(\hat{k}_2+m)\phi(\hat{k}_1-m)\\
&&=-\frac{m^2}{2\pi^3} \int
\frac{1}{4m^2(Q^2+\mu^2)}
(\hat{k}_2+m) O_2(\hat{k}'_2+m)\phi'(\hat{k}'_1-m)
\tilde{O}_1 (\hat{k}_1-m) \frac{d^3k'}{\varepsilon_{k'}}\nonumber
\end{eqnarray}
with $\tilde{O}=U_c O^t U_c$. Replacing here $\phi$ by its
decomposition (\ref{eq1_1}), multiplying  equation (\ref{eq8d}) by
$\bar{S}_{i}$  and using the orthogonality relations
(\ref{na5_1}), we end up with a two-dimensional integral equations system for components $f_{i}$:
\begin{equation}\label{eq_J0}
[4(\vec{k}\,^2+m^2)-M^2]\; f_i(k,\theta)
=-\frac{m^2}{2\pi^3} \sum_{j=1,2}\int K_{ij}(k,\theta;k',\theta')f_j(k',\theta')\frac{d^3k'}{\varepsilon_{k'}},
\end{equation}
Its solution will directly provide the mass of the $J^{\pi}=0^+$  state.

Kernels $K_{ij}$ appearing in (\ref{eq_J0}) result from integrating kernels $\kappa_{ij}$
over the azimuthal angle $\varphi$:
\begin{eqnarray}\label{nz1}
K_{ij}&=&  {1\over m^2\varepsilon_k\varepsilon_{k'}}\int_0^{2\pi}{\kappa_{ij} \over Q^2+\mu^2}\; {d\varphi'\over2\pi},
\end{eqnarray}
with $Q^2$ defined in (\ref{Q2}).
For  S, PS and PV couplings $\tilde{O}_1=U_c O^t_1 U_c=O_1$ and $\kappa_{ij}$ are given by
\begin{eqnarray}\label{kappa_S_J0}
\kappa_{ij}&=&{1\over4}\varepsilon_k \varepsilon_k'
Tr\left[\bar{S}_i(\hat{k}_2+m)O_2(\hat{k'}_2+m)S'_j
(\hat{k'}_1-m)O_1(\hat{k}_1-m)\right]
\end{eqnarray}
 We denote by $S'_j$ the quantities (\ref{eq1_2}) as a function of primed arguments.  For vector exchange:
\begin{equation}\label{kappa_V_J0}
\kappa_{ij}=-\frac{1}{4}\varepsilon_k \varepsilon_k'L_{\alpha\beta}
Tr\left[\bar{S}_i(\hat{k}_2+m)O_2^{\alpha}(\hat{k'}_2+m)S'_j
(\hat{k'}_1-m)O_1^{\beta}(\hat{k}_1-m)\right],
\end{equation}
Tensor $L_{\alpha\beta}$ is defined in (\ref{v4}) and we have
taken into account that for V coupling $\tilde{O}_1=U_c O^t_1 U_c=-O_1$.
The analytic expressions of $\kappa_{ij}$ for
S, PS, PV and V exchanges are given in appendix \ref{ap1}.

One would remark that we have kept, for convenience,  a three-dimensional
volume element in equation (\ref{eq_J0}) despite the fact that kernels $K_{ij}$
as well as amplitudes $f_j$ are independent of variable $\varphi'$.

\section{$J=1$ states}\label{J1}

In a similar way than in (\ref{Phi_J0}),
the $J=1^+$ two-fermion wave function can be written in the form \cite{VAK_81,CK_NPA581_95}:

\begin{equation}\label{nz1_1}
{\mit\Phi}^{\lambda}_{\sigma_2\sigma_1}(k_1,k_2,p,\omega \tau)=
\sqrt{m}\; e_{\mu}(p,\lambda)\;
\bar{u}^{\sigma_2}(k_2) \; \phi^{\mu}  \; U_c\bar{u}^{\sigma_1}(k_1)\ ,
\end{equation}
where $e_{\mu}(p,\lambda)$ is the polarization vector.
$\phi^{\mu}$ develops over the six spin structures
\begin{eqnarray}\label{eq12d}
&&S_{1\mu}=\frac{(k_1- k_2)^{\mu}}{2m^2},\quad
S_{2\mu}=\frac{1}{m}\gamma^{\mu},\quad
S_{3\mu}=\frac{\omega^{\mu}}{\omega\cd p},\quad
S_{4\mu}=\frac{(k_1-k_2)^{\mu}\hat{\omega}}{2m\omega\cd p},\nonumber\\
&&S_{5\mu}=-\frac{i}{m^2\omega\cd p}\gamma_5
\epsilon^{\mu\nu\rho\gamma}k_{1\nu}k_{2\rho}\omega_{\gamma},\quad
S_{6\mu}=\frac{m\omega^{\mu}\hat{\omega}}{(\omega\cd p)^2}
\end{eqnarray}
with components $\varphi_{i}$, invariant functions depending on the same scalar variables
than for $J=0$,
\begin{equation}\label{nz2}
\phi^{\mu}=
\varphi_1S_{1\mu}+\varphi_2S_{2\mu}+\varphi_3S_{3\mu}+\varphi_4S_{4\mu}+\varphi_5S_{5\mu}+\varphi_6S_{6\mu}\ .
\end{equation}

In the reference system $\vec{k}_1+\vec{k}_2=0$ this wave function takes the form:
\begin{equation}\label{nz7}
\vec{\mit\Psi}_{\sigma_2\sigma_1}(\vec{k},\vec{n}) =
\sqrt{m}w^\dagger_{\sigma_2} \vec\psi(\vec{k},\vec{n})\sigma_yw^\dagger_{\sigma_1}\ ,
\end{equation}
with
\begin{eqnarray}\label{nz8}
\vec{\psi}(\vec{k},\vec{n})
&=& f_1\frac{1}{\sqrt{2}}\vec{\sigma} +
    f_2\frac{1}{2}(\frac{3\vec{k}(\vec{k}\cd\vec{\sigma})}{\vec{k}^2}-\vec{\sigma})
    + f_3\frac{1}{2}(3\vec{n}(\vec{n}\cd\vec{\sigma})-\vec{\sigma}) \nonumber \\
&+& f_4\frac{1}{2k}(3\vec{k}(\vec{n}\cd\vec{\sigma}) +
   3\vec{n}(\vec{k}\cd\vec{\sigma}) - 2(\vec{k}\cd\vec{n})\vec{\sigma})
+f_5\sqrt{\frac{3}{2}}\frac{i}{k}[\vec{k}\times\vec{n}]
+f_6\frac{\sqrt{3}}{2k}(\vec{n}(\vec{k}\cd\vec{\sigma})-\vec{k}(\vec{n}\cd\vec{\sigma}))\ .
\end{eqnarray}
Contrary to the $J=0$ case,  components $f_{i}$ appearing in
(\ref{nz8}) are not the same than $\varphi_{i}$ from (\ref{nz2}).
Their relation is given in Appendix \ref{app3}.
Components $f_{3,4,5,6}$,
driving $\vec{n}$-dependent spin structures, are of relativistic
origin and are absent in a non relativistic approach.

\bigskip
As explained in section \ref{angl},  the system of equations
determining the six components $f_i$ is split in two subsystems,
corresponding to the eigenvalues $a=0,1$ of $\hat{A}^2$  (\ref{ac6}).
Like for the $J=0$ wave function, the $J=1,a=0$ eigenstate is
determined by two components whereas the remaining four correspond to $J=1,a=1$.
We would like to notice that the total number of components as well as the dimension of
decoupled subsystems (2+4) coincides with what is found in the standard approach \cite{GHPSW_PRD47_93}.

\bigskip
The components determining the eigenstates $\vec{\psi}^{(a)}$ of $A^2$
will be respectively denote  by $g_{i=1,2}^{(a=0)}$ and $g_{i=1,2,3,4}^{(a=1)}$.
They are indeed different from the $f_i$ appearing in
the wave function (\ref{nz8}) though  $g$'s
 fully determine $f$'s  by linear combinations.
In view of constructing the superposition (\ref{eqf1a})
it is convenient to represent the eigenfunctions $\vec{\psi}^{(a)}$ in the form (\ref{nz8}).
Only some of the six $f_i^{(a)}$ involved components will be independent --  two  for the
$a=0$ state and four for  $a=1$ -- but this way of doing will facilitate further analysis.

In the two coming subsections we will explicitly construct the eigenfunctions
$\vec{\psi}^{(a)}$ of the kinematical operator $A^2$,
obtain the corresponding mass equation (\ref{eq5d}) in terms of $g_{i}^{(a)}$ and
relate them with components $f_i^{(a)}$  defined in (\ref{nz8}).

\subsection{$a=0$}\label{a0}

One can check from equation (\ref{eqf1}) that $\vec{\psi}^{(0)}$ is parallel to
$\vec{n}$, i.e. it satisfies $\vec{\psi}^{(0)}=\vec{n}(\vec{n}\cd
\vec{\psi}^{(0)})$, and has the following general decomposition:
\begin{equation}\label{eq4a}
\vec{\psi}^{(0)}(\vec{k},\vec{n})=\sqrt{\frac{3}{2}}\left\{g^{(0)}_1{\vec{\sigma}\cd\vec{k}\over k}
+
g^{(0)}_2 \frac{\vec{\sigma}\cd(\vec{k}\cos\theta-k\vec{n})}{k\,\sin\theta}\right\}\vec{n},
\end{equation}
It can be written in the form  (\ref{nz8}) by defining the $f^{(0)}_i$ components
\begin{eqnarray}\label{ff0}
f^{(0)}_1&=&\frac{1}{\sqrt3}\cos\theta g^{(0)}_1-\frac{1}{\sqrt3}\sin\theta g^{(0)}_2\cr
f^{(0)}_2&=&0\cr
f^{(0)}_3&=&-\frac{\sqrt{2}}{\sqrt{3}\sin\theta}g^{(0)}_2\cr
f^{(0)}_4&=&\frac{1}{\sqrt{6}}g^{(0)}_1+\frac{1}{\sqrt{6}}\cot\theta g^{(0)}_2\cr
f^{(0)}_5&=&=0\cr
f^{(0)}_6&=&\frac{1}{\sqrt{2}}g^{(0)}_1+\frac{1}{\sqrt{2}}\cot\theta g^{(0)}_2.
\end{eqnarray}
that is four non-zero components,  with only two of them being independent.
It can also be represented in a four-dimensional form similar to (\ref{nz2})
\begin{equation}\label{nz2_0}
\phi^{(0)}_{\mu}= f^{(0)}_1 S^{(0)}_{1\mu}+ f^{(0)}_2 S^{(0)}_{2\mu}.
\end{equation}
by introducing the spin structures $S^{(0)}_{i\mu}$:
\begin{eqnarray}\label{nzf1}
S^{(0)}_{1\mu}&=&\frac{\sqrt{3}M}{2\sqrt{2}k}S_{3\mu},\\
S^{(0)}_{2\mu}&=&\frac{\sqrt{3}M}{m\sqrt{2}\sin\theta}
\left(\frac{m^2\cos\theta}{2\varepsilon_k k}S_{3\mu}+S_{6\mu}\right),
\end{eqnarray}
with $S_{i\mu}$ defined in (\ref{eq12d}).

The normalization condition is
\begin{eqnarray}\label{na4}
&&\frac{1}{3(2\pi)^3}\sum_{\lambda\sigma_2\sigma_1}\int
\left|{\mit\Phi}^{\lambda}_{\sigma_2\sigma_1}\right|^2
{d^3k\over\varepsilon_k}=
{m\over (2\pi)^3}\int \Pi^{\mu\nu}Tr\{\phi^{(0)}_\mu(\hat{k}_2+m)
\phi^{(0)}_\nu(\hat{k}_1-m)\}{d^3k\over\varepsilon_k} \\
&=&{m\over 3(2\pi)^3}\int Tr\{\vec{\psi}^{(0)\dag}(\vec{k},\vec{n})
\vec{\psi}^{(0)}(\vec{k},\vec{n})\}{d^3k\over\varepsilon_k}
={m\over (2\pi)^3}\int \left[(g^{(0)}_1)^2+ (g^{(0)}_2)^2\right]
{d^3k\over\varepsilon_k}=1.\nonumber
\end{eqnarray}
with
\begin{equation}\label{Pi}
\Pi^{\mu\nu}=\frac{1}{3}\sum_{\lambda}e^{\mu *}(p,\lambda)e^{\nu }(p,\lambda)=
\frac{1}{3}\left(\frac{p^{\mu}p^{\nu}}{M^2}-g^{\mu\nu}\right).
\end{equation}

The spin structures $S^{(0)}_{i\mu}$ are
orthonormalized relative to the trace operation in (\ref{na4}) (cf. eq. (\ref{na5_1})):
\begin{equation}\label{na5_2}
\Pi^{\mu\nu}Tr\{\bar{S}^{(0)}_{i\mu}(\hat{k}_2+m)S^{(0)}_{j\nu}(\hat{k}_1-m)\}=\delta_{ij}.
\end{equation}
Note that $\bar{S}^{(0)}_{i\mu}=\gamma_0S^{(0)\dagger}_{i\mu} \gamma_0=S^{(0)}_{i\mu}$.
Similarly to equation (\ref{eq8d}) we get:
\begin{eqnarray}\label{eq8}
&&\left[4(\vec{k}\,^2 + m^2)-M^2\right](\hat{k}_2+m)\phi^{(0)}_\mu(\hat{k}_1-m)\\
&&=-\frac{m^2}{2\pi^3} \int \frac{g^2}{4m^2(Q^2+\mu^2)}(\hat{k}_2+m) O_2(\hat{k}'_2+m)\phi'^{(0)}_\mu (\hat{k}'_1-m)
\tilde{O}_1 (\hat{k}_1-m) \frac{d^3 k'}{\varepsilon_{k'}}.\nonumber
\end{eqnarray}

In order to obtain the system of equations for components $g^{(0)}_{i}$, we  multiply equation (\ref{eq8}) by $\Pi^{\mu\nu}$
and $S^{(0)}_{i\nu}$. Taking the trace and using the orthogonality
condition (\ref{na5_2}) we obtain the system of equations:
\begin{equation}\label{eq_J1a0}
\left[4(\vec{k}\,^2 +m^2)-M^2\right] g^{(0)}_i(\vec{k},\vec{n})
=-\frac{m^2}{2\pi^3} \int \sum_{j=1}^2  K^{(0)}_{ij}(\vec{k},\vec{k}\,',\vec{n})g^{(0)}_j(\vec{k}\,',\vec{n})\frac{d^3k'}{\varepsilon_{k'}}.
\end{equation}
which provides the mass of the $J=1,a=0$ state.
They have the same form than (\ref{eq_J0}),
with kernels $K^{(0)}_{ij}$ given in terms of $\kappa^{(0)}_{ij}$ integrated over the azimuthal
angle $\varphi'$:
\begin{eqnarray}\label{eq10b}
K^{(0)}_{ij}&=& \frac{1}{m^2\varepsilon_k \varepsilon_{k'}}\int_0^{2\pi}{\kappa^{(0)}_{ij}\over Q^2+\mu^2}
{d\varphi'\over 2\pi}.
\end{eqnarray}
For S, PS and PV couplings they read
\begin{equation}\label{kappa_S_J1a0}
\kappa^{(0)}_{ij}={1\over4}\varepsilon_k \varepsilon_k'
\Pi^{\mu\nu} Tr\left[S^{(0)}_{i\mu}(\hat{k}_2+m)O_2
(\hat{k'}_2+m)\;S'^{(0)}_{j\nu}\;(\hat{k'}_1-m)\;\tilde{O}_1(\hat{k}_1-m)\right]
\end{equation}
where $S'^{(0)}_{j\nu}$ denotes (\ref{nzf1}) with primed arguments. For vector exchange
\begin{equation}\label{kappa_V_J1a0}
\kappa_{ij}^{(0)}=-\frac{1}{4}\varepsilon_k \varepsilon_k'
\Pi^{\nu\mu}L_{\alpha\beta}
Tr\left[S^{(0)}_{i\nu}(\hat{k}_2+m) O_2^{\alpha}(\hat{k}'_2+m)
{S'}^{(0)}_{j\mu}(\hat{k}'_1-m) O_1^{\beta} (\hat{k}_1-m)\right].
\end{equation}
Tensors $L_{\alpha\beta}$ and $\Pi^{\nu\mu}$ are defined in (\ref{v4}) and (\ref{Pi}).
The analytic expressions of $\kappa_{ij}^{(0)}$ for S, PS,
PV and V ($f_t=0$) exchanges are given in appendix \ref{ap1}.

\subsection{$a=1$}\label{a1}

It follows also from (\ref{eqf1}) that  $\vec{\psi}^{(1)}$, the $A^2$ eigenfunction corresponding to
$a=1$, is orthogonal to $\vec{n}$, i.e. satisfies $\vec{n}\cd \vec{\psi}^{(1)}=0$.
To fulfill this condition, it is convenient to introduce  two vectors
($\hat{\vec{k}}_\perp,\vec{\sigma}_\perp$) orthogonal to $\vec{n}$:
\begin{eqnarray*}
\hat{\vec{k}}_\perp &=& \frac{\hat{\vec{k}}-\cos\theta\vec{n}}{\sin\theta}\cr
\vec{\sigma}_\perp  &=& \vec{\sigma}-(\vec{n}\cd \vec{\sigma})\vec{n}.
\end{eqnarray*}
with  $\hat{\vec{k}}=\vec{k}/k$  and $\mid\hat{\vec{k}}_\perp\mid=1$.
Function $\vec{\psi}^{(1)}$ obtains then the decomposition, analogous of (\ref{eq4a}),
\begin{equation}\label{eq4b}
\vec{\psi}^{(1)}(\vec{k},\vec{n})=g^{(1)}_1\frac{\sqrt{3}}{2}\vec{\sigma}_\perp
+g^{(1)}_2\frac{\sqrt{3}}{2}\left(2\hat{\vec{k}}_\perp
(\hat{\vec{k}}_\perp \cd \vec{\sigma}_\perp)-\vec{\sigma}_\perp\right)
+g^{(1)}_3\sqrt{\frac{3}{2}}\hat{\vec{k}}_\perp (\vec{\sigma}\cd \vec{n})
+g^{(1)}_4\sqrt{\frac{3}{2}}i[\hat{\vec{k}}\times \vec{n}]
\end{equation}
in terms of the four scalar amplitudes $g^{(1)}_i$.
It can  also be represented in the form (\ref{nz8}) by defining components $f^{(1)}_i$
\begin{eqnarray}\label{eq4c}
f^{(1)}_1&=&\sqrt{\frac{2}{3}}g^{(1)}_1,\cr
f^{(1)}_2&=&\frac{2}{\sqrt{3}\sin^2\theta}g^{(1)}_2,\cr
f^{(1)}_3&=&-\frac{1}{\sqrt{3}}g^{(1)}_1
+ \frac{(1 + \cos^2\theta)}{\sqrt{3}\sin^2\theta}g^{(1)}_2
-   \frac{\sqrt{2}}{\sqrt{3}}\cot\theta g^{(1)}_3,\cr
f^{(1)}_4&=&-\frac{2\sqrt{3}\cos\theta}{3\sin^2\theta}g^{(1)}_2
+\frac{1}{\sqrt{6}\sin\theta}g^{(1)}_3,\cr
f^{(1)}_5&=&\frac{1}{\sin\theta}g^{(1)}_4,\cr
f^{(1)}_6&=&-\frac{1}{\sqrt{2}\sin\theta} g^{(1)}_3
\end{eqnarray}
and in the four-dimensional form $\phi^{(1)}_\mu$ similar to (\ref{nz7}):
\begin{equation}\label{nz2_1}
\phi^{(1)}_{\mu}=g^{(1)}_1 S^{(1)}_{1\mu}+ g^{(1)}_2 S^{(1)}_{2\mu}+
g^{(1)}_3 S^{(1)}_{3\mu}+g^{(1)}_4 S^{(1)}_{4\mu}.
\end{equation}

The four spin structures $S^{(1)}_{j\mu}$ are orthonormalized according to (\ref{na5_2}) and read:
\begin{equation}\label{nzf2}
S^{(1)}_{i\mu}=\sum_j h_{ij}S_{j\mu},\quad i=1,\ldots,4;\;j=1,\ldots,6,
\end{equation}
with $S_{j\mu}$ defined in (\ref{eq12d}) and $h_{ij}$ coefficients  given in appendix \ref{app3}.
The normalization condition in terms of $\phi^{(1)}_\mu$ and
$\vec{\psi}^{(1)}$ exactly coincides with (\ref{na4}).
In terms of  components $g^{(1)}_i$ it becomes:
\begin{equation}\label{nb4}
{m\over (2\pi)^3}\int \left[(g^{(1)}_1)^2+ (g^{(1)}_2)^2+
(g^{(1)}_3)^2+ (g^{(1)}_4)^2\right] {d^3k\over\varepsilon_k}=1.
\end{equation}

The system of equations for the scalar functions
$g^{(1)}_{i}$ is obtained similarly to (\ref{eq_J1a0}) and reads
\begin{equation}\label{eq_J1a1}
\left[4(\vec{k}\,^2 +m^2)-M^2\right] g^{(1)}_i(\vec{k},\vec{n})
=-\frac{m^2}{2\pi^3} \int \sum_{j=1}^4
K^{(1)}_{ij}(\vec{k},\vec{k}\,',\vec{n})
g^{(1)}_j(\vec{k}\,',\vec{n})\frac{d^3k'}{\varepsilon_{k'}}.
\end{equation}
It is the mass equation of the $J=1,a=1$ states.
Kernels $K^{(1)}_{ij}$ are calculated in a way similar than (\ref{eq10b}).
The corresponding $\kappa_{ij}^{(1)}$ are obtained with the replacement
$S^{(0)}_{i\mu}\rightarrow S^{(1)}_{i\mu}$ in (\ref{kappa_S_J1a0}) and  (\ref{kappa_V_J1a0}).
Their analytic expressions  for S and PS exchanges are given in appendix \ref{ap1}.

\subsection{Physical solution}\label{phys}

The solutions $\vec{\psi}^{(a)}$ constructed in the preceding
sections, although being exact eigenstates of the truncated Hamiltonian, are only auxiliary.
As explained in section \ref{angl}, the solution satisfying the angular condition (\ref{kt12})
is given by the superposition (\ref{eqf1a}) of states with different $a$.
The coefficients $c_a$ of the superposition can be obtained by solving the
angular condition in the truncated Fock space.
We will show in what follows that they can alternatively be determined
by imposing the independence of the wave function on the light-front vector $\vec{n}$ at $k=0$.

In order to do that, it is convenient to write down $\vec{\psi}^{(a)}$ in the form
(\ref{nz8}) with the components $f^{(a)}_i$ given by equations (\ref{ff0}) and (\ref{eq4c}).
Written in term of $f$'s the superposition (\ref{eqf1a}) reads
\begin{equation}\label{ph30}
f_i=c_0 f^{(0)}_i +c_1 f^{(1)}_i.
\end{equation}
The condition that $\vec{\psi}(\vec{k}=0,\vec{n})$  does not depend on $\vec{n}$ becomes:
\begin{eqnarray}
\partial_{\theta} f_{i}(k=0,\theta)&\equiv&0 \qquad i=1,2     \label{Cond1}\\
                  f_{j}(k=0,\theta)&\equiv&0 \qquad j=3,4,5,6.\label{Cond2}
\end{eqnarray}
Let us show that there exists two coefficients $c_{a}$, normalized to $c_0^2+c_1^2=1$, satisfying
the above six equations.
They are determined by the only values at $k=0$ of the first components $g^{(a)}_1$.

To this aim, we consider the behavior of  $f^{(a)}_i(k,z)$ in the $k\to 0$ limit.
The components in front of structures involving the unit vector $\hat{\vec{k}}$ are $f^{(a)}_{2,4,5,6}$.
By construction, they must vanish at $k=0$, i.e. satisfy:
\begin{equation}\label{cond}
f^{(a)}_{2,4,5,6}(k=0,\theta)\equiv0  \qquad a=0,1.
\end{equation}
Concerning $a=0$ states, this condition is trivially satisfied by $f^{(0)}_{2,5}$
since from (\ref{ff0}), they are identically zero whereas $f^{(0)}_{4,6}$ will satisfy (\ref{cond}) if:
\begin{eqnarray}\label{b_0}
g^{(0)}_1(k=0,\theta)&=&+b_0\cos\theta, \\
g^{(0)}_2(k=0,\theta)&=&-b_0\sin\theta,\nonumber
\end{eqnarray}
$b_0$ being {\it a priori} an arbitrary function of $\theta$
which later on will be shown to be constant.
The only components which are non zero at $k=0$ are
$f^{(0)}_{1,3}$. Inserting the values (\ref{b_0}) in (\ref{ff0}) we find:
\begin{eqnarray*}
f^{(0)}_1(0,\theta)&=&\frac{1}{\sqrt{3}}\cos\theta g^{(0)}_1(0,\theta)
-\frac{1}{\sqrt{3}}\sin\theta g^{(0)}_2(0,\theta)=\frac{1}{\sqrt{3}}b_0 \nonumber\\
f^{(0)}_3(0,\theta)&=&-\frac{\sqrt{2}}{\sqrt{3}\sin\theta} g^{(0)}_2(0,\theta)=\sqrt{\frac{2}{3}}b_0.
\end{eqnarray*}

Concerning  $a=1$ solutions, determined by four independent components $g^{(1)}_{i}$,
we see from  (\ref{eq4c}) that condition (\ref{cond}) implies  $g^{(1)}_{2,3,4}(0,\theta)\equiv0$.
The only non-vanishing component at $k=0$ is thus $g^{(1)}_1$ and we will denote by $b_1$ its value:
\begin{equation}\label{b_1}
g^{(1)}_1(0,\theta)=b_1.
\end{equation}
By inserting this values in (\ref{eq4c}) we get:
\begin{eqnarray*}
f_1^{(1)}(0,\theta)&=&\sqrt{\frac{2}{3}}g_1^{(1)}(0,\theta)=\sqrt{\frac{2}{3}}b_1,\nonumber\\
f^{(1)}_3(0,\theta)&=&-\frac{1}{\sqrt{3}}g^{(1)}_1(0,\theta)
+ \frac{(1+\cos^2\theta)}{\sqrt{3}\sin^2\theta}g^{(1)}_2(0,\theta)
-   \frac{\sqrt{2}}{\sqrt{3}}\cot\theta g^{(1)}_3(0,\theta)=-\frac{1}{\sqrt{3}}b_1.
\end{eqnarray*}

Components $f^{(0)}_3$ and  $f^{(1)}_3$ are the only
$\vec{n}$-dependent structures which gives non-zero contributions at $k=0$ in the
corresponding wave functions $\vec{\psi}^{(0)}$ and $\vec{\psi}^{(1)}$.
These contributions must cancel in the physical wave function $\vec{\psi}$, what gives the relation
\[ f_3(0,\theta)=c_0 \, f^{(0)}_3(0,\theta)+ c_1\,f^{(1)}_3(0,\theta)=
c_0\sqrt{\frac{2}{3}}b_0-c_1\frac{1}{\sqrt{3}}b_1=0 \]
This relation, together with the normalization condition $c_0^2+c_1^2=1$,
allows us to determine the coefficients $c_a$ of the superposition (\ref{eqf1a}). They read:
\begin{equation}\label{ph2}
c_0=\frac{b_1}{\sqrt{2b_0^2+b_1^2}},\qquad
c_1=\frac{\sqrt{2}b_0}{\sqrt{2b_0^2+b_1^2}}.
\end{equation}
We see from the above expressions
that conditions (\ref{Cond1}) and (\ref{Cond2}) will be satisfied if and only
if coefficients $b_a$ are actually independent of $\theta$.

It is worth noticing that if the wave function $\vec{\psi}$ does not depend on $\vec{n}$,
these coefficients becomes especially simple:
\begin{equation}\label{ph3}
c_0=\sqrt{\frac{1}{3}},\qquad c_1=\sqrt{\frac{2}{3}}.
\end{equation}
Indeed, from an $\vec{n}$-independent wave function $\vec{\psi}$ we can construct
normalized $\vec{n}$-dependent states with definite $a$ as follows:
\begin{eqnarray*}
\vec{\psi}^{(0)}(\vec{k},\vec{n})&=&\sqrt{3}\vec{n}\,(\vec{n}\cd \vec{\psi}(\vec{k})),\cr
\vec{\psi}^{(1)}(\vec{k},\vec{n})&=&\sqrt{\frac{3}{2}}\,\left[\vec{\psi}(\vec{k})-
\vec{n}\,(\vec{n}\cd \vec{\psi}(\vec{k}))\right].
\end{eqnarray*}
The initial function $\vec{\psi}(\vec{k})$ is reproduced by taking their
superposition with coefficients (\ref{ph3}).
In the case of scalar constituents, we found  \cite{MCK_Heid_00} that
coefficients $c_{a}$ are very close (with the accuracy $\approx 1\%$)
to the values  (\ref{ph3}), despite the fact that the wave function strongly
depended on $\vec{n}$ and the split between $M_0$ and $M_1$ masses was large.

\bigskip
Let us finally summarize the procedure followed to construct the physical wave function.
The solution of the mass equations (\ref{eq_J1a0}) and (\ref{eq_J1a1}),
provides the mass squared $M_a^2$ and the components $g^{(0)}_{1,2}$ and $g^{(1)}_{1-4}$
of the $A^2$ eigenstates.
The non-zero values of the first-components $g^{(a)}_1$ at $k=0$ determine --  by means of
(\ref{b_0}) and (\ref{b_1}) -- the coefficients $b_a$.
These are inserted in equation (\ref{ph2})
to provide $c_a$, coefficients of the linear combination  determining the physical
mass  $M^2$  from (\ref{M2}) and the components  (\ref{ph30}) of  the wave function (\ref{nz8}).
Components $f^{(a)}_i$ of this superposition are related to $g^{(a)}_{j}$ by (\ref{ff0}) and (\ref{eq4c}) correspondingly.

\section{Nonrelativistic limit}\label{NR}

In the forthcoming sections the LFD results will be compared to the corresponding non
relativistic limits.
We mean by that the zero order terms
in the  ${1/m}$ expansion of the LFD equations and kernels.
This section is devoted to precise how this limit is obtained in the different OBE
kernels, having in mind in each case
({\it i}) what are the LFD wave function components that should be retained  and
({\it ii}) what kind of equations will they satisfy.

In order to have some insight in the weak coupling limit, but also as a test for
numerical calculations, it is often useful to consider the LFD solutions
as a perturbation of the non relativistic wave functions.
This approximation was used in \cite{CK_NPA581_95,CK_NPA589_95} to calculate
the NN S-wave function  and deuteron electromagnetic form factors \cite{CK_EPJA_99}.
We will also present in what follows how these first order relativistic corrections can be obtained
in the different mass equations (\ref{psi_J0}), (\ref{eq_J1a0}) we consider.

\subsection{$J=0$ states}

For the scalar exchange the leading contribution in the kernel matrix is,
according to (\ref{eqap1}):
\begin{equation}\label{eqnr1}
K_{11}=-\frac{4\pi\alpha }{(\vec{k}-\vec{k'})^2+\mu^2}\equiv
V_S(\vec{k}-\vec{k'})
\end{equation}
Corrections to this kernel are of the $1/m^2$ order both in diagonal and non-diagonal terms.
It follows that the $J=0$ wave function (\ref{psi_J0}),
contains in the non-relativistic limit the $f_1$ component only, which is furthermore independent of $\theta$.
Introducing non-relativistic kinematics,  i.e.
 $4(\vec{k}\,^2+m^2)-M^2\approx 4(k^2+mB)$
where $B=2m-M<<m$ is the binding energy, the equation for  $f_1\equiv f_{NR}$ component becomes:
\begin{equation}\label{eqnr}
(k^2+mB)f_{NR}(k)=-m\int V_S(\vec{k}-\vec{k'})f_{NR}(k')\frac{d^3k'}{(2\pi)^3}
\end{equation}
with kernel (\ref{eqnr1}). This is the Schr\"odinger equation with
the Yukawa potential $V_S(r)=-\alpha\exp(-\mu r)/r$.

\bigskip
For vector exchange we obtain the same equation (\ref{eqnr}) with a kernel differing
from (\ref{eqnr1}) by a global sign.
This corresponds to the repulsion between two fermions ($e^- e^-$, for instance).

We see that for the scalar and vector couplings, the non relativistic limit of LFD
equations coincides with the one-component Schrodinger equation.

\bigskip
For pseudoscalar and pseudovector exchanges the leading diagonal kernels are of the
$1/m^2$ order, whereas the non-diagonal ones are of $1/m^3$.
Thus, for these couplings the non relativistic limit does not exist.
In the leading order and since the $K_{22}$ kernel is repulsive,
only the $f_1$ component remains.
The corrections due to $f_2$ are expected to be bigger than for scalar and vector cases.
Component $f_1$ satisfies at this order the Schrodinger equation (\ref{eqnr}) with a kernel
proportional to $1/m^2$:
\begin{equation}\label{eqnr1ps}
V_{PS}(\vec{k}-\vec{k'})=\frac{\pi\alpha}{(\vec{k}-\vec{k'})^2+\mu^2}\,\frac{(\vec{k}-\vec{k'})^2}{m^2}
=\frac{\pi\alpha}{m^2} \left[ 1 -
\frac{\mu^2}{(\vec{k}-\vec{k'})^2+\mu^2}  \right]
\end{equation}
In coordinate space it corresponds to
\begin{equation}\label{eqnr4}
V_{PS}(\vec{r})=  \frac{\pi\alpha}{m^2}
\left[\delta^{(3)}(\vec{r})-\frac{\mu^2}{4\pi} \frac{\exp(-\mu
r)}{r}\right].
\end{equation}
For these couplings the
leading term is of the same order as relativistic correction in the scalar and vector cases.
We will see  that a similar situation takes place for the $J=1$ state.
This fact makes an important difference between the couplings.
Pseudoscalar and pseudovector exchanges appear always as being relativistic corrections.

\bigskip
We would like to remark from the above results that in the non relativistic limit
the $\vec{n}$-dependent terms in the LFD wave function (\ref{psi_J0}) and kernels disappear.

\bigskip
For models involving the sum of all exchanges (like for the OBE $NN$ interaction) the
non-relativistic limit is determined only by the S and V exchanges.
First order corrections can be obtained by inserting the non-relativistic
component $f_1=f_{NR}$ into the r.h.-side of equations (\ref{eq_J0}).
\begin{equation}\label{eq10p}
[4(\vec{k}\,^2+m^2)-M^2]\; f_i(k,\theta)
=-\frac{m^2}{2\pi^3}\int K_{i1}(k,\theta;k',\theta')f_{NR}(k')\frac{d^3k'}{\varepsilon_{k'}}.
\end{equation}
They generate a perturbative solution for the two components
which incorporates the first order relativistic effects.
This approach was followed in \cite{CK_NPA589_95} to obtain the $^1S_0$ NN scattering
wave function.

\subsection{$J=1$ states}

For $J=1$ states, components $g_i^{(a)}$ obtained by solving the
mass equations differ from those appearing in the wave function ($f_i$).
Our first step is to determine the form of $g_i^{(a)}$ in case of a non relativistic wave function.
The non relativistic wave function components  do not
depend on $\vec{n}$ and, according to (\ref{ph3}), are given by:
\begin{equation}\label{eq4p}
f_i= \frac{1}{\sqrt{3}} f^{(0)}_i +\sqrt{\frac{2}{3}} f^{(1)}_i.
\end{equation}
Substituting (\ref{eq4p}) into (\ref{ff0}) and (\ref{eq4c}) we obtain a relation between  $f_{i}$ and $g^{(a)}_{j}$ components.
These equations are solved relative to $g^{(a)}_{j}$ and the result,
 expressed through $f_{i}$, reads:
\begin{eqnarray}\label{eqff}
g_1^{(0)}&=&f_1\cos\theta+f_2\sqrt{2}\cos\theta+f_3\sqrt{2}\cos\theta+f_4\frac{7+\cos 2\theta}{2\sqrt{2}}+f_6\sqrt{\frac{3}{2}}\sin^2\theta \nonumber\\
g_2^{(0)}&=&-f_1\sin\theta+f_2\frac{1}{\sqrt{2}}\sin\theta-f_3\sqrt{2}\sin\theta-f_4\frac{1}{2\sqrt{2}}\sin 2\theta+f_6\sqrt{\frac{3}{8}}\sin 2\theta \nonumber\\
g_1^{(1)}&=&f_1-f_2\frac{1+3\cos 2\theta}{4\sqrt{2}}-f_3\frac{1}{\sqrt{2}}-f_4\sqrt{2}\cos\theta \nonumber\\
g_2^{(1)}&=&f_2\frac{3}{2\sqrt{2}}\sin^2\theta\nonumber\\
g_3^{(1)}&=&f_2\frac{3}{4}\sin 2\theta+f_4\frac{3}{2}\sin \theta-f_6\frac{\sqrt{3}}{2}\sin \theta\nonumber\\
g_4^{(1)}&=&f_5\sqrt{\frac{3}{2}}\sin \theta
\end{eqnarray}
As previously discussed, in the non relativistic limit
there are no $\vec{n}$-dependent terms in the LFD wave function (\ref{nz8})  and
only $f_1$ and $f_2$ components among the six $f_i$ survive.
We have shown in \cite{CK_NPA581_95} that one actually has $f_1\approx u_S, f_2\approx-u_D,f_{3-6}\approx0$
where $u_S$ and $u_D$ are respectively the usual S- and D-wave non relativistic components.
Inserting their expressions in (\ref{eqff}) we obtain the form of the non-relativistic  functions $g$:
\begin{eqnarray}\label{eq6af}
g^{(0)}_{1}&=&(u_S-\sqrt{2}\;u_D)\cos\theta,\nonumber\\
g^{(0)}_{2}&=&-\left(u_S+\frac{1}{\sqrt{2}}\;u_D\right)\sin\theta,\nonumber\\
g^{(1)}_{1}&=&u_S+\frac{1}{4\sqrt{2}}(1+3\cos2\theta)\;u_D,\nonumber\\
g^{(1)}_{2}&=&-\frac{3}{2\sqrt{2}}\sin^2\theta\; u_D,\nonumber\\
g^{(1)}_{3}&=&-\frac{3}{4}\sin2\theta\;u_D,\nonumber\\
g^{(1)}_{4}&=&0.
\end{eqnarray}
We see here that the $\theta$-dependence of the auxiliary components $g^{(a)}_i$
remains even in the non relativistic limit. It will disappear
only in the linear combination  giving the physical components $f_{1,2}$.

\bigskip
Let us first consider the scalar exchange.
The  mass equation  for $a=0$ eigenstate  (\ref{eq_J1a0})
and the scalar kernels (\ref{app_sc}), becomes in the leading order $(1/m)^0$:
\begin{eqnarray}\label{rh1}
C(k)g^{(0)}_1(k,\theta)&=&-4\alpha\pi\cos\theta\int\left[g^{(0)}_1(k',\theta')\cos\theta'- g^{(0)}_2(k',\theta')\sin\theta'\right]
\left\{\ldots\right\} \frac{d^3k'}{\varepsilon_{k'}}\\
C(k)g^{(0)}_2(k,\theta)&=&+4\alpha\pi\sin\theta\int\left[g^{(0)}_1(k',\theta')\cos\theta'-g^{(0)}_2(k',\theta')\sin\theta'\right]
\left\{\ldots\right\} \frac{d^3k'}{\varepsilon_{k'}}\nonumber
\end{eqnarray}
For shortness we denote by $C(k)$ the kinematical part and by $\{\ldots\}$ the
kernel contributions which are common to all couplings and states.
\[ \left\{\ldots\right\} =\frac{1}{m^2\,\varepsilon_{k}\varepsilon_{k'}} \frac{1}{Q^2+\mu^2} \]
These factors contain $1/m$ and $1/m^2$ terms but  we do not write them explicitly
and analyze only the kernel contributions resulting from $\kappa_{ij}$.

Since the integrals in the right hand sides of (\ref{rh1}) are the same, its solution has the form:
\begin{eqnarray}\label{rh5}
g^{(0)}_1(k,\theta) &=& +g^{(0)}(k) \cos\theta,\nonumber\\
g^{(0)}_2(k,\theta) &=& -g^{(0)}(k) \sin\theta.
\end{eqnarray}
with $g^{(0)}$ an unknown function to determine.
Substituting (\ref{rh5}) into (\ref{rh1}) we find the equation for $g^{(0)}$:
\begin{equation}\label{rh6}
C(k) g^{(0)}(k)  =-4\alpha\pi \int g^{(0)}(k') \left\{\ldots\right\} \frac{d^3k'}{\varepsilon_{k'}}
\end{equation}
For $a=1$ state, we found in a similar way that only  $g^{(1)}_1$ survives and satisfies
to the same $(1/m)^0$ order, the equation
\begin{equation}\label{rh3}
C(k)g^{(1)}_1(k,\theta) = -4\alpha\pi \int g^{(1)}_1(k',\theta') \left\{\ldots\right\}\frac{d^3k'}{\varepsilon_{k'}}
\end{equation}
It coincides with the equation (\ref{rh6}) for $a=0$ and, hence, provides the same mass.
We see in this way that, in the leading order, $a=0$ and $a=1$ states are degenerate.
The coefficients $c_a$ of the superposition (\ref{ph30}) are
calculated in terms of $b_a$ given by (\ref{b_0}) and (\ref{b_1}).
Since in the leading order,  $g^{(0)}_1$ and $g^{(1)}_1$ equals
$g^{(0)}$, one has $b_0=b_1=g^{(0)}(0)$ and, from (\ref{ph2}), the values (\ref{ph3}).

In  next to leading order -- $1/m^2$ -- we get for $a=0$:
\begin{eqnarray}\label{rh2}
C(k)g^{(0)}_1&=&\alpha\pi\frac{1}{m^2}\int\left[2g'^{(0)}_1kk'-3(k^2+k'^2)\cos\theta
(g'^{(0)}_1\cos\theta'-g'^{(0)}_1\sin\theta')\right]\left\{\ldots\right\}\frac{d^3k'}{\varepsilon_{k'}}\nonumber\\
C(k)g^{(0)}_2&=&\alpha\pi \sin\theta\frac {1}{m^2}\int 3(k^2 + k'^2)(g'^{(0)}_1\cos\theta' - g'^{(0)}_1\sin\theta')
\left\{\ldots\right\}\frac{d^3k'}{\varepsilon_{k'}}
\end{eqnarray}
and for $a=1$:
\begin{eqnarray}\label{rh4}
C(k)g^{(1)}_1 &=&-\alpha\pi \frac{1}{m^2}\int [3(k^2 + k'^2) g'^{(1)}_1 +  k k'\cos\theta (-2 g'^{(1)}_1\cos\theta'
     + \sqrt{2} g'^{(1)}_3\sin\theta')] \left\{\ldots\right\}\frac{d^3k'}{\varepsilon_{k'}}\nonumber\\
C(k)g^{(1)}_3 &=&
     -\alpha\pi\sin\theta \frac {1}{m^2}\int k k'(\sqrt{2} g'^{(1)}_1\cos\theta'- g'^{(1)}_3\sin\theta')
 \left\{\ldots\right\}\frac{d^3k'}{\varepsilon_{k'}}.
\end{eqnarray}
These systems of equations -- (\ref{rh2}) and (\ref{rh4}) --
are already different and the masses of the two eigenstates are split.

\bigskip
For vector exchange, the situation  is quite similar.
The equations in the leading order $(1/m)^0$ differ
from (\ref{rh1}) an (\ref{rh3}) only by a global sign in their right hand sides.
Thus for these two couplings, as it was the case for $J=0$, the leading order is $(1/m)^0$.

\bigskip
For pseudoscalar exchange, the leading contribution in the kernel has order $1/m^2$.
Indeed, from the analytic expressions given in \ref{app_ps}, we found for the $a=0$ state:
\begin{eqnarray}\label{rh7}
g^{(0)}_1&=&\phantom{-}\alpha\pi\cos\theta\frac{1}{m^2}\int
k'^2\left(g^{(0)}_1\cos\theta'+ g^{(0)}_2\sin\theta'\right) \left\{\ldots\right\}\frac{d^3k'}{\varepsilon_{k'}}\nonumber\\
g^{(0)}_2&=&-\alpha\pi\sin\theta\frac{1}{m^2}\int
k'^2\left(g^{(0)}_1\cos\theta'+ g^{(0)}_2\sin\theta'\right) \left\{\ldots\right\}\frac{d^3k'}{\varepsilon_{k'}}
\end{eqnarray}
Like for the scalar coupling, the solution of (\ref{rh7}) has the form (\ref{rh5})
with $g^{(0)}$ satisfying the equation:
\begin{equation}\label{rh6a}
g^{(0)}= \alpha\pi\frac{1}{m^2}\int {k'}^2(\cos^2\theta'-\sin^2\theta')g^{(0)}\left\{\ldots\right\}\frac{d^3k'}{\varepsilon_{k'}},
\end{equation}

For $a=1$ the leading order equation reads:
\begin{equation}\label{rh8}
g^{(1)}_1=-\alpha\pi\frac{1}{m^2}\int k'^2 \cos^2\theta' g^{(1)}_1  \left\{\ldots\right\}\frac{d^3k'}{\varepsilon_{k'}}
\end{equation}
which is now different from (\ref{rh6a}). The masses $M_0$ and
$M_1$ calculated with pseudoscalar exchange are therefore always different.
Their difference remains even in systems having small binding energies
or when the large momentum
contributions are removed using small cutoff parameter $\Lambda$ in form factors (\ref{BFF}).

Pseudovector exchange kernel differs from the pseudoscalar one by the replacement
$\gamma_5\to\gamma_5 - \frac{\hat{\omega}\gamma_5\tau}{2m}$
or by $\gamma_5\to\gamma_5+ \frac{\hat{\omega}\gamma_5\tau'}{2m}$ (see eq. (4.18) in \cite{CDKM_PR_98}).
There is so an extra term proportional to
$\frac{\hat{\omega}\gamma_5\tau'}{2m}\propto \frac{k'^2+m|B|}{m^2}$
which does not contain $(1/m)^0$ terms.
The situation is therefore the same as for the pseudoscalar case.

\bigskip
To summarize, we have shown analytically that in the
non-relativistic limit for scalar and vector exchanges, the
energies $B(a=0)$ and  $B(a=1)$ coincide with each other and the
coefficients $c_0,c_1$ tend to
$\sqrt{\frac{1}{3}},\sqrt{\frac{2}{3}}$ respectively. On the
contrary, for pseudoscalar and pseudovector couplings this is not
the case. In this sense, for the pseudoscalar and pseudovector
exchanges, the non-relativistic limit does not exist. If the
kernel is the sum of all the exchanges, like $NN$ kernel, the
situation is the same as for the scalar and vector exchanges,
since in non-relativistic limit the $(1/m)^0$ order dominates,
resulting from these exchanges. The existence of deuteron, for
example, as a nonrelativistic system (with a reasonable accuracy)
is due to contribution of the scalar and vector exchanges in NN
interaction.

\bigskip
Perturbative solutions are obtained by substituting the zero-order
functions (\ref{eq6af})  into the right hand sides of LFD
equations (\ref{eq_J1a0}) and (\ref{eq_J1a1}). If D-wave is
neglected, the six perturbative components are given  in terms of
the only non relativistic wave function $u_S$ simply by:
\begin{eqnarray}
\left[4(\vec{k}\,^2 +m^2)-M^2\right] g^{(0)}_i(\vec{k},\vec{n})
&=&-\frac{m^2}{2\pi^3}\int\left(K^{(0)}_{i1}\cos\theta'-K^{(0)}_{i2}\sin\theta'\right)u_S(k')\frac{d^3k'}{\varepsilon_{k'}}  \label{eq0p}\\
\left[4(\vec{k}\,^2 +m^2)-M^2\right] g^{(1)}_i(\vec{k},\vec{n})
&=&-\frac{m^2}{2\pi^3}\int K^{(1)}_{i1}u_S(k')\frac{d^3k'}{\varepsilon_{k'}} \label{eq1p}
\end{eqnarray}

We would like to mention here that one appreciable advantage
of the LFD formalism  with respect to other relativistic approaches
is the clear link it has with the non relativistic dynamics.
On one hand because LFD wave functions have the same physical
meaning of probability amplitudes.
On the other hand, because their components
$f_i$ split in two families: those
which in the non relativistic limits become negligible
and those  which tend to the usual non relativistic wave functions.

\bigskip
Next sections are devoted to show the numerical solutions  obtained with
differents couplings. Their very different behaviour motivates to be treated separately.

\section{Results for scalar coupling}\label{Res_S}

Our first results concerning the Yukawa model have been reported
in \cite{MCK_PRD_01,KMC_Prague_01}. The main interest in these
papers concerned the stability of the $J=0,1$ solutions with
respect to the cut-off, i.e. the possibility of getting stable
results without any vertex form factor. We showed in particular
that  $J=0^+$ states were stable for coupling constant smaller
than some critical value $\alpha\leq\alpha_c\simeq3.72$ and
unstable above. On the contrary  the $J=1^+$ states were found to
be unstable for any value of the coupling constant and both
projections $a=0,1$. This instability manifests in the logarithmic
decrease of $M^2(k_{max})$ for a given value of $\alpha$ -- or
equivalently of $\alpha(k_{max})$ for a given value of $M$ -- and
imposes the use of form factors.


\begin{figure}[h]
\begin{center}  
\mbox{\epsfxsize=8.5cm\epsffile{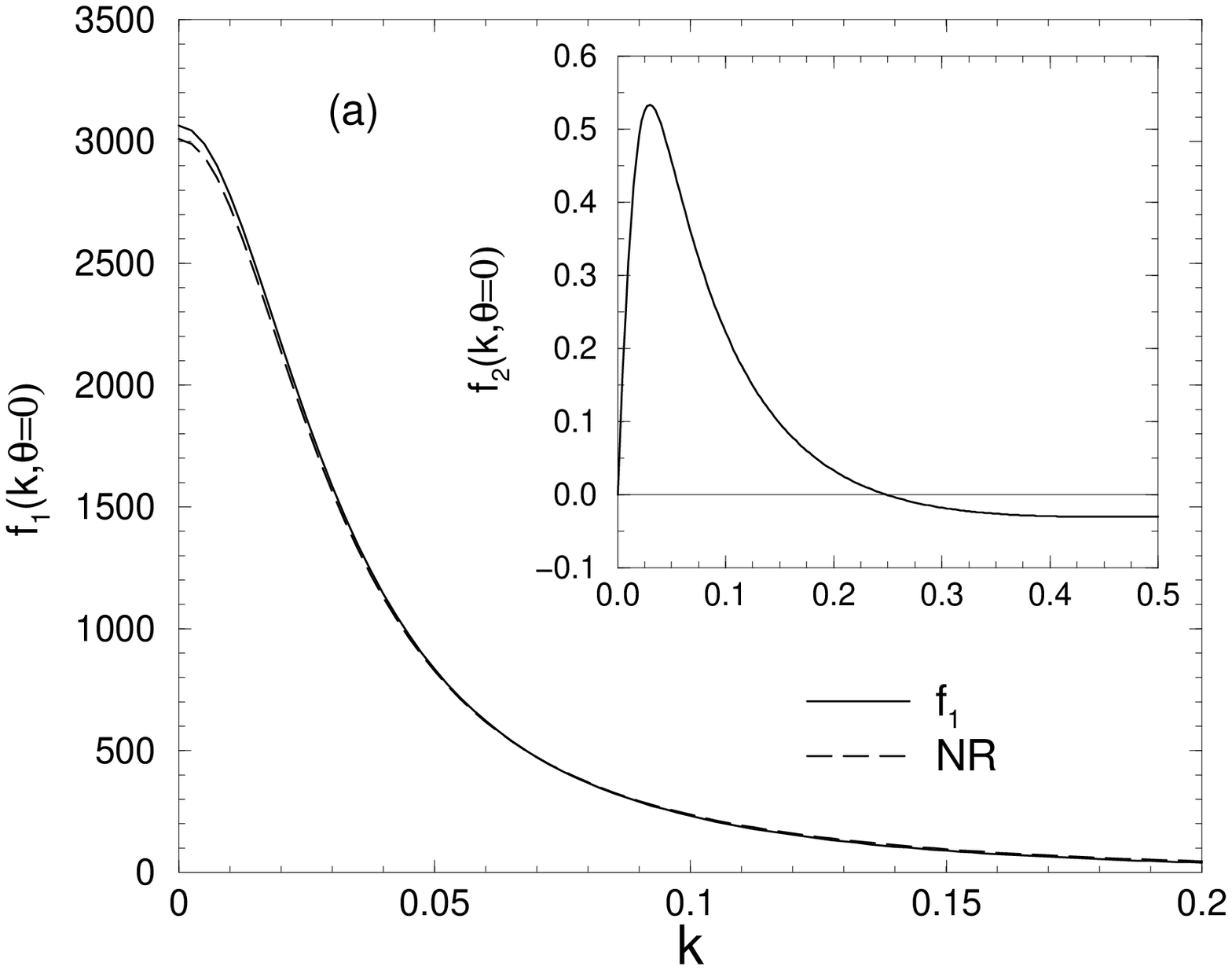}}\hspace{0.5cm}  
\mbox{\epsfxsize=8.5cm\epsffile{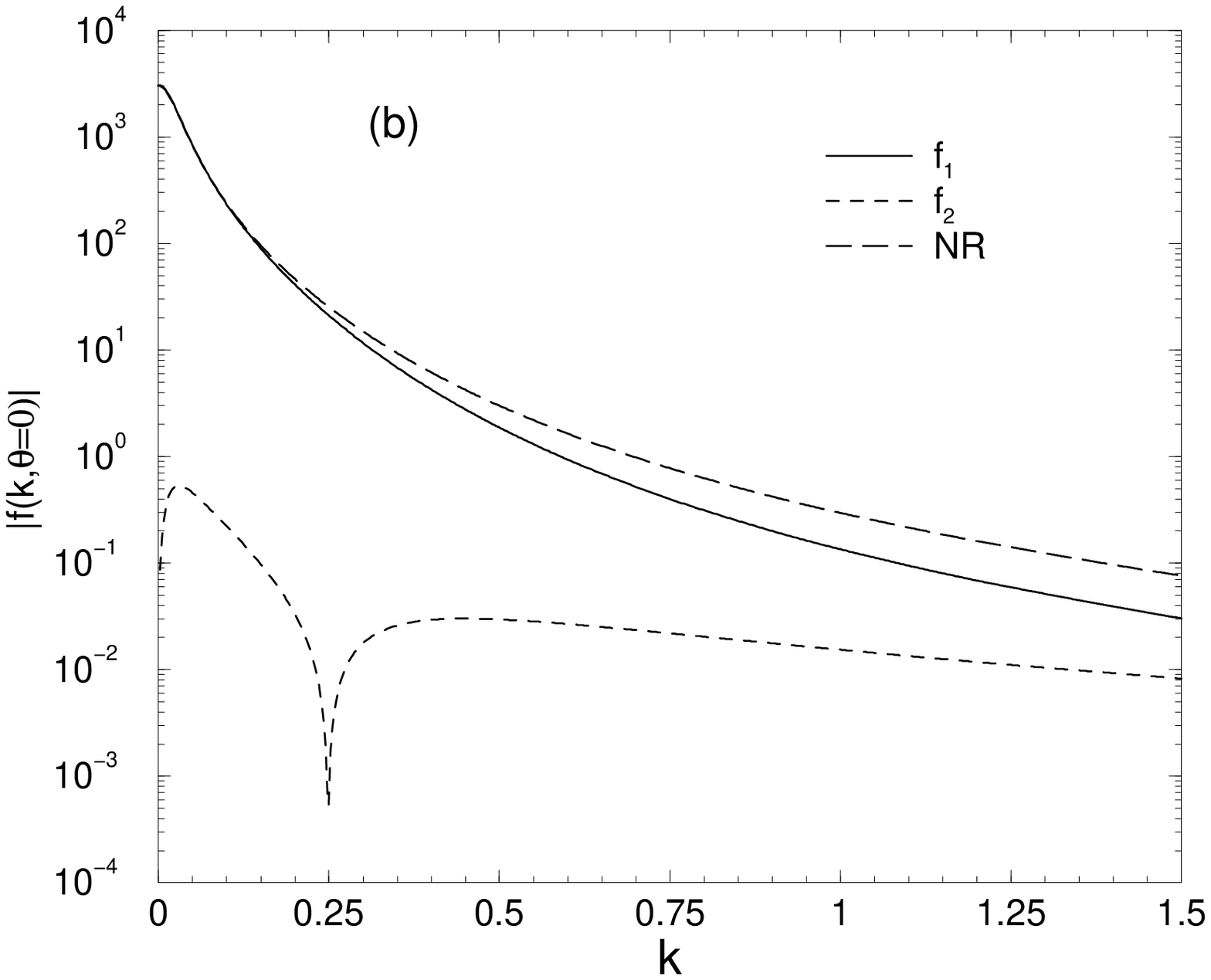}}
\caption{LFD wave function components $f_i$ for scalar coupling (B=0.001, $\mu=0.15$)
in linear (a) and  logarithmic (b) scale compared with the non relativistic solutions}\label{S_f12_1_1MeV_sf}
\end{center}
\end{figure}

\bigskip
We first consider the $J=0^+$ state.
Its wave function is determined by two components $f_i$.
Although the use of vertex form factors (FF) is not required \cite{MCK_PRD_01},
we would like to notice that  the convergence as a function of $k_{max}$ is very slow.
Unless otherwise specified the results that follow correspond to $\mu=0.15$.

For a weakly bound system (B=0.001),
the coupling constant found solving LFD equations is $\alpha_{Yuk}$=0.331 whereas
the non relativistic (NR) value is $\alpha_{NR}$=0.323.
By the latter we understand, the results obtained
by inserting into the Schrodinger equation (\ref{eqnr})
the static potential (\ref{eqnr1}) resulting from the leading order approximation
as has been discussed  in section \ref{NR}.
Like in the Wick-Cutkosky (WC) model -- scalar particles interacting by scalar exchange --
relativistic effects are repulsive \cite{MC_PLB_00}.
They account for only a 3\% difference
in the coupling constants whereas in WC they are sizeably bigger ($\alpha_{WC}$=0.364).

Corresponding wave functions  are displayed in Figs. \ref{S_f12_1_1MeV_sf}a and
\ref{S_f12_1_1MeV_sf}b.
One can see that component $f_1$ dominates over $f_2$ in all the interesting momentum range
and  that $f_2$ has a zero at $k\approx0.25$.
One also notices in Fig. \ref{S_f12_1_1MeV_sf}b that $f_1$
is very close to the NR wave function in
the small momentum but it sensibly departs with increasing $k$; for
$k\sim 1.5$ the differences represents more than one order of magnitude
in the probability densities.
The coupling between the two relativistic amplitudes has a very small (0.1\%)
attractive effect in the binding energy.

\begin{figure}[!h]
\begin{center}
\mbox{\epsfxsize=8.5cm{\epsffile{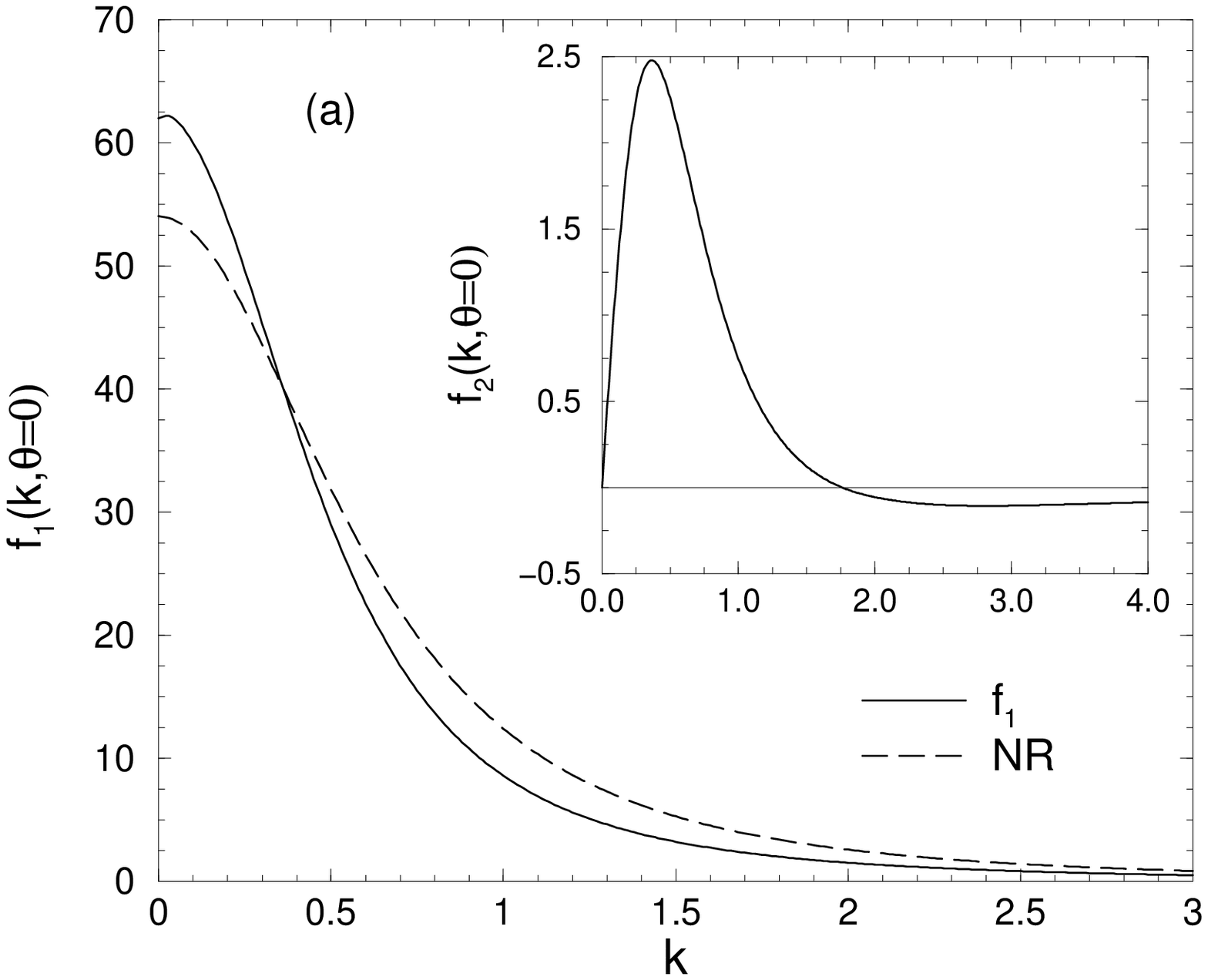}}}
\hspace{0.5cm}
\mbox{\epsfxsize=8.5cm{\epsffile{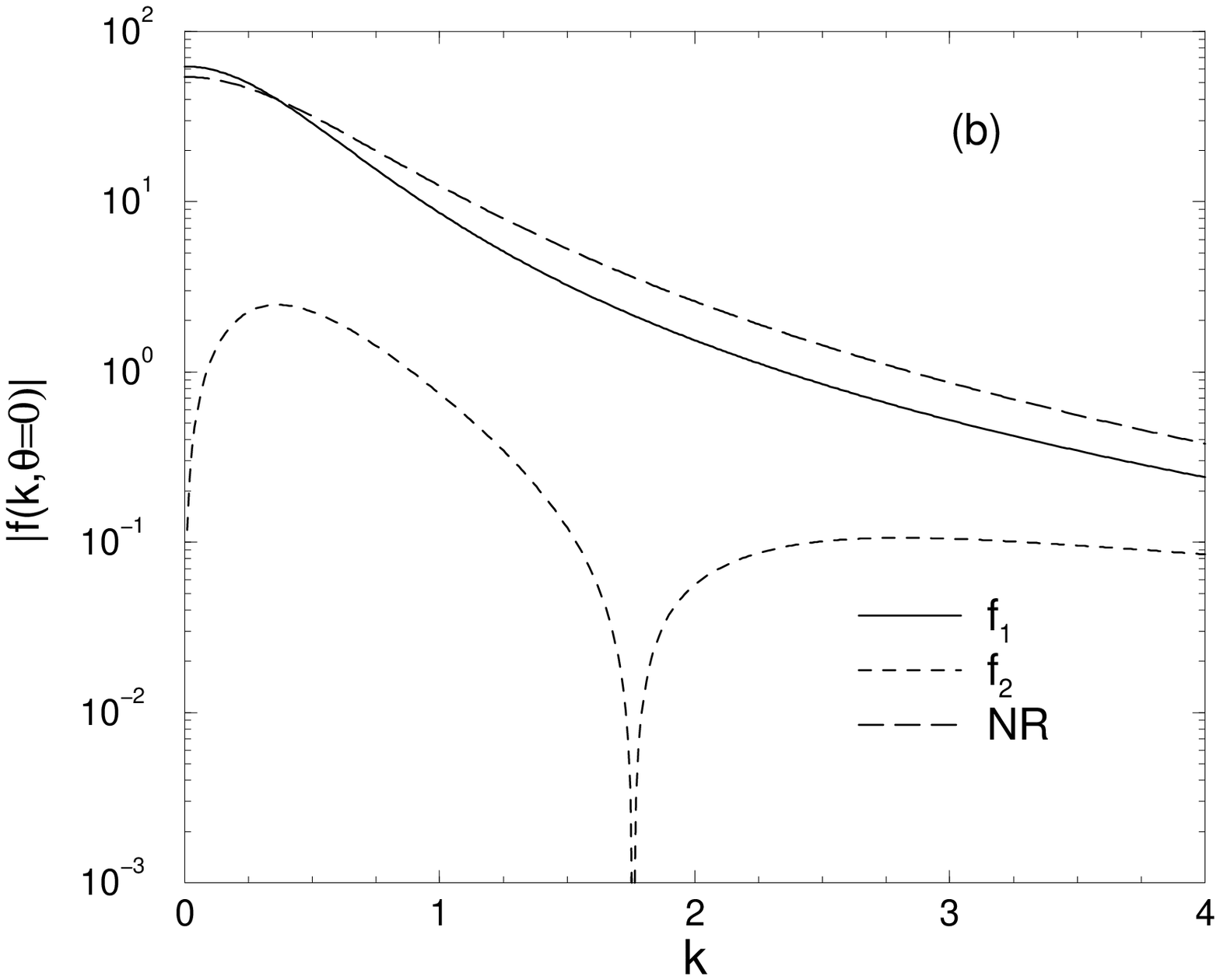}}}
\caption{LFD wave function components $f_i$ for scalar coupling
(B=0.5, $\mu=0.15$) in linear (a) and  logarithmic (b) scale
compared with the non relativistic solutions}\label{S_f12_2_500MeV_sf}
\end{center}
\end{figure}

In the strong binding limit (B=0.5), the situation is quite
similar with enhanced relativistic effects in binding energies and
wave functions. One has $\alpha_{Yuk}$=2.44 for $\alpha_{NR}$=1.71
and the differences in the wave functions - displayed in Figs.
\ref{S_f12_2_500MeV_sf}a and \ref{S_f12_2_500MeV_sf}b - are
already visible at $k=0$ momentum (Fig. \ref{S_f12_2_500MeV_sf}).
One can see however in Fig. \ref{S_f12_2_500MeV_sf}b that -- even
for deeply bound systems -- $f_1$ component still dominates over
$f_2$.

\begin{figure}[!ht]
\begin{center}
\mbox{\epsfxsize=11.cm\epsffile{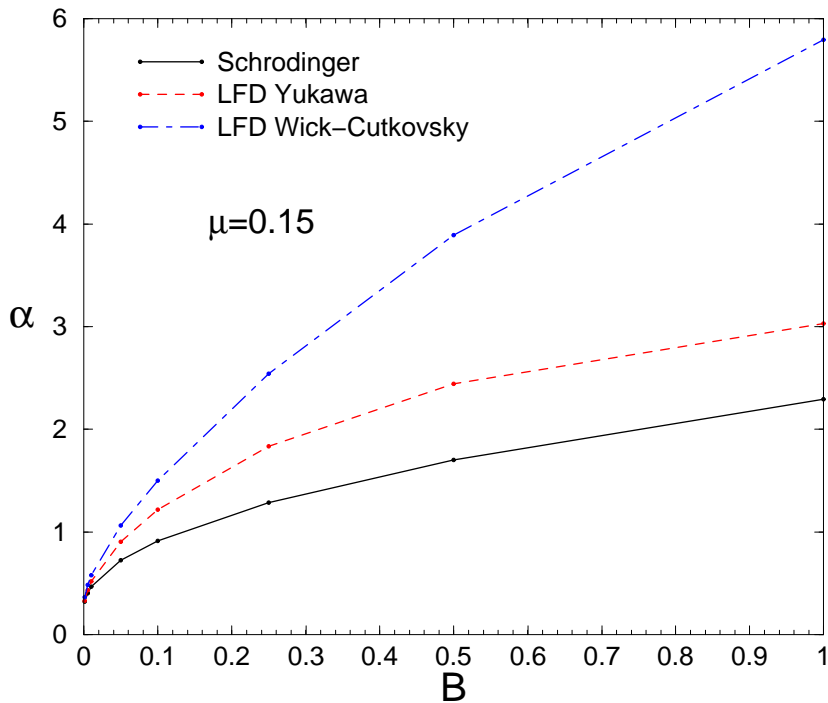}}
\caption{Comparison of $B(\alpha)$ between the Yukawa (dashed line) and Wick-Cutkosky
(dot dashed lines) models in LFD and non relativistic (solid line) results in $J=0^+$ state}\label{Fig_B_Y_WC_NR}
\end{center}
\end{figure}

It has some interest to compare the LFD results for Yukawa
(two-fermion) and WC (two-scalar) models with the NR results. We
have displayed in Fig. \ref{Fig_B_Y_WC_NR} the corresponding
coupling constants for different values of the binding energy. One
can see that the Yukawa results ($\alpha_{Yuk}$) are
systematically closer to the non relativistic values than
$\alpha_{WC}$ are, as if the fermionic character of the
constituents generates closer binding energies to the NR ones but
larger differences in the high momentum components of the wave
function, due to the different asymptotic of interaction kernels.

Though not necessary
to get stable solutions, form factors they have been widely used
in most of the preceding OBEP calculations performed in momentum space \cite{Bonn}.
It is thus interesting to estimate their influence  in the predictions.
To this aim we have considered the vertex form factors used in the Bonn model (\ref{BFF})
with, for the scalar coupling, n=1 and $\Lambda$=2.0.
Their effects are found to be repulsive.
For B=0.001 they remain relatively small ($\alpha_{Yuk}$=0.376 instead
of $\alpha_{Yuk}$=0.331)
but for B=0.5 the differences reach already a factor two
($\alpha_{Yuk}$=5.32 instead of $\alpha_{Yuk}=2.44$).
It is worth emphasizing that whatever will be the degree of refinement in the dynamics,
the results of a relativistic calculation
will be strongly influenced by this phenomenological and not well controlled trick.
\begin{figure}[!ht]
\begin{center}\epsfxsize=11.cm\mbox{\epsffile{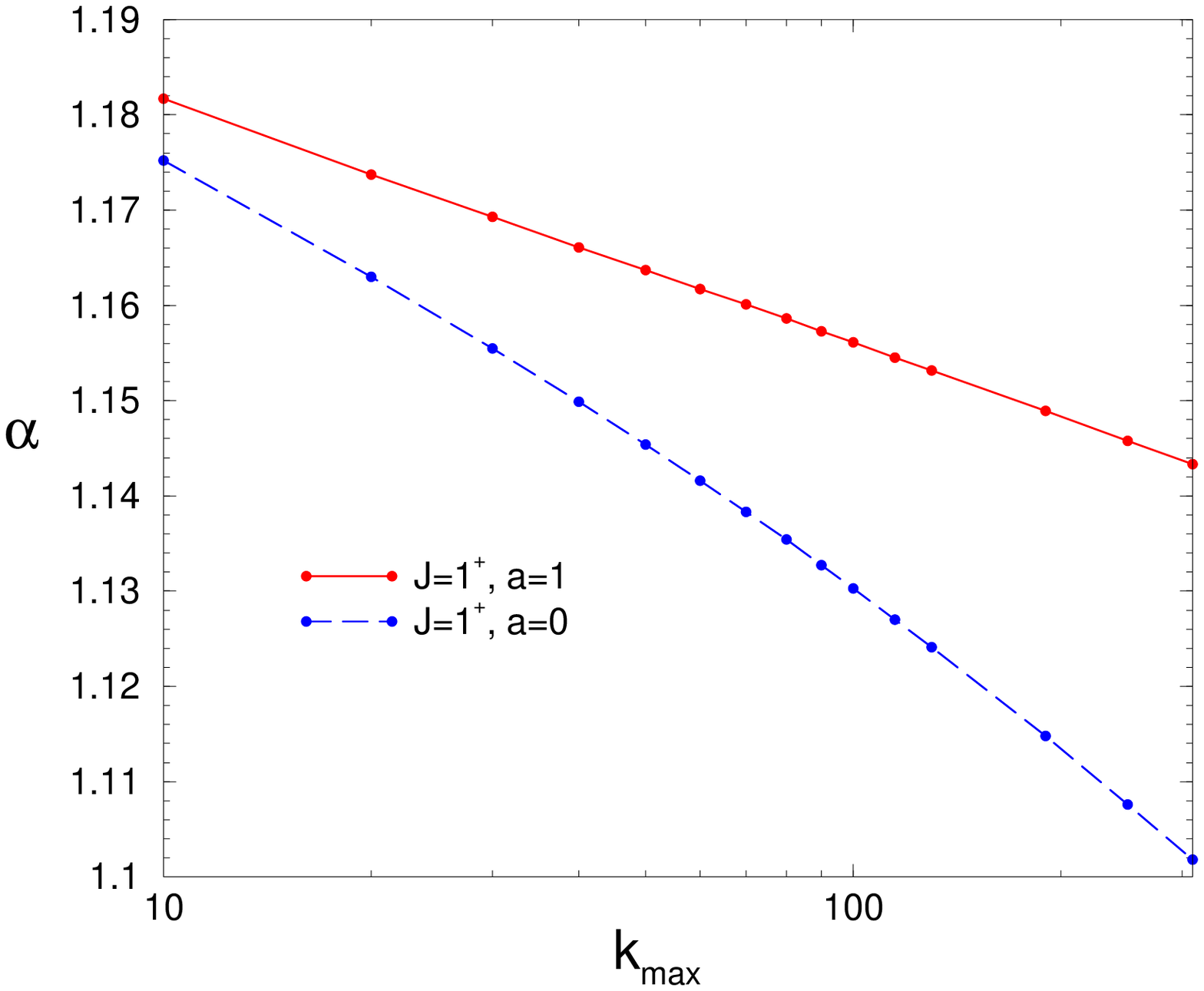}}
\caption{Logarithmic dependence of the coupling constant as a function of cutoff for the $J=1^+$ a=0 and a=1 states.
 Calculations correspond to B=0.05 and $\mu=0.25$}\label{alpha_kmax_J1a0a1_50}
\end{center}
\end{figure}

\bigskip
The system of equations for determining the $J=1^+,a=0$
($g^{(a=0)}_{i=1,2}$) and $a=1$ ($g^{(a=1)}_{i=1,2,3,4}$)
solutions are both unstable and require  cutoff regularization
\cite{GHPSW_PRD47_93,MCK_PRD_01}. This can be seen in Fig.
\ref{alpha_kmax_J1a0a1_50} where the $\alpha(k_{max})$ variation
for $a=0$ and $a=1$ cases displays a logarithmic dependence. One
can also see in this figure the non degeneracy of both states due
to the Fock space truncation discussed in section \ref{angl}. We
remark however that if the binding energies -- or equivalently
coupling constants -- of states with different projections $a$ are
not equal, they are almost-degenerated in a wide range of
$k_{max}$ values. For instance, at $k_{max}=10$ one has
$\alpha_{a=0}$=1.17 and $\alpha_{a=1}$=1.18 while at $k_{max}=90$
one has $\alpha_{a=0}$=1.14 and $\alpha_{a=1}$=1.16. These weak
splitting -- of less than 1\% -- for a noticeably bound system
($B=0.05$), are rather surprising in view of the results obtained
in the purely scalar WC case \cite{MCK_Heid_00,KCM_Taiw_01}, in
which the difference  in  coupling constants for the same binding
energy is rather 20\%, what corresponds to $\Delta B\approx B$.

\begin{figure}
\mbox{\epsfxsize=8.5cm\epsffile{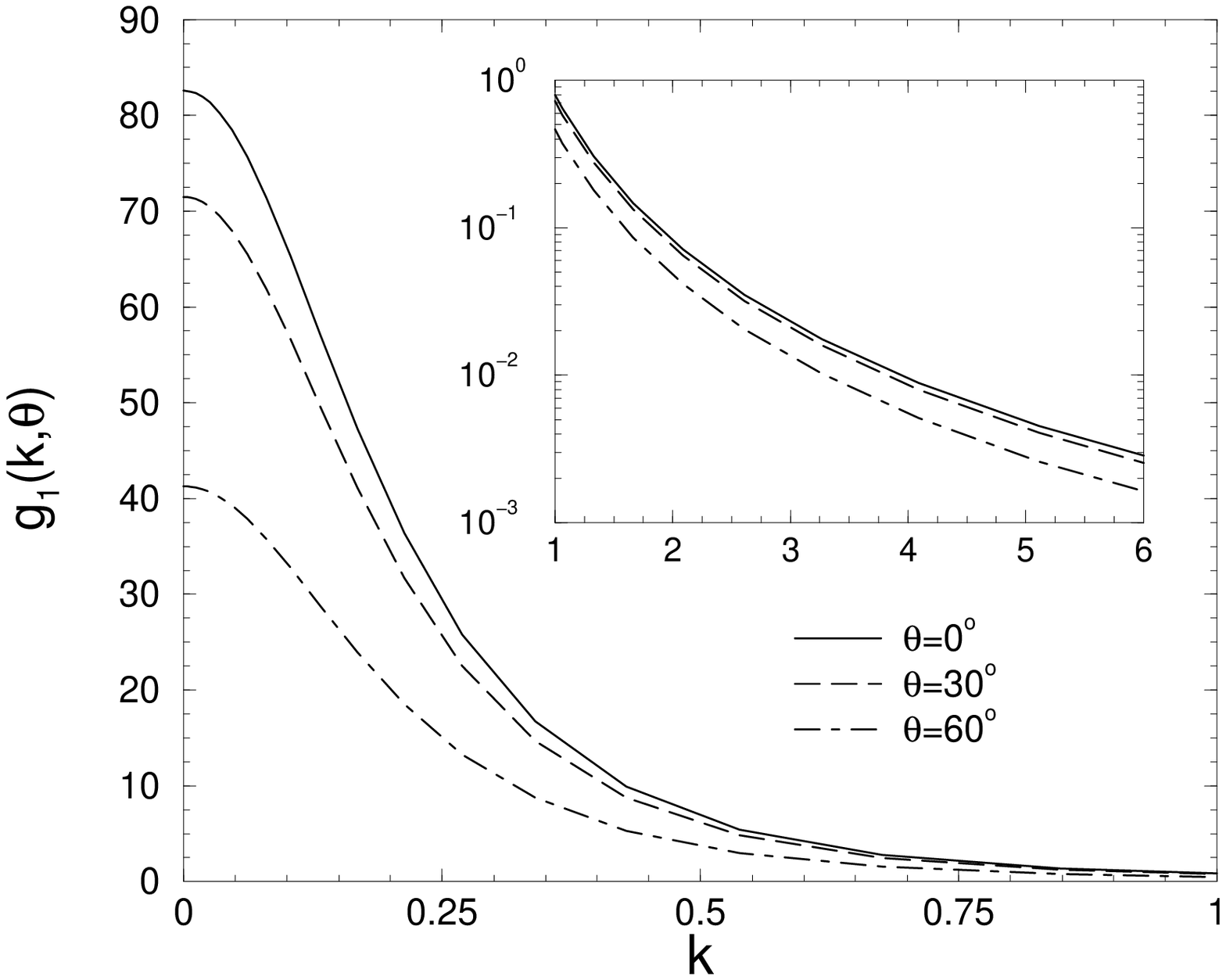}}
\hspace{0.5cm}
\mbox{\epsfxsize=8.5cm\epsffile{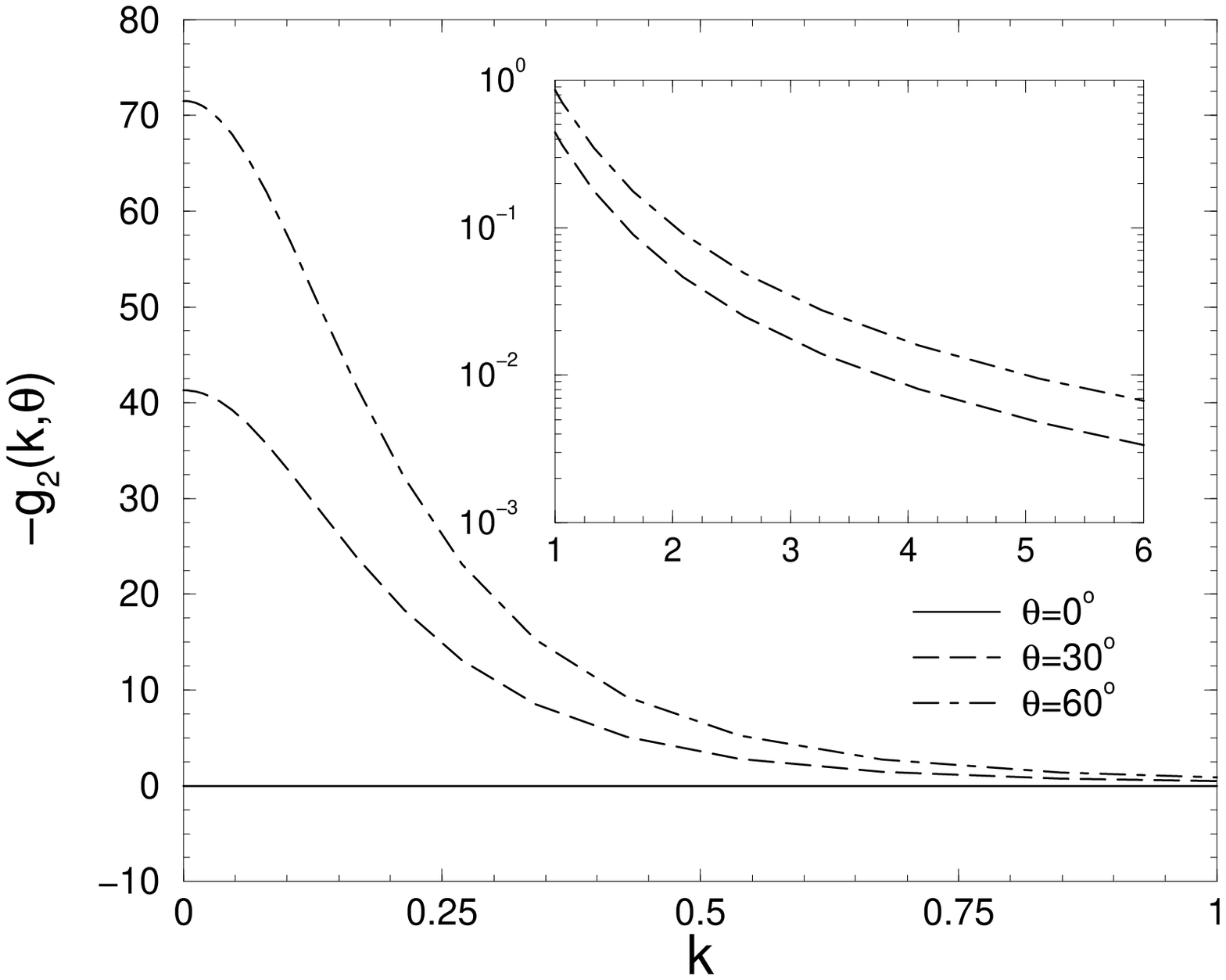}}
\caption{{$g^{a=0}_{i=1,2}$ solutions for scalar
coupling with $\alpha=1.18$,
$\mu=0.25$ and sharp cutoff at $k_{max}=10$. Binding energy is $B=0.0506$.}}\label{g1g2_J=1_a=0}
\end{figure}

\begin{figure}[!hp]
\begin{center}
\mbox{\epsfxsize=8.cm\epsffile{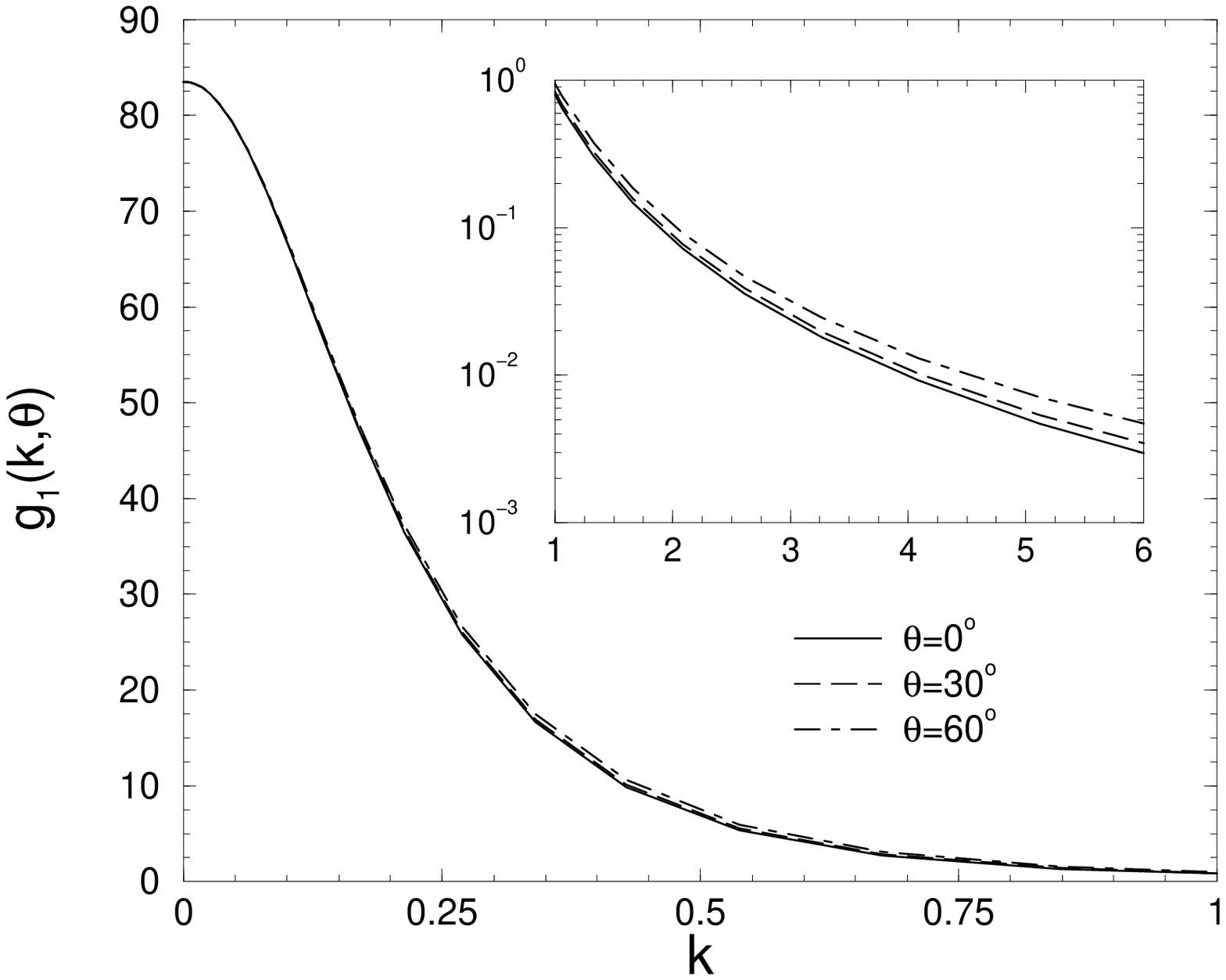}}
\mbox{\epsfxsize=8.cm\epsffile{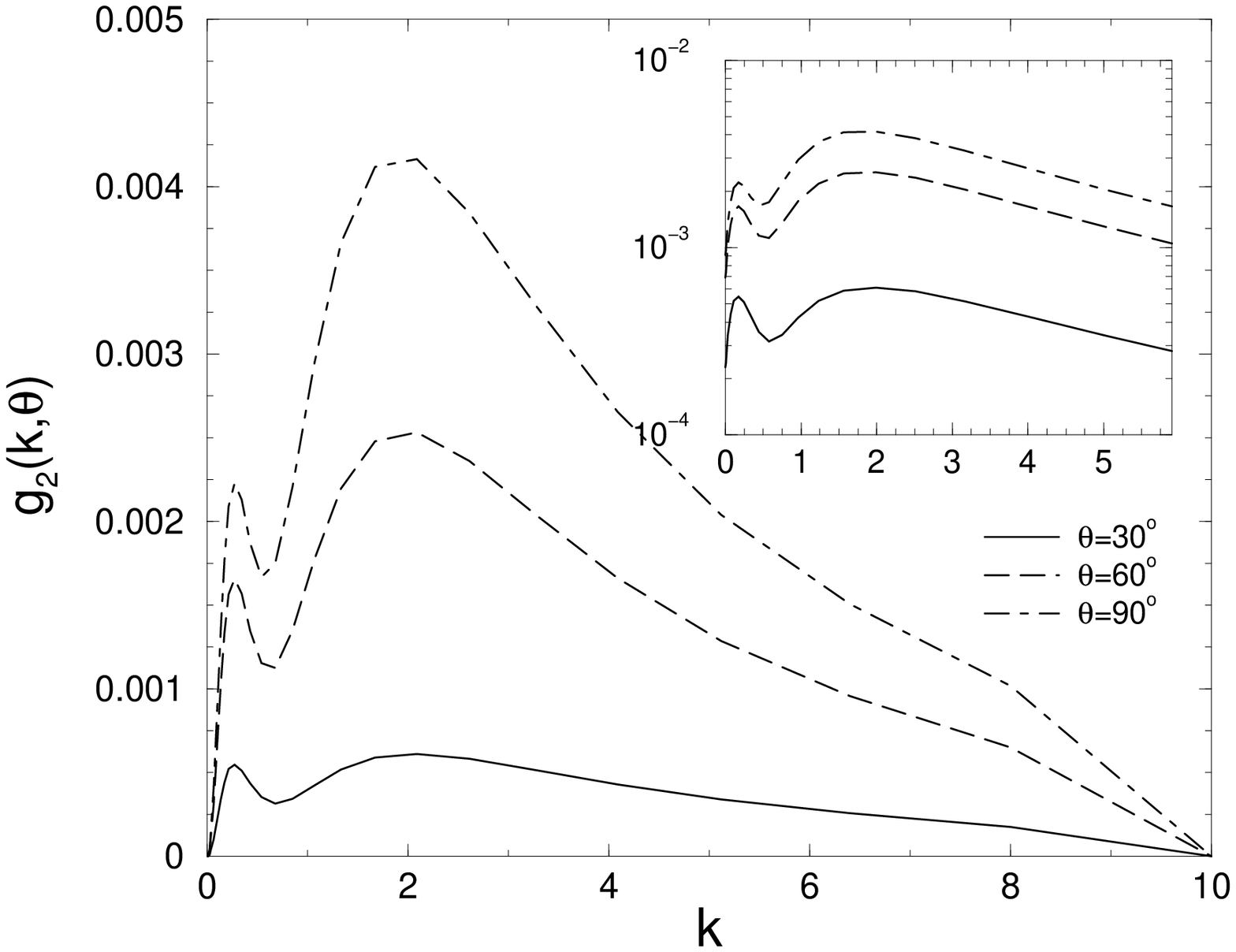}}
\mbox{\epsfxsize=8.cm\epsffile{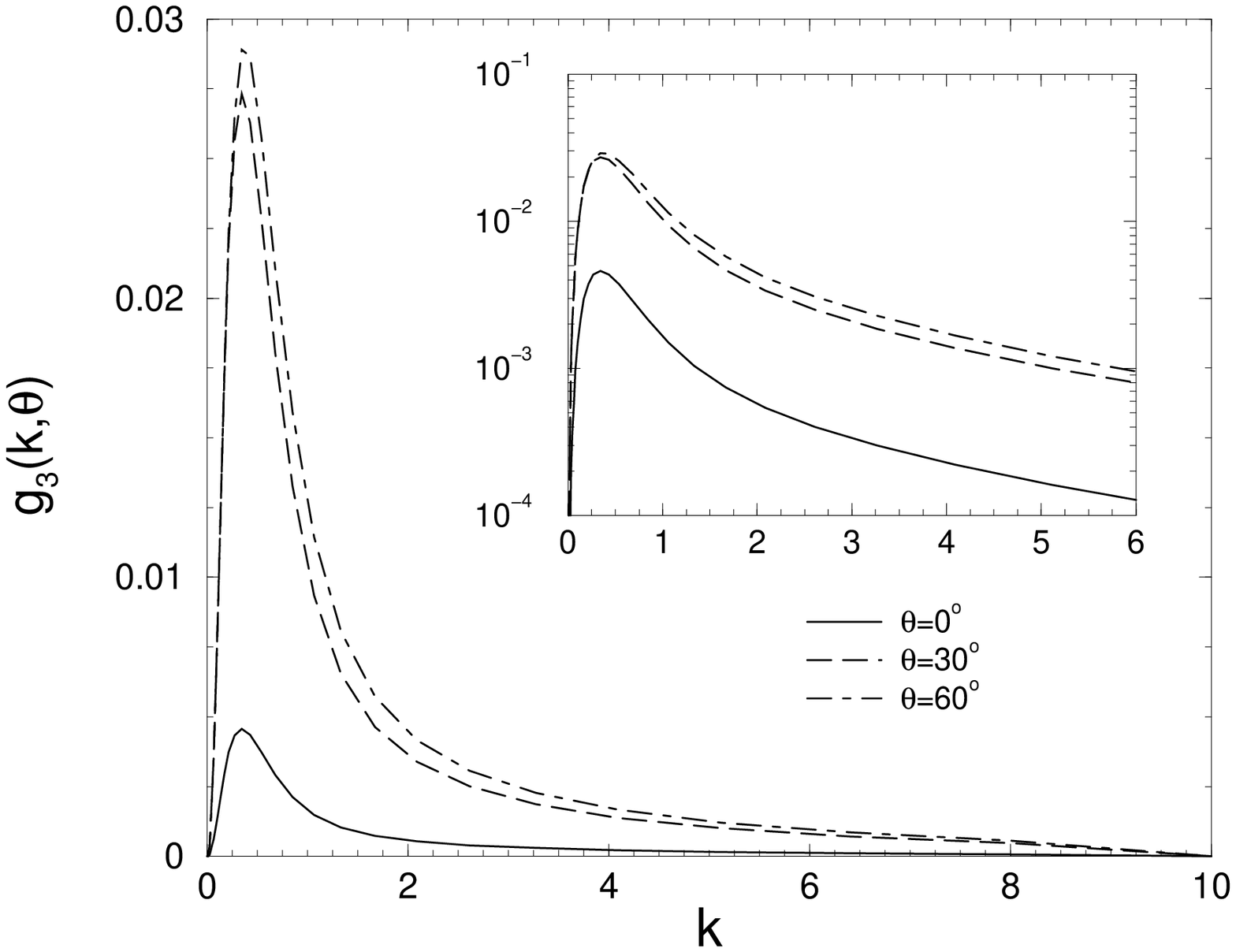}}
\mbox{\epsfxsize=8.cm\epsffile{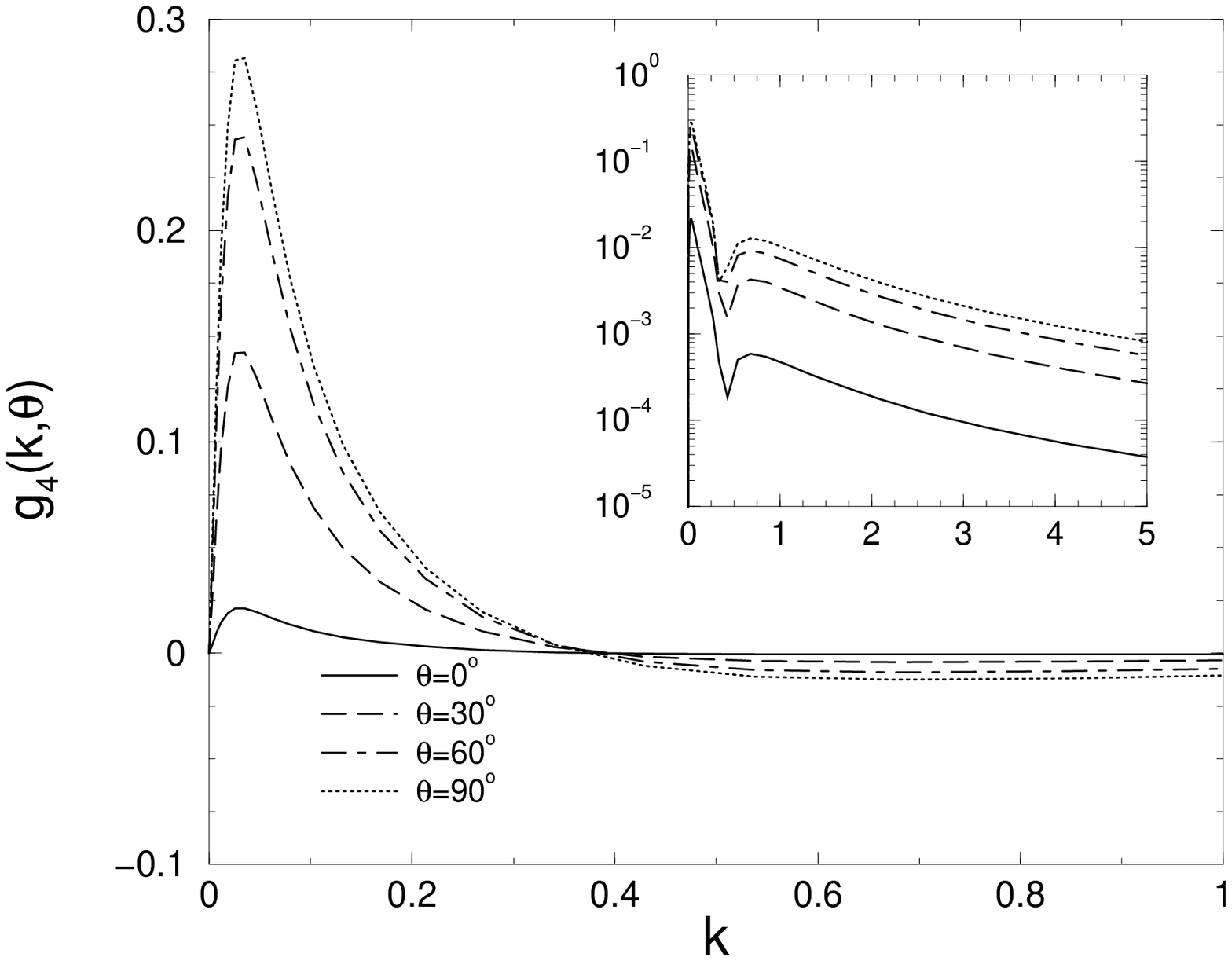}}
\caption{$g^{a=1}_{i=1,\ldots,4}$ solutions for scalar coupling with $\alpha=1.18$,
$\mu=0.25$ and sharp cutoff at $k_{max}=10$.
Corresponding binding energy is $B=0.0498$}\label{g1g2g3g4_J=1_a=1}
\end{center}
\end{figure}

\begin{figure}[!h]
\begin{center}
\epsfxsize=10cm{\epsffile{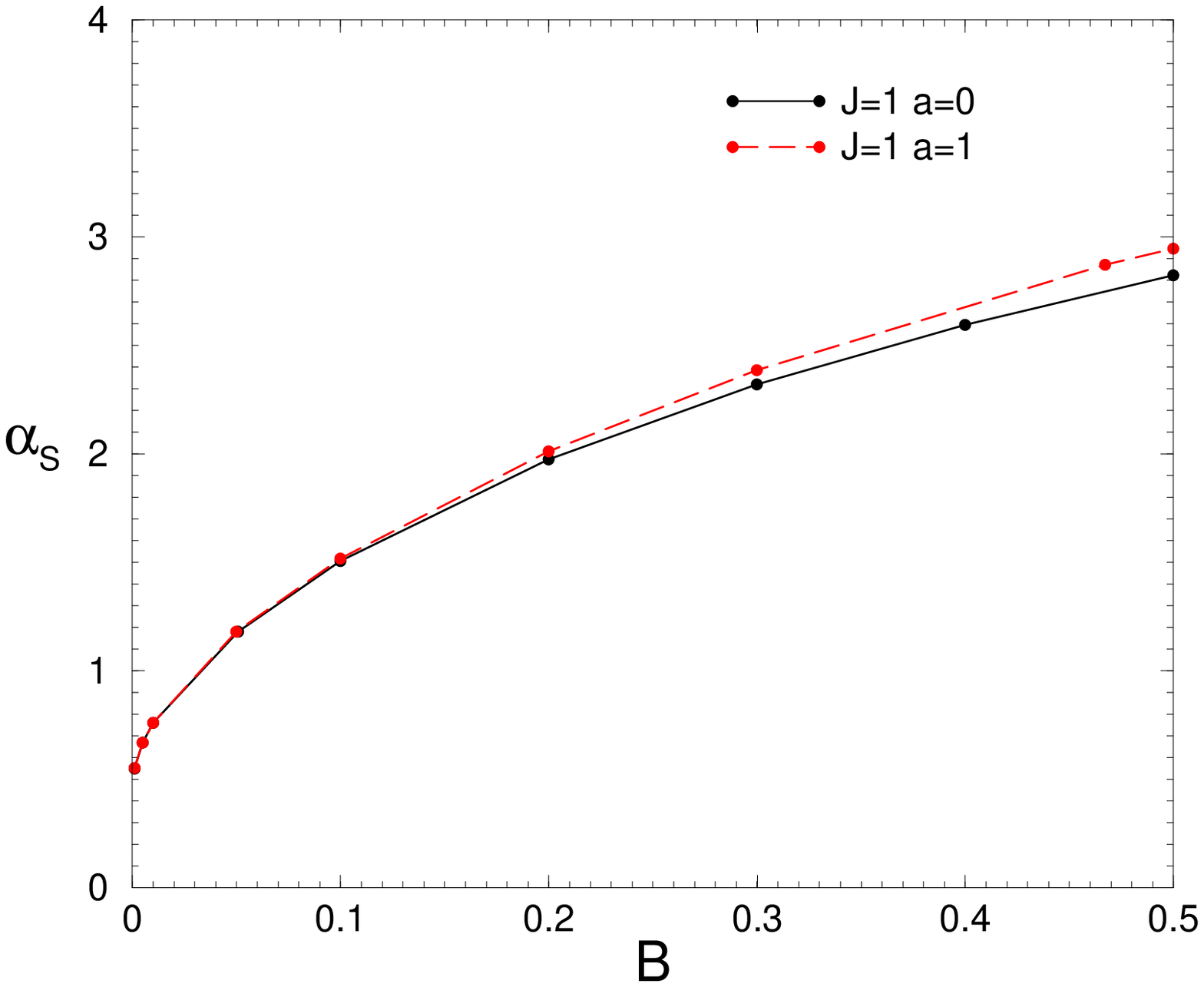}}
\caption{Splitting of the J=1 solutions for the
scalar coupling. Results correspond to $\mu=0.25$ and sharp cut-off at
kmax=10}\label{alpha_B_J1_S}
\end{center}
\end{figure}

The $g^{(a)}_{i}$  solutions for a=0  and a=1 states are
respectively represented in Figs. \ref{g1g2_J=1_a=0} and
\ref{g1g2g3g4_J=1_a=1} for several values of $\theta$. They were
obtained with a coupling constant $\alpha_s=1.18$ and a sharp
cutoff at $k_{max}=10$. We remark that with the conventions used
$g^{(0)}_{2}(k,0)=0$ and on has
$g^{(0)}_{1}(0,0)=-g^{(0)}_{2}(0,90^\circ)$, as expected from (\ref{b_0}).
In addition:
$g^{(0)}_{1}(0,0)=-g^{(0)}_{2}(0,90^\circ)\approx
g^{(1)}_{1}(0,\theta)$, as expected from  (\ref{b_1})
and from the fact that  coefficient $c_i$, defined in
(\ref{ph2}), are very close to the values (\ref{ph3}).
Corresponding binding energies
are $B_{a=0}=0.0506$ and $B_{a=1}=0.0498$, values which are 1\%
close to each other. The splitting of the binding energies is
an increasing function of the coupling constant. Figure
\ref{alpha_B_J1_S} shows the calculated $B_{a}(\alpha)$ dependence
for both J=1 eigenstates. For $\alpha_s=0.55$ the values are
respectively $B_{a=0}=9.7\,10^{-3}$ and $B_{a=1}=9.6\,10^{-3}$
whereas for $\alpha_s=2.87$ $B_{a=0}=0.523$ and $B_{a=1}=0.467$.
The non degeneracy remains reasonably small even for strongly
bound systems.

\begin{figure}[!htp]
\begin{center}
\mbox{\epsfxsize=8.5cm\epsffile{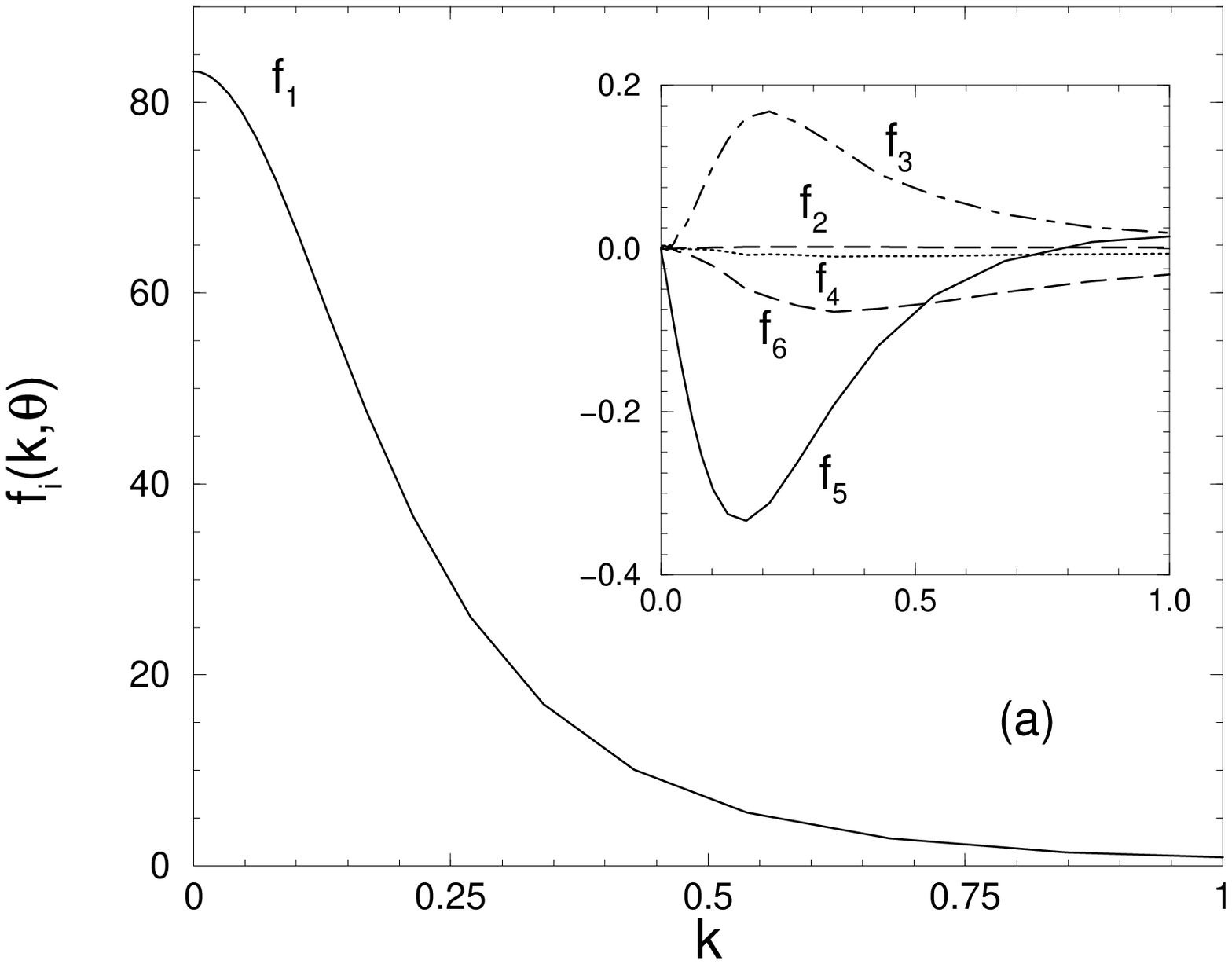}}
\hspace{0.5cm}
\mbox{\epsfxsize=8.5cm\epsffile{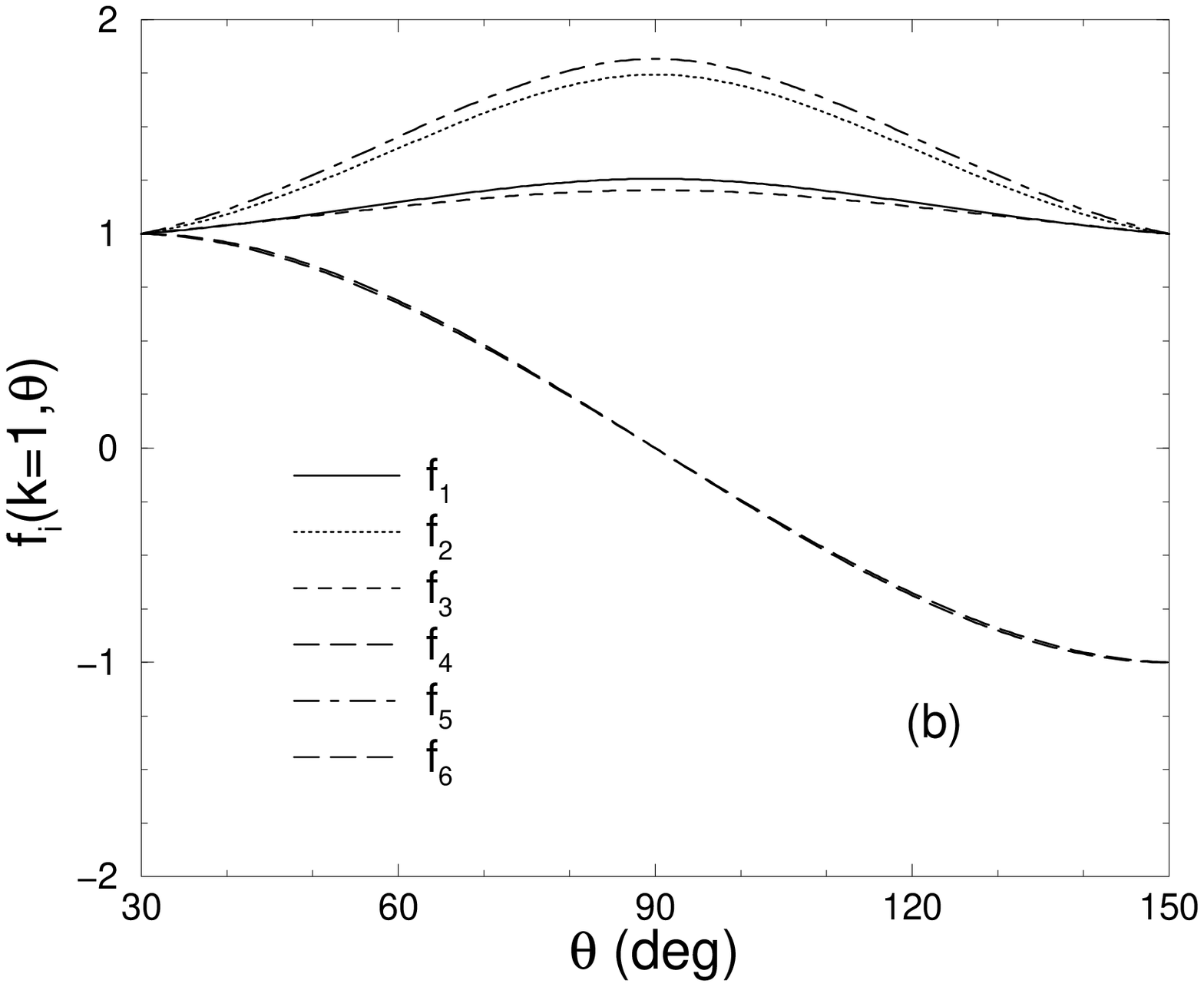}}
\caption{Wave function components $f_i$ of the physical solutions (a) as a function of $k$
at $\theta=30^o$ and (b) their $\theta$-dependence at fixed $k=1$ value.
Calculations are for the scalar coupling with $\alpha=1.18$, $\mu=0.25$ and sharp
cut-off $k_{max}=10$. Binding energy is B=0.0501.}\label{fi_phys}
\end{center}
\end{figure}

The six components $f_i$ of the $J=1^+$ physical wave function are determined by
a linear combination (\ref{ph30}) of functions $f_i^{(a)}$
which  in their turn are expressed in terms  of $g_i^{(a)}$  by (\ref{ff0}) and (\ref{eq4c}).
We remind that coefficients $c_a$ of this linear combination
are computed from components $g_1^{(a=0,1)}$ only.
For the solutions presented in Figs. \ref{g1g2_J=1_a=0} and \ref{g1g2g3g4_J=1_a=1}
they are found to be $c_0=0.582$ and $c_1=0.813$
 and the corresponding energy is B=0.0501.
Note that these values are very close to  those  obtained in case
of $\vec{n}$-independent interactions (\ref{ph3}): $c_0={1\over\sqrt3}=0.577$ and
$c_1=\sqrt{\frac{2}{3}}=0.816$.
They become even closer to these values
for smaller binding energies and they smoothly depart for strongly bound
systems. For a state with $B\approx 0.5$ and the same sharp cutoff $k_{max}=10$
one has for instance  $c_0=0.610$  and $c_0=0.793$.
Components $f_i$ thus obtained are displayed in Fig. \ref{fi_phys}
for $\theta=30$ degrees in linear (a) and logarithmic (b) scales.
One can see that component $f_1$ dominates over all remaining five in all the momentum range.
Among the components of relativistic origin there is not a clear dominance.
 Notice the very small value of $f_2$ component,
corresponding to the tensor D-wave, that would be absent in a non
relativistic approach. These components have a definite parity in
variable $\cos\theta$, $f_{1,2,3,5}$ being even and $f_{4,6}$ odd,
as shown in Fig. \ref{fi_phys}b  for a fixed value $k=1$.

\section{Results for pseudoscalar coupling}\label{Res_Ps}

For pseudoscalar coupling, the stability analysis was performed using the same methods
than for the scalar one  \cite{MMB_These_01,MCK_LCM_01} and presents some peculiarities.

Equations for J=$0^+$ states are found to be stable without any
regularization. The asymptotic behavior of the pseudoscalar kernel
is the same than the scalar one it has a repulsive character which
do not generates instability. The results lead to a quasidegeneracy of the coupling
constants for binding energies which vary over all the physical
range $[0,2m]$. One gets for instance, $\alpha=55.4$ for $B=0.001$
whereas $\alpha=58.5$ for a binding energy 500 times bigger
$B=0.5$, showing an extreme sensibility of this model to small
variations of the coupling constant. The origin of this behavior
was found to lie in the second channel equation ($\kappa_{22}$)
and has been understood analytically \cite{MCK_LCM_01} with a
simple model. The use of form factors -- though not  required for
the convergence of solutions -- is necessary if one wishes to
eliminate this unusual $\alpha(B)$ dependence. Calculations have
thus been performed using form factors (\ref{BFF}) with n=1 and
$\Lambda$=1.3 as in the Bonn model.

\begin{figure}[!hp]
\begin{center}
\mbox{\epsfxsize=12.cm\epsffile{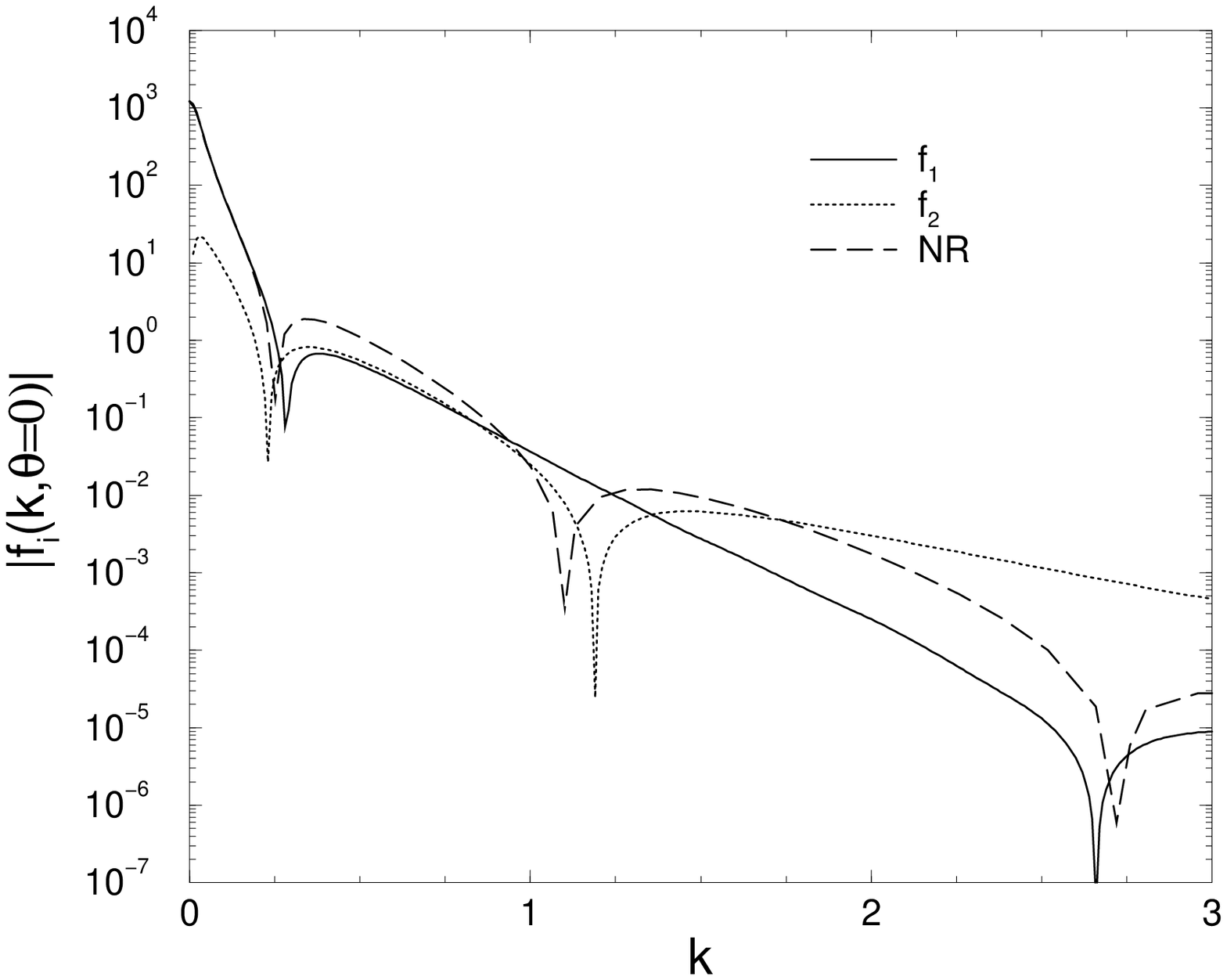}}
\caption{Wave function components (in logarithmic scale)  for $J=0^+$ state with B=0.001,
$\mu=0.15$ obtained with pseudoscalar coupling and form factor $\Lambda=1.3$.}\label{PS_f12NR_I=1_1MeV_ff_log}
\end{center}
\end{figure}

In the weak binding limit (B=0.001) one has  $\alpha_{LFD}=190$ and $\alpha_{NR}=166$,
a repulsive effect much stronger (15\%)  than in the scalar coupling.
Corresponding wave functions are shown in Fig. \ref{PS_f12NR_I=1_1MeV_ff_log}.
One can see that the component of relativistic origin
$f_2\approx f_1$ at $k\sim0.3$ and dominates above $k$=1.
A similar result was found in the np $^1S_0$ scattering wave function
calculated perturbatively with all the OBEP kernel in \cite{CK_NPA589_95}.
Contrary to the Yukawa model, the role of relativistic components
is crucial already for such a loosely bound system.
The coupling between components is also very important:
by switching off the non diagonal
kernels $K_{12}=K_{21}=0$ the coupling constant moves
from $\alpha_{LFD}=190$ to $\alpha_{LFD}=251$. It has thus
an attractive effect which tends to minimize the difference between LFD and NR results.
The comparison between $f_1$ and the non relativistic solution $f_{NR}$ shows
a very good agreement in the small $k$.
When  $k$ increases, large differences appear
and $f_{NR}$ has even an additional zero at $k=1.1$.

\begin{figure}[!hp]
\begin{center}
\mbox{\epsfxsize=12.cm\epsffile{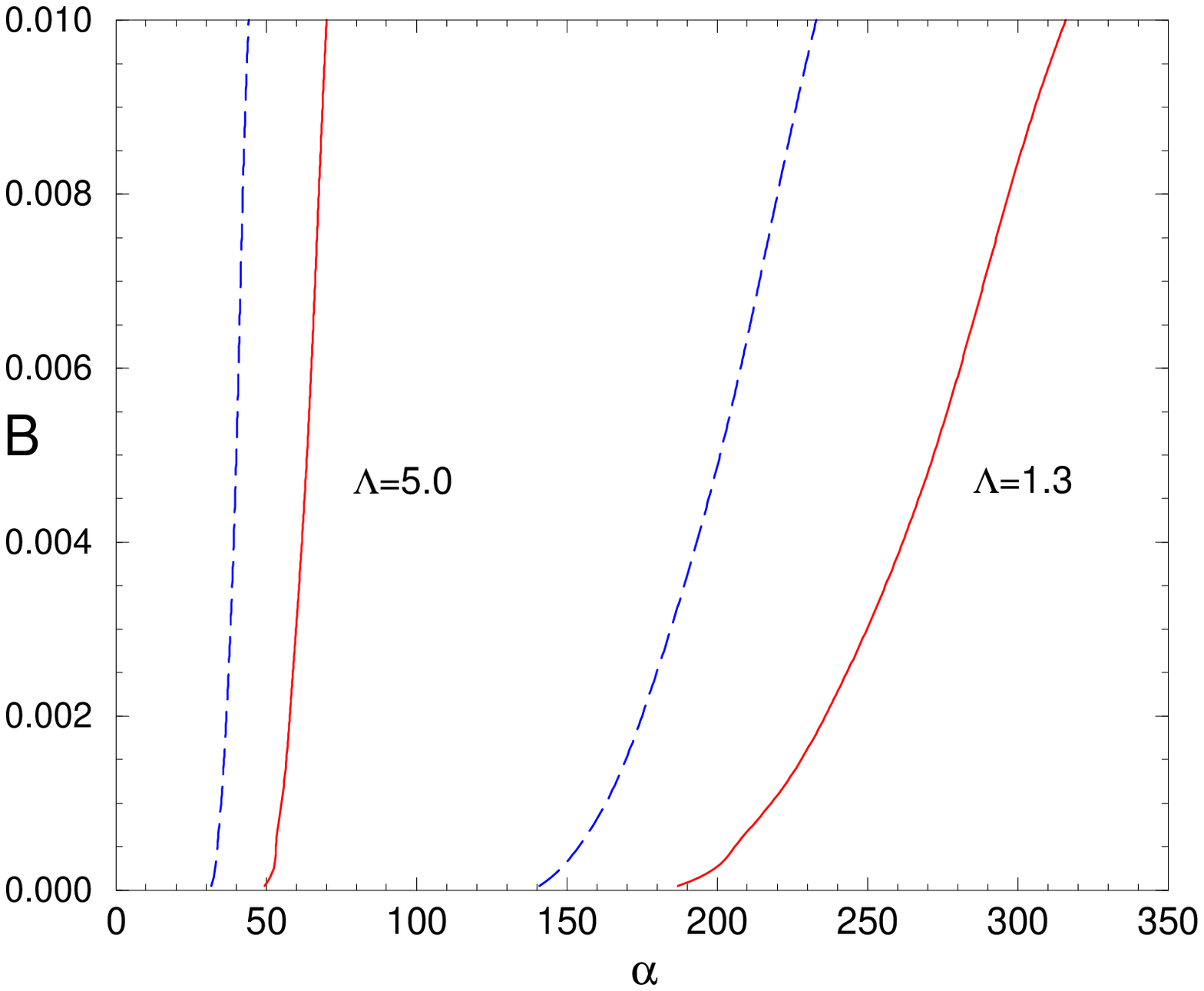}}
\caption{B($\alpha$) for pseudoscalar coupling and $J^{\pi}=0^+$ state
with $\mu=0.5$ and two different form factors compared to non relativistic results}\label{B=0_ps}
\end{center}
\end{figure}

It is worth noticing the dramatic influence of the form factor in all these calculations.
One has for instance
$\alpha_{LFD}$=103  for $\Lambda=5$ and
$\alpha_{LFD}$=1725 for $\Lambda$=0.3. We remind that the value used in the Bonn model
for this coupling is $\Lambda_{Bonn}=1.3$.

Quite surprisingly, in the strong binding limit (B=0.5)
we have found $\alpha_{LFD}$=1462 and $\alpha_{NR}$=3065.
Relativistic effects become now strongly  attractive ($\alpha_{LFD}<\alpha_{NR}$).
An essential part of  this attraction is due to the coupling  of the two $f_1-f_2$
components in the LFD wave function.
By performing one channel calculations, one has indeed $\alpha_{LFD}$=3001,
what represents a strong reduction in the effect though it remains slightly attractive.
We have checked if this attractive effect happens for
different values of the exchange mass $\mu$.
For the same binding energy ($B=0.5$) and $\mu=0.5$ we have found
$\alpha_{LFD}$=1728 and $\alpha_{NR}$=1400, repulsive once again.
It is worth noticing that for this coupling $\alpha_{NR}$
is a decreasing function of $\mu$ whereas $\alpha_{LFD}$ increases, at least in this energy region.
This tells us the difficulty of talking about the "sign of relativistic effects"
in general. They turn to depend not only
on the kind of coupling but also on the binding energy
of the system and - furthermore - on the mass of the exchanged particle.

\bigskip
It is interesting to study the zero binding limit of the LFD
results and compare them with the non relativistic ones. The NR
potential (\ref{eqnr4}) has been modified by including the Bonn
form factor (\ref{BFF}). The results are given in Fig.
\ref{B=0_ps} for an exchange mass  $\mu=0.5$ and with two
different cutoff parameters $\Lambda$ in the form factors. They
show the same behavior that was found in the scalar case
\cite{MC_PLB_00} i.e. that the relativistic and non relativistic
approaches do not coincide even when describing systems with zero
binding energies as far as they interact with massive exchanges.

\begin{figure}[!hp]
\begin{center}
\epsfxsize=12.cm\epsffile{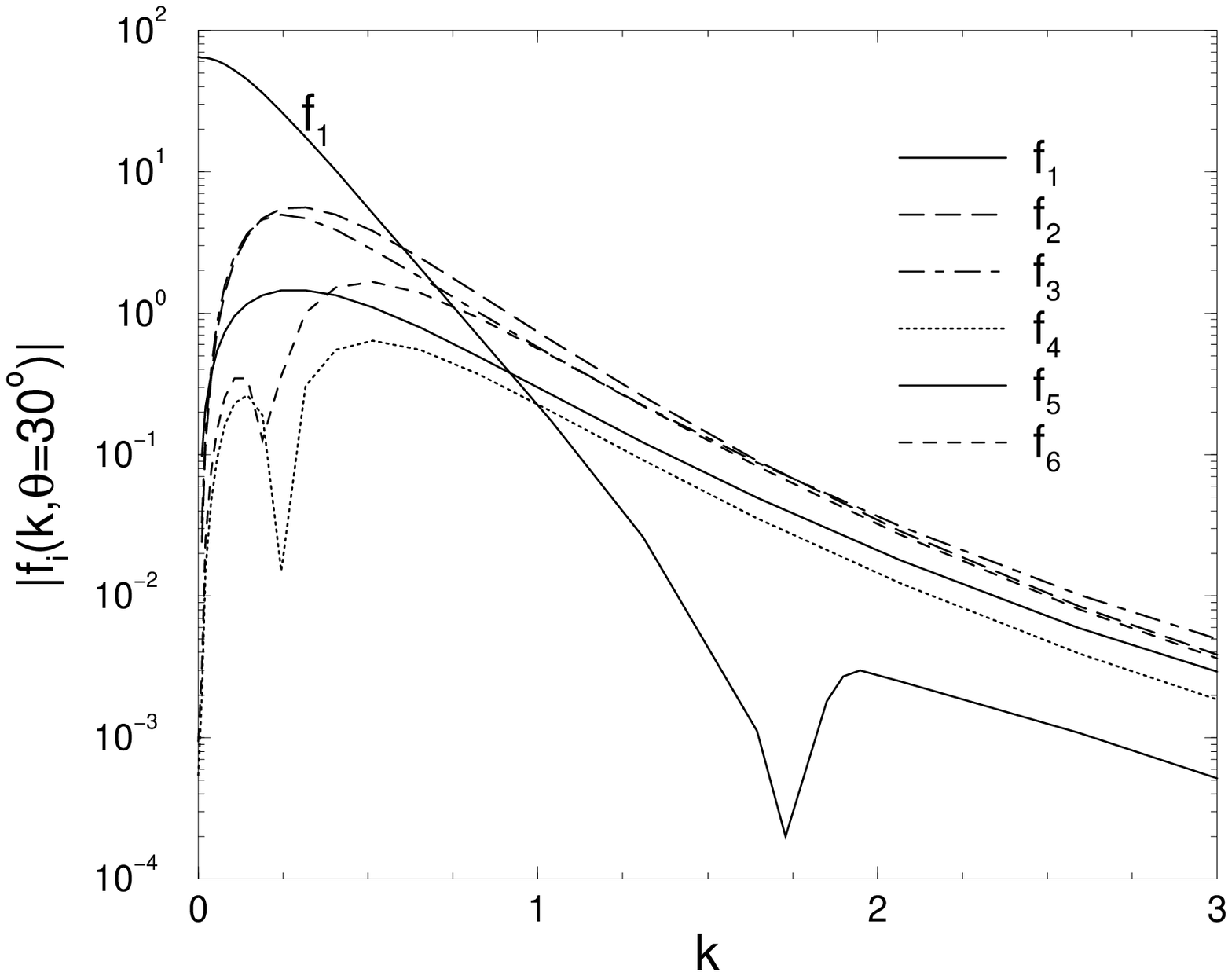}
\caption{Physical solutions for $J=1^+$ state with PS coupling.
Parameters are $\alpha=60$, $\mu=0.25$, $\Lambda=1.3$.
Corresponding binding energy is $B=0.079$ and components are
plotted for $\theta=30^o$.}\label{fi_phys_PS}
\end{center}
\end{figure}

\bigskip
The $J=1$ state displays the same kind of departures from the
scalar case than $J=0$. Functions $g^{(a)}_{i}$ for $a=0,1$ have
been calculated using the values  $\alpha_{PS}=60$, $\mu=0.25$ and
$\Lambda=1.3$. Contrary to the scalar case, binding energies are
sizeably different: $B_{a=0}=0.103$ whereas $B_{a=1}=0.0494$. The
physical wave function is obtained using the same procedure than
for the scalar case, i.e. compute $f^{a=0,1}_{i=1,6}$ and extract
from them the coefficients $c_i$. Their values, $c_0=0.749$ and
$c_1=0.662$, are different from $c_0\approx
{1\over\sqrt3},c_1\approx\sqrt{2\over3}$ with $c_0$ bigger than
$c_1$. The averaged binding energy is $B=0.0793$. The
corresponding solutions are plotted in Figs. \ref{fi_phys_PS}. One
can see that $f_1$ dominates at small momenta ($k<<1$) but
starting from $k\sim1$, the components of relativistic origin
become larger than $f_1$.

\begin{figure}[!h]
\begin{center}
\epsfxsize=12.cm{\epsffile{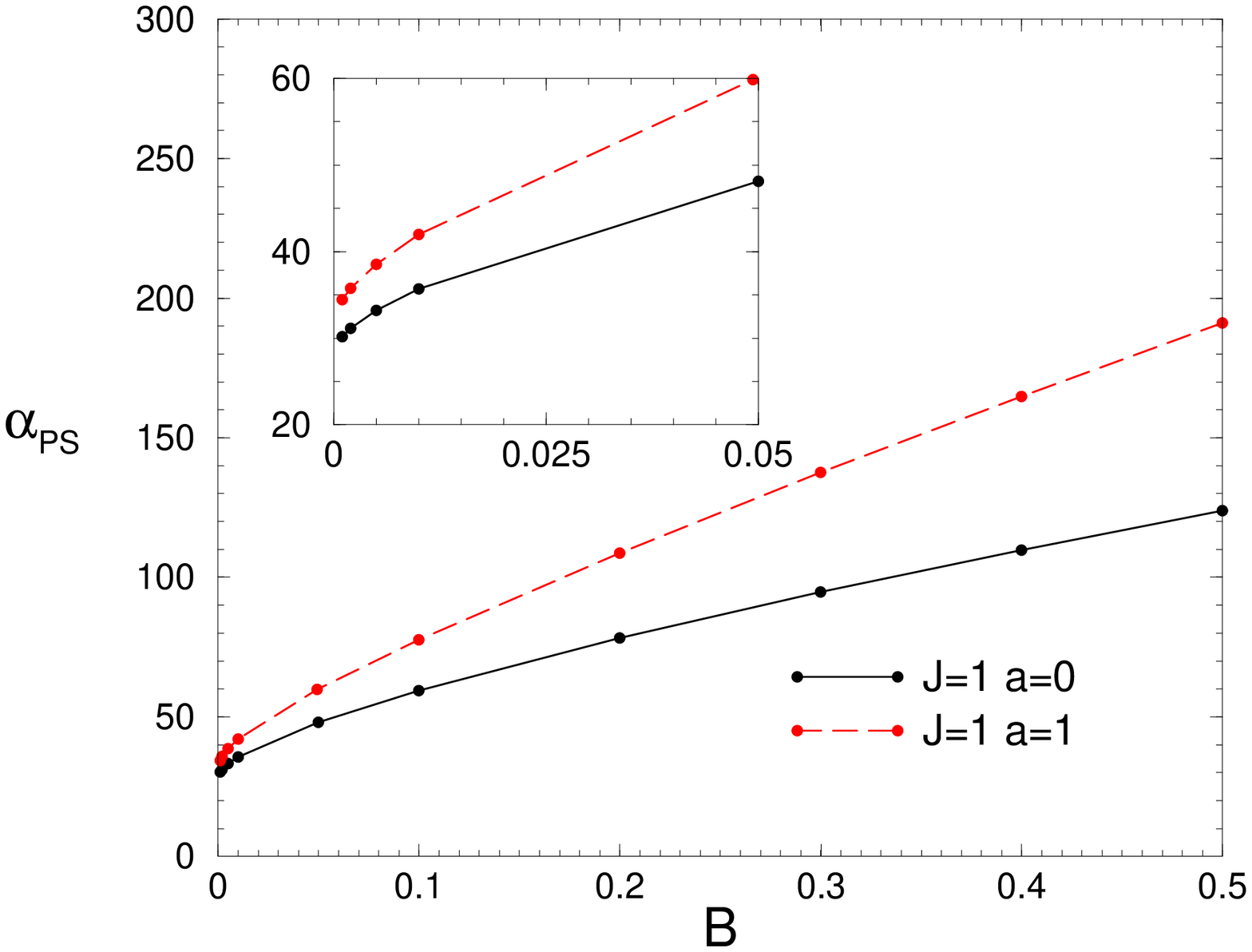}}
\caption{Splitting of the J=1 solutions for  pseudoscalar
coupling. Results correspond to $\mu=0.25$ and $\Lambda$=1.3, $n$=1}\label{alpha_B_J1_Ps}
\end{center}
\end{figure}

The splitting in binding energies is much bigger than for the scalar coupling.
It can be seen in Figure \ref{alpha_B_J1_Ps}a
where the results of $B_{a}(\alpha)$ for both $\hat{A}^2$ eigenstates are plotted.
The energy differences  remain important
even in the  $B\to0$ limit -- Figure \ref{alpha_B_J1_Ps}b --
in accordance with the analytical considerations in section \ref{NR}.

In summary, as it was noticed in Section \ref{NR},
pseudoscalar coupling displays the largest deviations with respect
to the non relativistic dynamics. Small and large spinor
components are mixed to the first order. The coupling between
$f_1$ and $f_2$ is essential even for very weakly bound systems,
the components of relativistic origin dominates already at
moderates values of $k$ and the splitting of the binding energies
for the different projections of the $J\neq0$ states are of the
same order than the energies themselves.

\section{Results for vector coupling}\label{Res_V}

The stability analysis applied to vector kernels shows that vertex form factors
are required for both $J=0^+$ and $J=1^+$ states to obtain stable solutions.

\begin{figure}[htbp]
\begin{center}
\mbox{\epsfxsize=12cm\epsffile{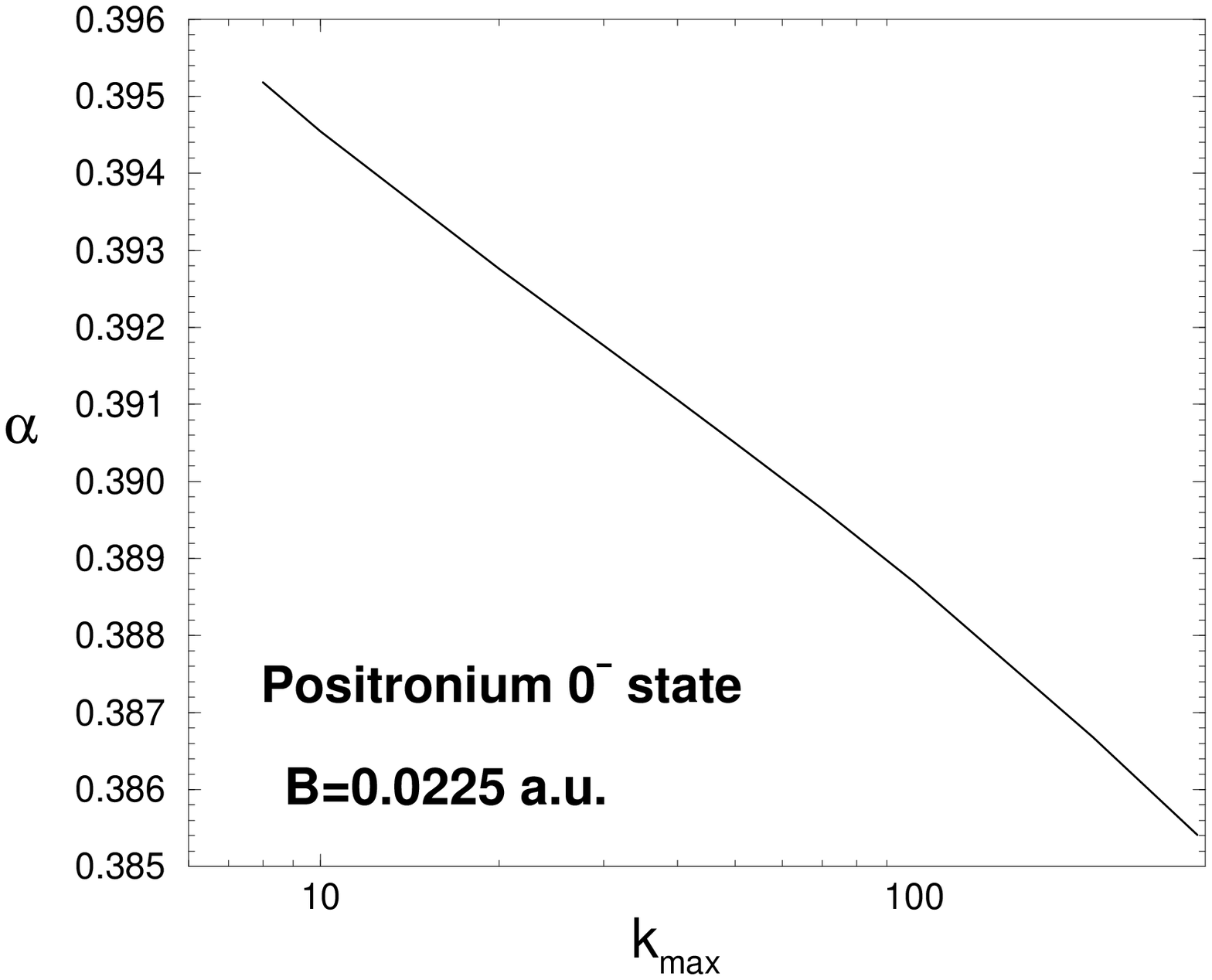}}
\caption{Coupling constant $\alpha$ as a function of the sharp
cut-off $k_{max}$ for the $J=0^-$ positronium state with binding
energy $B=0.0225$ a.u.}\label{alpha_kmax_positr}
\end{center}
\end{figure}

\bigskip
This is true in particular in the simplest application of vector
coupling: the positronium $J=0^-$ state. The negative parity of
the state comes from the intrinsic positron parity so that the
corresponding kernels are those of the $J^{\pi}=0^+$ two-fermion
system already given in Appendix \ref{ap1}. In Table \ref{pos2}
are presented the values of the coupling constant $\alpha$ as a
function of the sharp cut-off $k_{max}$ and for a fixed binding
energy $B=0.0225$. The dependence is very slow -- 0.3\% variation
for $k_{max}\in[10,300]$ -- but it actually corresponds to a
logarithmic divergence of $\alpha(k_{max})$ as it can be seen in
Fig. \ref{alpha_kmax_positr}. The origin of this instability is
the coupling to the second component, whose kernel matrix element
$\kappa_{22}$ has an attractive, constant asymptotic limit. If one
removes this component -- which has a very small contribution in
norm -- calculations become stable and give for $\alpha_{NR}=0.30$
the value $\alpha_{LFD}=0.3975$.
\begin{table}[htbp]
\caption{Coupling constant $\alpha$ as a function of the sharp
cut-off $k_{max}$ for the $J=0^-$ positronium state with binding
energy $B=0.0225$ a.u.}\label{pos2}
\begin{tabular}{|c||c|c|c|c|c|c|c|c|c|}\hline
$k_{max}$ &  10  & 20   & 30   &  40  &  50  &  70  & 100  & 200
& 300  \\\hline\hline $\alpha$
&0.3945&0.3928&0.3918&0.3911&0.3905&0.3896&0.3887&0.3867&0.3854\\\hline
\end{tabular}
\end{table}

 The comparison of LFD ladder results with those obtained in
perturbative QED or to the physical energies is meaningless due to
the instability of the solutions themselves. The use of vertex
form factors in a system of pointlike particles would be hazardous
and the introduction of renormalizable counterterms seems to be a
more appropriate cure.

First positronium results in Light Front Dynamics were obtained in
\cite{GPW_PRD45_92,TP_NP90_00}. These authors introduced a large
number of states in the Fock expansion but observed the same
instability of the solutions. For a fixed value of the cut-off,
the results become finite and can be compared.  By
taking $k_{max}=10$ and $\alpha=0.3$ -- which corresponds to
$B_{NR}=0.0225$ -- we found $B_{LFD}=0.0132$, i.e. repulsive
relativistic effects.
The leading order  QED corrections \cite{BS_QM_77} reads
\[ B_{QED}=  {\alpha^2\over 4}  \left[1 + \frac{21}{16}\alpha^2 + o(\alpha^4) \right]\approx 0.02516,\]
and are so attractive. Equation (10) from \cite{TP_NP90_00} gives
for $k_{max}=10$ the value $B_{DLC}=0.0308$, in qualitative
agreement -- thought still sizeably different -- with $B_{QED}$.
We should notice that a recent work \cite{Tritmann_IJMP_01}
analyzes the results of \cite{TP_NP90_00} in terms of flow
equations and obtains a closer value $B_{DLC}=0.02341$. We
conclude from that, that the ladder LFD predictions for such a
genuine system  are unable to reproduce even the sign of first
order relativistic corrections. Because the lowest order
corrections of the singlet state are not affected by the
annihilation channels, the differences could be due to cross
ladder graphes.

\bigskip
For $\mu\neq0$, the two fermion system is  bound due to the $\mu$-dependent
terms ($\sim {tt'\over\mu^2} \upsilon_{ij}$) in the vector kernel (\ref{vkern}),
since the $\mu$-independent ones ($\chi_{ij}$) are repulsive.
This binding disappear in the non relativistic limit.

When solving  the equations for $J=0^+$ state,
the standard form factors (\ref{BFF}) -- depending on $Q^2$ and local
in the non relativistic limit -- were found to be insufficient for
any power $n$ to ensure stable solution. A $Q^2$ dependent gaussian form
factor failed as well.
This unstability comes from the $\mu$-dependent terms.
These are off-shell corrections depending on variables $t,t'$ defined by
\begin{equation}\label{ttp}
4m^2t =4\varepsilon^2_{k}-M^2,\qquad
4m^2t'=4\varepsilon^2_{k'}-M^2
\end{equation}
and are not regularized by  a form factor depending on variable $Q^2$.
Such a function  cuts off the high
$|\vec{k}-\vec{k'}|$ components, but not the $|\vec{k}+\vec{k'}|$ ones.
A similar situation is encountered in the framework
of chiral perturbation theory \cite{EGM_NPA67_00} and was solved by
the replacement $\kappa(k,k')\to F(k)\kappa(k,k')F(k')$.

Our way of doing is the following.
Variable $Q^2=-(k_{meson}-\omega\tau_1)^2$ entering
$F(Q^2)$  is associated with the off-energy shell
effects in the intermediate state containing one massive meson ($\mu$).
In a similar way, we introduce  the variable
$\eta=m^2-(k_1-\omega\tau)^2$ --  see vertex 2 in the first graph of Fig. \ref{fkern} --
and correspondingly $\eta'=m^2-(k'_2-\omega\tau')^2$ from  vertex 1.
Variables $\eta,\eta'$ control the  off-energy shell contribution to the fermion states
and have been regularized by means of a cut-off function
\[H(\eta)=\left(\Lambda^2\over \Lambda^2 + \eta\right)^n. \]
This corresponds to a non-local form factor even in the non relativistic limit.
On energy shell one has $\eta=\eta'=0$.

Thus, for instance, the  total form factor associated with vertex 2 in of Fig. \ref{fkern} reads:
\begin{equation}\label{NLFF}
F_{nloc}(Q^2,\eta)=F(Q^2)H(\eta)
\end{equation}

In center of mass variables (\ref{cdm}) the expressions for $\eta,\eta'$ are:
\begin{equation}\label{eta}
\eta=\left\{
\begin{array}{ll}
(1-{\vec{k}\cd \vec{n}\over \varepsilon_{k}})\;
2m^2t &  \rm{if \;\; {\vec{k}\,'\cd \vec{n}\over \varepsilon_{k'}}-{\vec{k}\cdot \vec{n}\over \varepsilon_{k}} >0} \\
(1+{\vec{k}\cd \vec{n}\over \varepsilon_{k}})\;
2m^2t & \rm{if \;\; {\vec{k}\,'\cd \vec{n}\over \varepsilon_{k'}}-{\vec{k}\cd \vec{n}\over \varepsilon_{k}} <0}
\end{array}
\right.
\end{equation}
and
\begin{equation}\label{etap}
\eta'=\left\{
\begin{array}{ll}
(1+{\vec{k}\,'\cd \vec{n}\over \varepsilon_{k'}})\;
2m^2t' &  \rm{if \;\; {\vec{k}\,'\cdot \vec{n}\over \varepsilon_{k'}}-{\vec{k}\cd \vec{n}\over \varepsilon_{k}} >0} \\
(1-{\vec{k}\,'\cd \vec{n}\over \varepsilon_{k'}})\;
2m^2t' &  \rm{if \;\; {\vec{k}\,'\cd \vec{n}\over \varepsilon_{k'}}-{\vec{k}\cd \vec{n}\over \varepsilon_{k}} <0} \end{array}
\right.
\end{equation}
with $t,t'$ given defined in (\ref{ttp}).

Each coupling constant is  replaced by $g\to g F(Q^2)H(\eta)$ -- or $g\to g
F(Q^2)H(\eta')$ --
and the kernel is  multiplied by $F^2(Q^2)H(\eta)H(\eta')$.
The values for $\Lambda$ and $n$ in $H$ are taken the same than for $F(Q^2)$,
but could in principle be different.

\begin{figure}[!htbp]
\begin{center}
\mbox{\epsfxsize=8.5cm\epsffile{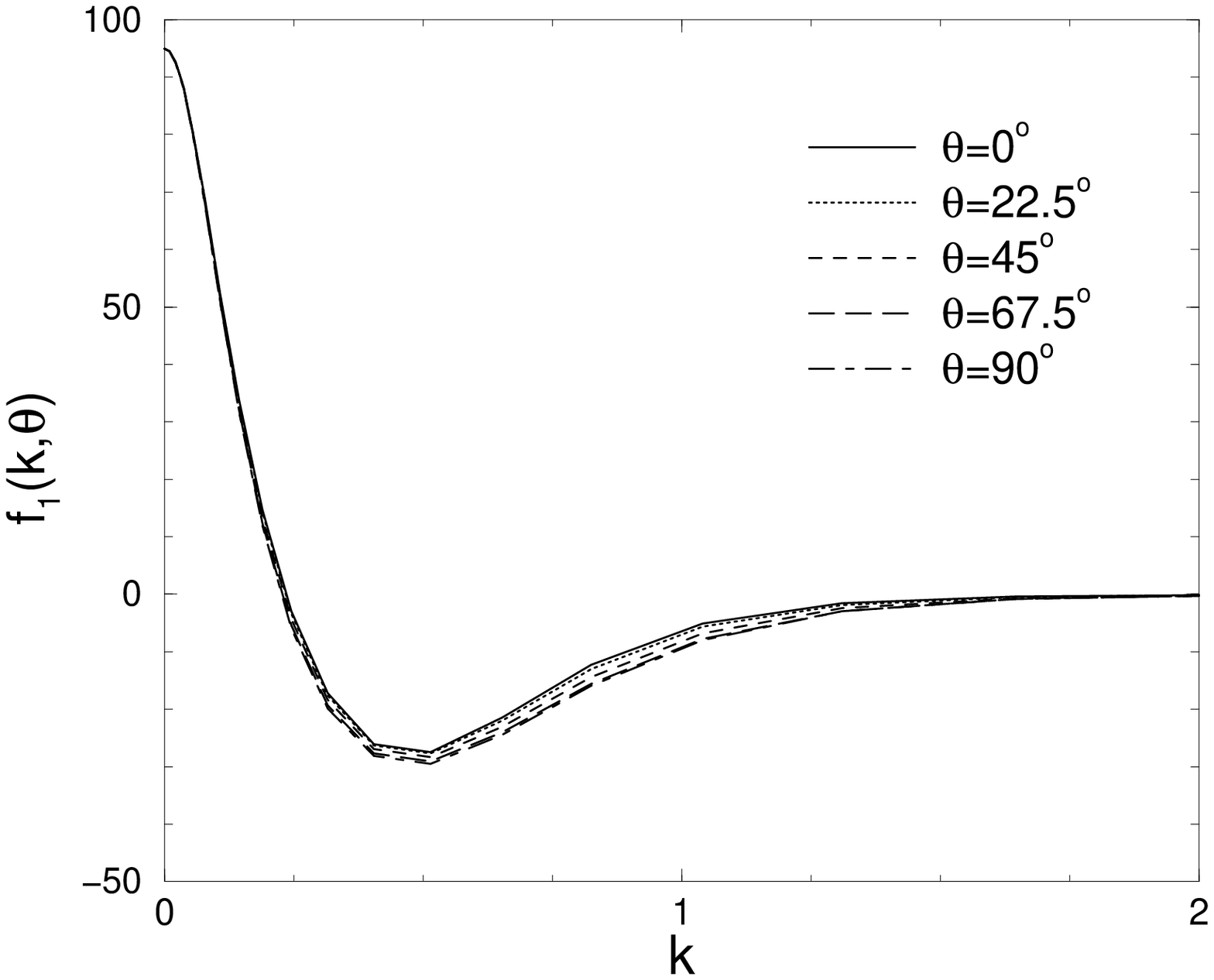}}\hspace{0.5cm}
\mbox{\epsfxsize=8.5cm\epsffile{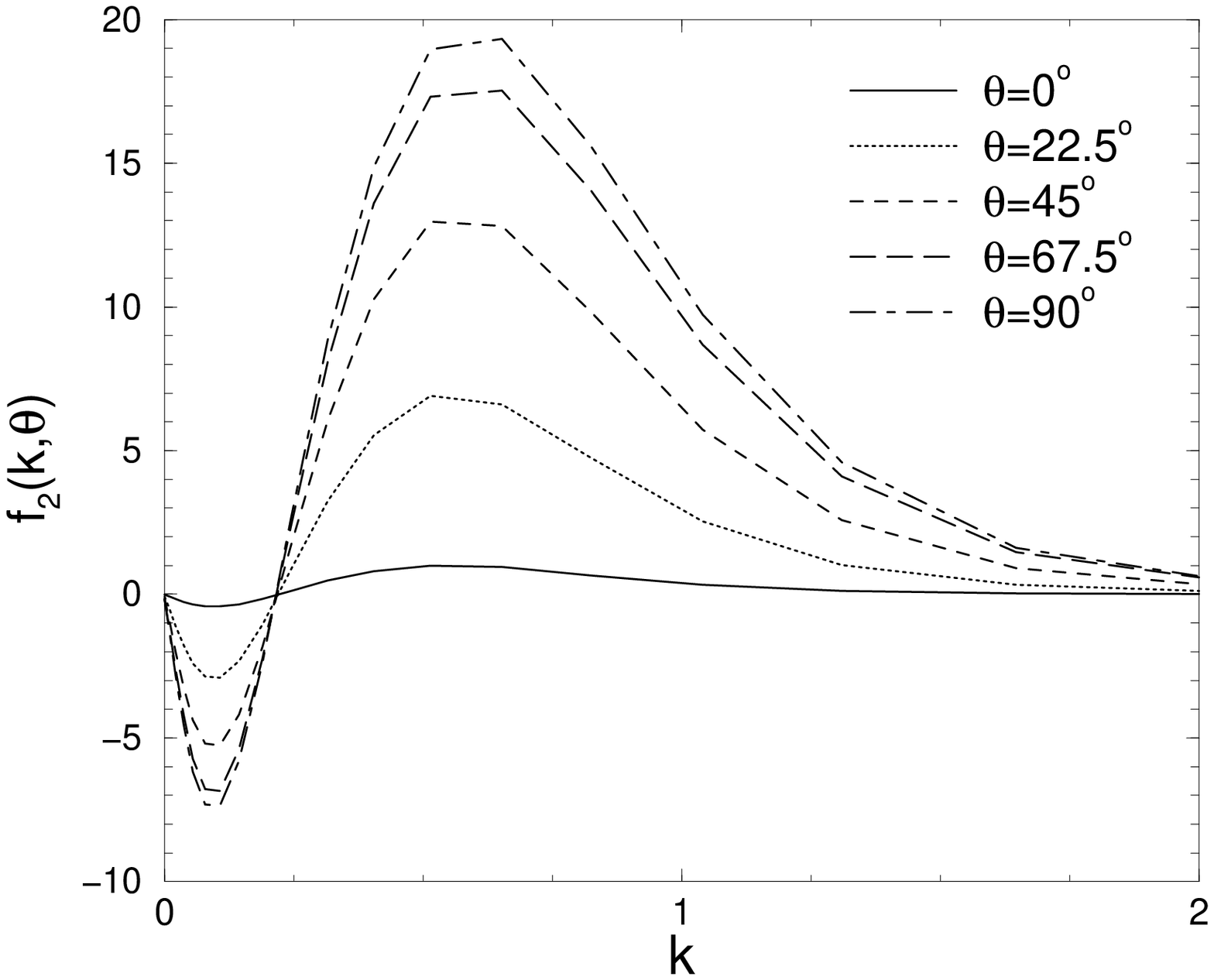}}
\caption{Wave functions $f_i$ for a $J^{\pi}=0^+$ state
in the vector coupling  with $\mu=0.15$
and using the non-local form factor (\protect{\ref{NLFF}}) with n=1 and $\Lambda=1.3$.
The coupling constant is $\alpha=1.485$ and the binding energy $B=0.0225$.}\label{new_ff_1}
\end{center}
\end{figure}

By means of (\ref{NLFF}), the  solutions become stable but
we notice that the use of only one kind of form factor is not enough to ensure the stability.
Wave functions corresponding to $\mu=0.15$ obtained with $n=1$ and $\Lambda=1.3$ in  (\ref{NLFF})
are displayed in Fig. \ref{new_ff_1}.
Binding energy is B=0.0225 and $\alpha_v=1.485$.
They have normal behavior and one remarks sizeable relativistic
component $f_2$ starting from  $k\approx0.5$ with
a strong $\theta$-dependence despite the small binding energy of the state.

\bigskip
Let us now consider the $J^{\pi}=1^+$ state. Solving the
$J^{\pi}=1^+,a=0$ equations with  the $F(Q^2)$ form factor only,
leads to the same anomalies than for $J^{\pi}=0^+$. With the
non-local form factor the situations is regularised. With
parameters $B=0.050$, $\mu=0.25$,  $\Lambda=1.3$ and $n=1$ for instance,
one has a coupling constant  $\alpha=6.18$ and a well behaved wave function.
The same happens for the $J^{\pi}=1^+, a=1$ state. When
using, with the same parameters,  the non-local form factor
(\ref{NLFF}), we get $\alpha=6.01$.

\begin{figure}[!h]
\begin{center}
\epsfxsize=12cm{\epsffile{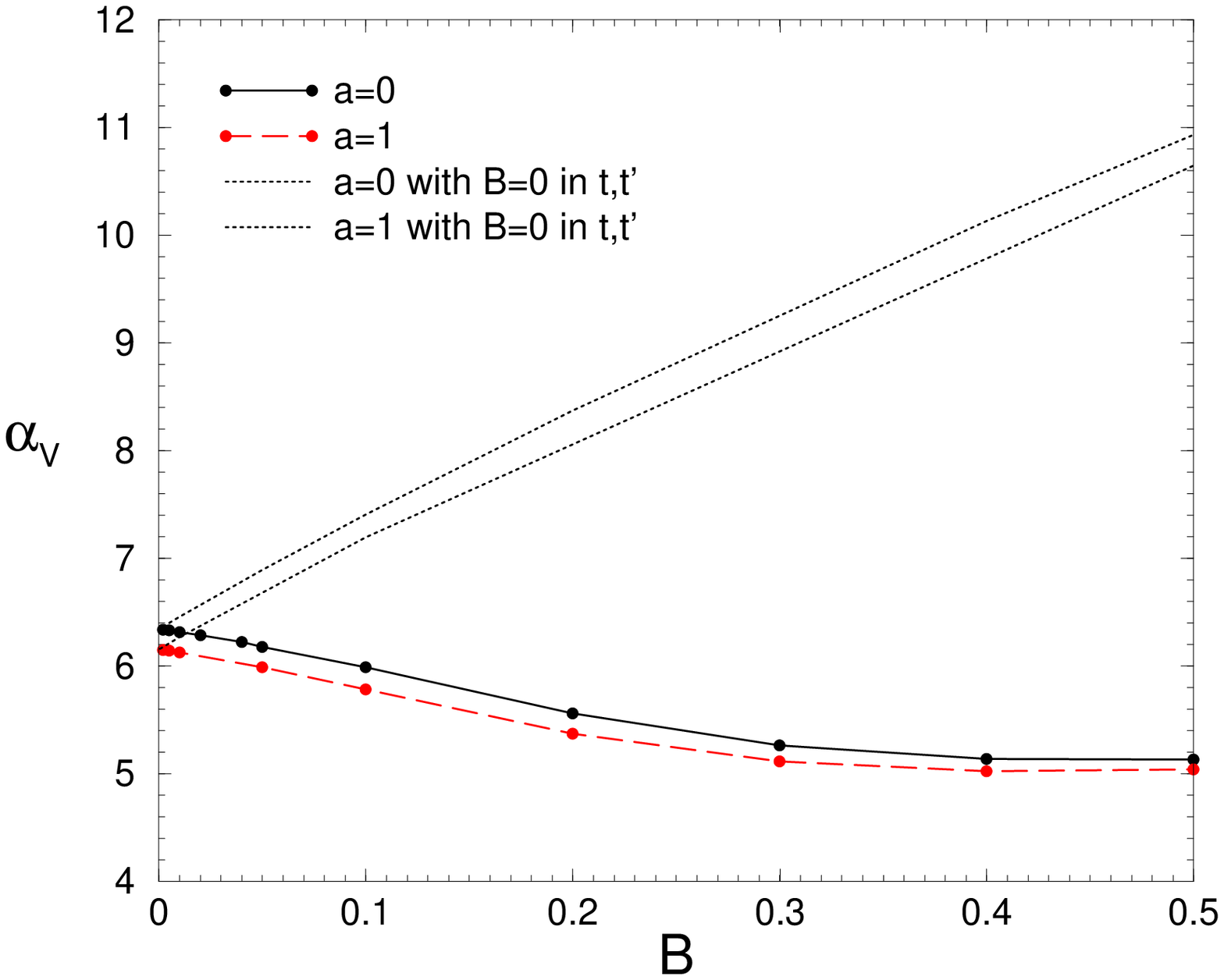}}
\caption{Splitting of the $J=1^+$ solutions for
vector coupling with $\mu=0.25$ and form factors (\protect{\ref{NLFF}}) with $\Lambda=1.3$ n=1.
Dotted lines correspond to a fixed binding energy (B=0)
in $t,t'$ off-shell variables of kernel (\protect{\ref{vkern}})}\label{alpha_B_J1_V}
\end{center}
\end{figure}

The mass splitting between the two $a=0,1$ projections is
shown in Fig. \ref{alpha_B_J1_V}. One first remark the striking
behavior of $\alpha_a(B)$ curves, i.e. larger binding energies
correspond to smaller values of the coupling constant $\alpha$.
This fact -- which takes place also for $J=0^+$ states -- is a consequence of the $M^2$-dependence of the
$\frac{tt'}{\mu^2}$ terms driving the vector kernel
$\kappa_{ij}^{(\mu)}$ in (\ref{vkern}). Its contribution is large,
because of $\mu^2$ in denominator. Increasing the binding energy
-- i.e. decreasing $M^2$ -- increases $t,t'$ factors, and results
into smaller values of $\alpha$. When the $M^2$ dependence in
$t,t'$ kernel is frozen -- setting e.g. $M^2=4m^2$  -- the usual
$\alpha(B)$ variation is recovered (dotted curve in Figure
\ref{alpha_B_J1_V}). When including the full dynamics, both
$\alpha_a(B)$ curves get close each other in all the variation
domain $B=[0,0.5]$, as it was the case in the scalar coupling.
However due to their peculiar behavior -- flat and almost parallel
-- the splitting in binding energies corresponding to a fixed
value of the coupling constant, can be very large.
 One can also remark in  Fig. \ref{alpha_B_J1_V}
the different values of $\alpha_a$ at B=0 despite the fact that the
systems of equations for a=0 and a=1 have -- as in the scalar coupling -- the same non-relativistic limit.
This difference is due to the $1/\mu^2$ terms in the kernel. They are
not relevant at the $(1/m)^0$ order but are crucial
for binding a relativistic two-fermion system by vector exchange.
For a fermion-antifermion system with massless exchange,
e.g. positronium, the splitting at B=0 disappears.

\section{Conclusion}\label{concl}

We have presented the explicitly covariant LFD solutions for the bound
state of two fermion systems in the ladder approximation.
 A method for constructing non zero angular momentum states
has been proposed and illustrated by numerical examples.
It is based on satisfying the angular condition
by a linear superposition of eigenstates
of an operator commuting with the LFD ladder hamiltonian.

\bigskip
We have separately examined the different types
of OBE couplings and found  very different behaviours
concerning the stability of the  solutions themselves
and their relation with the corresponding non relativistic reductions.

\bigskip
Scalar coupling (Yukawa model) is found to be stable  without any kernel
regularization for the $J^{\pi}=0^+$ state and coupling constants below
some critical value $\alpha<\alpha_c=3.72$. For values above $\alpha_c$ the system collapses.
For $J^{\pi}=1^+$ state the solutions of both a=0 and a=1 projections are unstable.
Their energy splitting is very small
even for binding energies (B) of the same order than the constituents mass and vanishes at B=0.
The physical solution, satisfying the angular condition, has
been constructed by a suitable linear combination of a=0,1 states.
LFD binding energies are found to be close to those given by their non relativistic limit,
even closer than the case of purely scalar particles (Wick Cutkosky model extended to $\mu\neq0$).
The comparison with the non relativistic solutions shows always repulsive effects.
The LFD wavefunction is dominated by the components which has
non-relativistic counterpart. Extra components of relativistic
origin remain negligible even at large values of the relative momentum ($k>m$).

\bigskip
Pseudoscalar coupling is also found stable for $J^{\pi}=0^+$ state.
It displays a very strong dependence of binding energies
as a function of the coupling constant: they vary from B=0.001 to
B=0.500 (in constituent mass units) while the coupling constant
changes from $\alpha=55.5$ to $\alpha=58.5$.
This dependence is due to the coupling to the  wave function component of relativistic origin.
Vertex form factors are required for $J^{\pi}=1^+$ states.
LFD solutions, obtained with regularized kernels, presents large deviations with respect to non relativistic
case, even for weakly bound states, and display a big sensitivity to the cut-off parameters.
The LFD wave function is dominated by relativistic component at relatively small
momenta ($k<m$).
The coupling between different components is strongly attractive
and can compensate the repulsive effects observed in the Yukawa model.
Thus, relativistic corrections  can be attractive or repulsive
depending on the quantum number of state, the value of the binding
energy and even the mass $\mu$ of the exchanged meson.
The energy splitting between different projections of $J=1$ states
is large and remains at B=0.

\bigskip
Vector coupling presents the stronger anomalies.
For $\mu=0$ it has been applied to positronium  $0^+$ state. It is found to be unstable
and, once regularized by means of sharp cut-off, the ladder approximation gives relativistic
corrections of opposite sign compared to QED perturbative results.
This failure shows the poorness of the ladder approximation
in one of the rare cases in which it can be confronted to experimental results.
For $\mu\neq0$ the LFD solutions collapse even using local cutoff form factors. The reason lies
in the strong non-localities of the $\mu$-dependent terms in the LFD kernel.
These terms have their origin in the massive vector propagator and
manifest as off-shell corrections of the $\mu=0$ kernels. They
have been regularized using appropriate vertex form factors.
The $J=1^+$ state has thus been calculated.
This state is not bound in the non relativistic limit
and its existence in a relativistic approach is entirely due to the $\mu$-dependent terms in the kernel
The importance of this off-shell terms is thus dramatic.
In particular their energy dependence generates a decreases of the binding energy as a
function of the coupling constant, what questions the very meaning of the interaction strength.
The $\alpha_a(B)$ dependence for different projections of $J=1$ states
remain very close to each other even for $B\sim m$ but their particular
form -- smooth and almost parallel variation --
can give rise to large energy splitting for a fixed value of the coupling constant.

\bigskip
Some general additional remarks concerning the relativistic calculations are given in order.
\begin{itemize}
\item [({\it i})]
Contrary to the non relativistic case, vertex form factors
are unavoidable in any realistic calculation.
The full spinor structure  generates highly singular kernels
which are not regularized by local vertex form factors.
It is clear that specially at large k-values, the obtained
wave function  and consequently the electromagnetic
form factors will crucially depend on the way the regularization is performed.
The large momentum components will thus be determined
not by the dynamics but by uncontrolled parameters.
We believe that here is the main drawback of relativistic approaches.

\item [({\it ii})] The consequences of implementing the Lorentz invariance in a
quantum mechanical description of a system are not only
kinematical but mainly dynamical. Large differences with respect
to the non relativistic solutions appear even in the zero binding
limit for systems with ${k\over m}\ll 1$ as far as the exchanged
mass is non zero. We have explicitly shown for scalar and
pseudoscalar couplings that the behavior of $\alpha(B)$ at
$B\approx0$ differs from their non relativistic counterparts, a
result already found in the Wick-Cutkosky model \cite{MC_PLB_00}.

\item [({\it iii})]
The question about the sign of relativistic effects has no simple answer.
They can be different, following:
the nature of the constituents, the kind of interaction,
the quantum numbers of the state, its binding energy,
and even the mass of the exchanged particle.
This shows that there are no simple recipes to perform {\it a priori} evaluations.

\item [({\it iv})] The splitting of different projections of J=1 states
is very different following the kind of coupling.
In nuclear physics -- where the weight of scalar mesons in the binding
energy is dominating -- is expected to be very small.
The same is true for the massless vector coupling like one-photon or one-gluon exchange.
It can be however very large in relativistic models where pseudoscalar exchange plays an important role.

\end{itemize}

Finally we would to emphasize one of the interest of using LFD in
describing the relativistic composite systems. It lies in the fact
that wave functions components appearing in this approach are
closely related to their non relativistic counterparts. Some of
these components are the formal equivalent of the usual non
relativistic solutions while others are of pure relativistic
origin. Relativity manifests both in modifying the former and in
giving a sizeable weight to the latter ones. We have found that
the coupling between these components plays an essential role,
even in determining the stability of the solutions. In addition --
except for the scalar exchange -- the total wave function is
dominated by the relativistic components at moderate values of its
arguments ($k< m$) and that, even for loosely bound systems.

\bigskip
{\bf Acknowledgements.} One of the authors (V.A.K.) is sincerely
grateful for the warm hospitality of the theory group at the
Institut des Sciences Nucl\'{e}aires de Grenoble, where this work
was performed. Numerical calculations were  carried out  at
CGCV (CEA Grenoble) and  IDRIS (CNRS). We thank the staff members
of these  organizations for their constant support. This work is
partially supported by the French-Russian PICS and RFBR grants
Nos. 1172 and 01-02-22002 as well as by the RFBR grant 02-02-16809.

\newpage
\appendix

\section{Kernels}\label{ap1}

Kernels $\kappa_{ij}$ are obtained from equations (\ref{kappa_S_J0}), (\ref{kappa_V_J0}),
(\ref{kappa_S_J1a0}) and (\ref{kappa_V_J1a0})  as traces of 4x4 matrices.
To calculate these traces, it is useful to express the scalar products between all the
concerned four-vectors in terms of variables $(k,k',\theta,\theta')$. They read:
\begin{eqnarray}\label{scal}
\omega^2      &=& 0   ,\nonumber\\
k_1^2         &=& k_2^2=k'^2_1=k'^2_2=m^2,\nonumber\\
\omega\cd k_1 &=& x\,\omega\cd p,\nonumber\\
\omega\cd k_2 &=& (1-x)\,\omega\cd p,\nonumber\\
\omega\cd k'_1&=& x'\,\omega\cd p,\nonumber\\
\omega\cd k'_2&=& (1-x')\,\omega\cd p,\nonumber\\
k_1\cd k_2    &=& 2\varepsilon_k^2 - m^2,\nonumber\\
k'_1\cd k'_2  &=& 2\varepsilon_{k'}^2 - m^2,\nonumber\\
k_1\cd p      &=& 2\varepsilon_k^2(1 - x) + \frac{1}{2}M^2 x,\nonumber \\
k_2\cd p      &=& 2\varepsilon_k^2 x+ \frac{1}{2}M^2 (1 - x) ,\nonumber\\
k'_1\cd p     &=& 2\varepsilon_{k'}^2 (1 - x') + \frac{1}{2} M^2 x',\nonumber \\
k'_2\cd p     &=& 2\varepsilon_{k'}^2 x+ \frac{1}{2}M^2 (1 - x') ,\nonumber\\
k_1\cd k'_1   &=& -kk'\sin\theta\sin\theta'\cos\phi+ 2 \varepsilon_{k'}^2 x
+2\varepsilon_k^2x'-2\varepsilon_k^2 xx'-2\varepsilon_{k'}^2 x x',\nonumber\\
k_2\cd k'_2   &=& -kk'\sin\theta\sin\theta'\cos\phi+ 2 \varepsilon_k^2 x+ 2 \varepsilon_{k'}^2 x' -
  2 \varepsilon_k^2 x x' - 2 \varepsilon_{k'}^2 x x',\nonumber\\
k_1\cd k'_2   &=& kk'\sin\theta\sin\theta'\cos\phi+2\varepsilon_k^2(1-x)(1-x')+2\varepsilon_{k'}^2xx',\nonumber\\
k_2\cd k'_1   &=& kk'\sin\theta\sin\theta'\cos\phi+2\varepsilon_{k'}^2(1-x)(1-x')+2\varepsilon_k^2xx',
\end{eqnarray}
where
\begin{equation}\label{x}
x=\frac{1}{2}(1-\frac{k}{\varepsilon_k}\cos\theta)\qquad
x'=\frac{1}{2}(1-\frac{k'}{\varepsilon_{k'}}\cos\theta')
\end{equation}

Using the above result, we have obtained the  analytical
expressions of $\kappa_{ij}$ kernels for $J^{\pi}=0^+,1^+$ states.
They  are written below, coupling by coupling,  in the form
\begin{equation}\label{cde}
\kappa_{ij}(k,\theta,k',\theta',\varphi')=
c_{ij}(k,\theta,k',\theta')+d_{ij}(k,\theta,k',\theta')\cos\varphi'+e_{ij}(k,\theta,k',\theta')\cos^2\varphi'
\end{equation}
with  coefficients $c_{ij},d_{ij},e_{ij}$
invariant under the transformation $(i,k,\theta)\leftrightarrow(i',k',\theta')$.
We introduce for shortness the notations
\[s\theta=\sin\theta\quad c\theta=\cos\theta \quad
S\theta=k\sin\theta \quad C\theta=k\cos\theta \quad c\varphi'=\cos\varphi'\]
-- plus corresponding primed -- and  the following quantities
\begin{eqnarray*}
b_{\pm}^2&=&m^2(\varepsilon_{k}^2+\varepsilon_{k'}^2)\pm2\varepsilon_{k}^2\varepsilon_{k'}^2\cr
\varepsilon_{k}^{\pm}&=&\varepsilon_{k}\pm m\cr
\Delta^{\pm}&=&\varepsilon_{k}^2\pm\varepsilon_{k'}^2
\end{eqnarray*}
Coupling constants appear through $\alpha={g^2\over 4\pi}$.

\subsection{Scalar}\label{app_sc}

Kernels for the scalar coupling were already given in \cite{MCK_PRD_01} and are included here for completeness.

$J=0^+$:
\begin{eqnarray}
{\kappa_{11}\over\alpha\pi}&=&
-[m^2\Delta^++2\varepsilon_k \varepsilon_{k'}(\varepsilon_k\varepsilon_{k'}-C\theta C\theta')]+\Delta^+S\theta S\theta' c\varphi'\nonumber \\
{\kappa_{12}\over\alpha\pi}&=& -m\Delta^- (S\theta' +S\theta c\varphi')\nonumber  \\
{\kappa_{21}\over\alpha\pi}&=& +m\Delta^-(S\theta+ S\theta' c\varphi')\label{eqap1}\\
{\kappa_{22}\over\alpha\pi}&=&
\Delta^+S\theta S\theta'-[m^2\Delta^++2\varepsilon_k\varepsilon_{k'}(\varepsilon_k\varepsilon_{k'}-C\theta C\theta')]c\varphi'\nonumber
\end{eqnarray}

$J=1^+,a=0$:
\begin{eqnarray}\label{eq11}
{\kappa_{11}\over\alpha\pi}&=&[2kk'\varepsilon_k \varepsilon_{k'} -b_+^2c\theta c\theta']
-\varepsilon_k \varepsilon_{k'}\Delta^+s\theta s\theta' c\varphi'  \\
{\kappa_{12}\over\alpha\pi}&=&m\varepsilon_{k'}(2\varepsilon_{k}^2+\Delta^+)c\theta s\theta'
-m\varepsilon_k (2\varepsilon_{k'}^2+\Delta^+)s\theta c\theta' c\varphi' \nonumber\\
{\kappa_{21}\over\alpha\pi}&=&m\varepsilon_{k}(2\varepsilon_{k'}^2+\Delta^+)s\theta c\theta'
-m\varepsilon_{k'}(2\varepsilon_{k}^2+\Delta^+)c\theta s\theta' c\varphi'\nonumber\\
{\kappa_{22}\over\alpha\pi}&=&-\varepsilon_k \varepsilon_{k'}\Delta^+s\theta s\theta'
+\left[2kk'\varepsilon_k \varepsilon_{k'}-b_+^2 c\theta c\theta'\right]c\varphi'\nonumber
\end{eqnarray}

$J=1^+,a=1$:
\begin{eqnarray*}\label{eq11_J1a1}
{2\kappa_{11}\over\alpha\pi}  &=&-\left\{
m\varepsilon_{k}s^2\theta(\Delta^++2\varepsilon_{k'}^2) + m\varepsilon_{k'}s^2\theta'(\Delta^++2\varepsilon_{k}^2)
+ (c^2\theta+c^2\theta')b_+^2 - 4\varepsilon_{k}\varepsilon_{k'}C\theta C\theta'\right\}\cr
&-& \left\{\Delta^+(\varepsilon_{k}^- \varepsilon_{k'}^- c\theta c\theta'-kk')s\theta s\theta' \right\}c\varphi'
- \left\{\Delta^+\varepsilon_{k}^-\varepsilon_{k'}^-s^2\theta s^2\theta'\right\} c^2\varphi'\cr
{2\kappa_{12}\over\alpha\pi}&=&
[m\varepsilon_{k}(\Delta^+ + 2\varepsilon_{k'}^2)-b_+^2]s^2\theta-
[m\varepsilon_{k'}(\Delta^++2\varepsilon_{k}^2)-b_+^2
]s^2\theta'\\
&-&\left\{kk'(\varepsilon_{k}-\varepsilon_{k'})^2
+(\varepsilon_{k}+\varepsilon_{k'})^2\varepsilon_{k}^-\varepsilon_{k'}^-c\theta c\theta'\right\}s\theta s\theta'c\varphi'
+\left\{\varepsilon_{k'}^-(\varepsilon_{k}+\varepsilon_{k'})^2 c^2\theta'
-\varepsilon_{k'}^+(\varepsilon_{k}-\varepsilon_{k'})^2
\right\}\varepsilon_k^- s^2\theta c^2\varphi'\cr
{\sqrt2\kappa_{13}\over\alpha\pi}&=&
-2\varepsilon_{k}\varepsilon_{k'}C\theta S\theta'
-\varepsilon_{k'}^-(m\Delta^+ -2\varepsilon^2_{k}\varepsilon_{k'}) c\theta's\theta'\cr
&+& \left\{  2\varepsilon_{k}\varepsilon_{k'}kC\theta' +\varepsilon_{k}^- c\theta
\left[\varepsilon_{k'}(\Delta^+-2m\varepsilon_{k})
-\varepsilon_{k'}^-(\varepsilon_{k}+\varepsilon_{k'})^2 c^2\theta'\right] \right\} s\theta c\varphi'
-(\varepsilon_{k}+\varepsilon_{k'})^2 \varepsilon_{k}^-\varepsilon_{k'}^- s^2\theta s\theta'c\theta' c^2\varphi'\cr
{\sqrt2\kappa_{14}\over\alpha\pi}&=&\Delta^-\left\{(m+\varepsilon_{k}^-s^2\theta)S\theta'
+mS\theta c\varphi' - \varepsilon_{k}^- s^2\theta S\theta'c^2\varphi'\right\}\cr
{2\kappa_{21}\over \alpha \pi}&=&
[m\varepsilon_{k'}(\Delta^+ + 2\varepsilon_{k}^2)-b_+^2]s^2\theta'-
[m\varepsilon_{k}(\Delta^++2\varepsilon_{k'}^2)-b_+^2 ]s^2\theta\\
&-&\left\{kk'(\varepsilon_{k}-\varepsilon_{k'})^2
     +(\varepsilon_{k}+\varepsilon_{k'})^2 \varepsilon_{k'}^- \varepsilon_{k}^- c\theta c\theta'\right\}s\theta s\theta'c\varphi'
+\left\{\varepsilon_{k}^-(\varepsilon_{k'}+\varepsilon_{k})^2c^2\theta
-\varepsilon_{k}^+(\varepsilon_{k'}-\varepsilon_{k})^2\right\}\varepsilon_{k'}^-s^2\theta'c^2\varphi'\\
{2\kappa_{22}\over\alpha\pi}&=&
m(\varepsilon_{k}+\varepsilon_{k'})^3
-\varepsilon_{k}^-(m\Delta^+ - 2\varepsilon_{k}\varepsilon_{k'}^2) c^2\theta
-\varepsilon_{k'}^-(m\Delta^+ - 2\varepsilon_{k}^2\varepsilon_{k'})c^2\theta'-4\varepsilon_{k}\varepsilon_{k'}C\theta C\theta'\\
&-&(\varepsilon_{k}+\varepsilon_{k'})^2(\varepsilon_{k}^-\varepsilon_{k'}^--C\theta C\theta')s\theta s\theta'c\varphi'\\
&-&\left\{ 2b_+^2(c^2\theta+c^2\theta')
+2m[\varepsilon_{k}(\Delta^++2\varepsilon_{k'}^2)s^2\theta +\varepsilon_{k'}(\Delta^++2\varepsilon_{k}^2)s^2\theta']
+(\varepsilon_{k}+\varepsilon_{k'})^2 \varepsilon_{k}^-\varepsilon_{k'}^- s^2\theta s^2\theta'
-8\varepsilon_{k}\varepsilon_{k'}C\theta C\theta'
\right\} c^2\varphi' \\
{\sqrt2\kappa_{23}\over\alpha\pi}&=&
\left\{2k'\varepsilon_{k} \varepsilon_{k'}C\theta
+\varepsilon_{k'}^- (m\Delta^+ -2\varepsilon^2_{k}\varepsilon_{k'})c\theta' \right\}s\theta'
-\left\{\varepsilon_{k}^- c\theta \left[\varepsilon_{k'}^-(\varepsilon_{k}+\varepsilon_{k'})^2 c^2\theta'
-\varepsilon_{k'} (\Delta^+-2m\varepsilon_{k})\right]-2k\varepsilon_{k}\varepsilon_{k'}C\theta' \right\}
s\theta c\varphi' \\
&-&\left\{\varepsilon_{k'}^- c\theta'\left((\varepsilon_{k}-\varepsilon_{k'})^2
\varepsilon_{k}^+ -(\varepsilon_{k}+\varepsilon_{k'})^2 \varepsilon_{k}^- c^2\theta\right)
+4k'\varepsilon_{k}\varepsilon_{k'}C\theta\right\}s\theta' c^2\varphi' \\
{\sqrt{2}\kappa_{24}\over\alpha\pi}&=&+\Delta^-
\left\{ (\varepsilon_{k}-\varepsilon_{k}^-c^2\theta)S\theta' - mS\theta c\varphi'
-(\varepsilon_{k}^+ -\varepsilon_{k}^- c^2\theta)S\theta'c^2\varphi'  \right\}\cr
{\sqrt2\kappa_{31}\over\alpha\pi}&=&-2\varepsilon_{k}\varepsilon_{k'}S\theta C\theta'  -\varepsilon_{k}^-
(m\Delta^+ -2\varepsilon_{k}\varepsilon_{k'}^2)s\theta c\theta     \cr
&+& s\theta'\left\{   \varepsilon_{k}^-c\theta'
\left[ (m\Delta^+ - 2\varepsilon_{k}^2\varepsilon_{k'})c^2\theta
+\varepsilon_{k}(\Delta^+-2m\varepsilon_{k'})s^2\theta\right]+2\varepsilon_{k} \varepsilon_{k'}kC\theta' \right\}c\varphi'
- \Delta^+ \varepsilon_{k}^-\varepsilon_{k'}^- s\theta c\theta s^2\theta'c^2\varphi'\cr
{\sqrt2\kappa_{32}\over\alpha\pi}&=&
\left\{2k\varepsilon_{k}\varepsilon_{k'}C\theta'+\varepsilon_{k}^-(m\Delta^+-2\varepsilon_{k}\varepsilon_{k'}^2)c\theta\right\}s\theta
-\left\{\varepsilon_{k'}^-c\theta'
\left[\varepsilon_{k}^-(\varepsilon_{k}+\varepsilon_{k'})^2  c^2\theta-\varepsilon_k(\Delta^+-2m\varepsilon_{k'})\right]
-2k'\varepsilon_{k}\varepsilon_{k'}C\theta\right\}s\theta'c\varphi'  \cr
&-&\left\{\varepsilon_{k}^- c\theta
\left[(\varepsilon_{k}-\varepsilon_{k'})^2\varepsilon_{k'}^+-(\varepsilon_{k}+\varepsilon_{k'})^2\varepsilon_{k'}^-c^2\theta'  \right]
+4k\varepsilon_{k}\varepsilon_{k'}C\theta'\right\}s\theta c^2\varphi'\cr
{\kappa_{33}\over\alpha\pi}&=&\left\{\varepsilon_{k}\varepsilon_{k'}^-(\Delta^+{\rm-2m}\varepsilon_{k'})c^2\theta'
{\rm+} \varepsilon_{k'}\varepsilon_{k}^-(\Delta^+{\rm-2m}\varepsilon_{k} )c^2\theta
{\rm-}\varepsilon_{k}\varepsilon_{k'}(\Delta^+{\rm +2m^2})
{\rm-}\left[(\varepsilon_{k}{\rm+}\varepsilon_{k'})^2\varepsilon_{k}^-\varepsilon_{k'}^-c\theta c\theta'{\rm-}2kk'\varepsilon_{k}\varepsilon_{k'}
\right]c\theta c\theta' \right\}c\varphi'\\
&-&\left[(\varepsilon_{k}+\varepsilon_{k'})^2\varepsilon_{k}^-\varepsilon_{k'}^-c\theta c\theta'-2kk'\varepsilon_{k}\varepsilon_{k'}
\right]s\theta s\theta' c^2\varphi'\\
{\kappa_{34}\over\alpha\pi}&=&+\Delta^-\varepsilon_{k}^- s\theta c\theta S\theta' (1- c^2\varphi') \cr
{\sqrt2\kappa_{41}\over\alpha\pi}&=&-\Delta^-\left\{ (m+\varepsilon_{k'}^-s^2\theta')S\theta
+mS\theta' c\varphi' - \varepsilon_{k'}^- s^2\theta' S\theta c^2\varphi'\right\}\cr
{\sqrt2\kappa_{42}\over\alpha\pi}&=&-\Delta^-\left\{(\varepsilon_{k'}-\varepsilon_{k'}^-c^2\theta')S\theta-mS\theta'c\varphi'
- (\varepsilon_{k'}^+-\varepsilon_{k'}^-c^2\theta')S\theta  c^2\varphi' \right\}\cr
{\kappa_{43}\over\alpha\pi}&=&-\Delta^-\varepsilon_{k'}^-s\theta'c\theta'S\theta (1-c^2\varphi')\cr
{\kappa_{44}\over\alpha\pi}&=& \left[2\varepsilon_{k}\varepsilon_{k'}(C\theta C\theta'-\varepsilon_{k}\varepsilon_{k'})-m^2\Delta^+\right]c\varphi'
+\Delta^+ S\theta S\theta' c^2\varphi'
\end{eqnarray*}

\subsection{Pseudoscalar}\label{app_ps}

$J=0^+$:
\begin{eqnarray}
{\kappa_{11}\over\alpha\pi}&=&
-[m^2\Delta^+ - 2\varepsilon_k\varepsilon_{k'}(\varepsilon_k\varepsilon_{k'}-C\theta C\theta')]
+ \Delta^+S\theta S\theta' c\varphi'  \nonumber\\
{\kappa_{12}\over\alpha\pi}&=& -m\Delta^-\left(S\theta'-S\theta c\varphi'\right) \nonumber \\
{\kappa_{21}\over\alpha\pi}&=& +m\Delta^- \left(S\theta-S\theta' c\varphi'\right)\label{eq1_ps}\\
{\kappa_{22}\over\alpha\pi}&=&\Delta^+ S\theta S\theta' +
[m^2\Delta^+-2\varepsilon_k\varepsilon_{k'}(\varepsilon_k\varepsilon_{k'}-C\theta C\theta')]c\varphi' \nonumber
\end{eqnarray}

$J=1,a=0:$
\begin{eqnarray}
{\kappa_{11}\over\alpha\pi}&=&-(2kk'\varepsilon_k \varepsilon_{k'} + b_-^2 c\theta c\theta')
+\varepsilon_k \varepsilon_{k'}(\Delta^+ -2m^2)s\theta s\theta' c\varphi'\nonumber\\
{\kappa_{12}\over\alpha\pi}&=&-m\Delta^-\left(\varepsilon_{k'}c\theta s\theta'-\varepsilon_k s\theta c\theta' c\varphi'\right)\nonumber\\
{\kappa_{21}\over\alpha\pi}&=&+m\Delta^-\left(\varepsilon_{k}s\theta c\theta'-\varepsilon_{k'}c\theta s\theta' c\varphi'\right)\label{eq11_ps}\\
{\kappa_{22}\over\alpha\pi}&=&-\varepsilon_k \varepsilon_{k'}(\Delta^+ -2m^2)s\theta s\theta'
+(2kk'\varepsilon_k \varepsilon_{k'}+ b_-^2 c\theta c\theta')c\varphi'\nonumber
\end{eqnarray}

$J=1^+,a=1$:
\begin{eqnarray*}
{2\kappa_{11}\over\alpha\pi}&=&
4C\theta C\theta' \varepsilon_{k}\varepsilon_{k'}
+m\Delta^- ( \varepsilon_{k}s^2\theta-\varepsilon_{k'}s^2\theta') + b_-^2(c^2\theta+ c^2\theta')\\
&+&(\varepsilon_{k}-\varepsilon_{k'})^2(kk'-\varepsilon_{k}^-\varepsilon_{k'}^-c\theta c\theta')s\theta s\theta'c\varphi'
+(\varepsilon_{k}-\varepsilon_{k'})^2 \varepsilon_{k}^-\varepsilon_{k'}^-  s^2\theta s^2\theta'c^2\varphi'   \\
%
{2\kappa_{12}\over\alpha\pi}&=&
(c^2\theta-c^2\theta')b_-^2-m\Delta^-(\varepsilon_{k}s^2\theta+\varepsilon_{k'}s^2\theta')
-\left\{kk'(\varepsilon_{k}+\varepsilon_{k'})^2+
c\theta c\theta'(\varepsilon_{k}-\varepsilon_{k'})^2\varepsilon_{k}^-\varepsilon_{k'}^-\right\}s\theta s\theta'c\varphi'\\
&-&\left\{(b_-^2-m\varepsilon_{k}\Delta^-)(1+c^2\theta')
+ \varepsilon_{k'}\varepsilon_{k}^-(\Delta^++2m \varepsilon_{k}) s^2\theta'\right\}s^2\theta c^2\varphi'\\
{\sqrt2\kappa_{13}\over\alpha\pi}&=&\left\{\left[\varepsilon_{k'}(\Delta^-+2m\varepsilon_{k})
-\varepsilon_{k'}^-(\varepsilon_{k}-\varepsilon_{k'})^2 c^2\theta'\right]\varepsilon_{k}^- c\theta
-2k\varepsilon_{k}\varepsilon_{k'}C\theta'\right\}s\theta c\varphi'\\
&-&\left\{(b^2_-+m\varepsilon_{k'}\Delta^-)c\theta'+2kk'\varepsilon_{k}\varepsilon_{k'}c\theta \right\}s\theta'
+ \varepsilon_{k}^- \varepsilon_{k'}^- (\varepsilon_{k}-\varepsilon_{k'})^2  s^2\theta s\theta'c\theta'c^2\varphi' \\
{\sqrt2\kappa_{14}\over\alpha\pi}&=&\Delta^-\left\{ -(\varepsilon_{k}-\varepsilon_{k}^+c^2\theta)S\theta'
+ mS\theta c\varphi' + \varepsilon_{k}^-  s^2\theta S\theta'c^2\varphi' \right\}\\
{2\kappa_{21}\over\alpha\pi}&=& (c^2\theta-c^2\theta')b_-^2+m\Delta^-(\varepsilon_{k}s^2\theta+\varepsilon_{k'}s^2\theta')
-\left\{kk'(\varepsilon_{k}+\varepsilon_{k'})^2+
c\theta c\theta'(\varepsilon_{k}-\varepsilon_{k'})^2\varepsilon_{k}^-\varepsilon_{k'}^-\right\}s\theta s\theta'c\varphi'\\
&-&\left\{(b_-^2+m\varepsilon_{k'}\Delta^-)(1+c^2\theta)
+ \varepsilon_{k}\varepsilon_{k'}^-(\Delta^++2m \varepsilon_{k'}) s^2\theta\right\}s^2\theta' c^2\varphi'\\
{2\kappa_{22}\over \alpha\pi}&=&-m\Delta^-(\varepsilon_{k}s^2\theta-\varepsilon_{k'}s^2\theta')-b_-^2(c^2\theta+c^2\theta')
-4\varepsilon_{k}\varepsilon_{k'}C\theta C\theta'  \\
&+&\left\{kk'(\varepsilon_{k}-\varepsilon_{k'})^2+c\theta c\theta'
\left[ m\Delta^-(\varepsilon_{k}-\varepsilon_{k'})-b_-^2-\varepsilon_{k}\varepsilon_{k'}(\Delta^+-2m^2)\right]
\right\} s\theta s\theta'c\varphi'\\
&+&\left\{8\varepsilon_{k}\varepsilon_{k'}C\theta C\theta'+ 2b_-^2(c^2\theta+c^2\theta')
+2m\Delta^-(\varepsilon_{k}s^2\theta-\varepsilon_{k'}s^2\theta')
+s^2\theta s^2\theta'(\varepsilon_{k}-\varepsilon_{k'})^2\varepsilon_{k}^-\varepsilon_{k'}^-\right\} c^2\varphi' \\
{\sqrt{2}\kappa_{23}\over\alpha\pi}&=&\left\{2k'\varepsilon_{k}\varepsilon_{k'}C\theta
+(b_-^2+m\varepsilon_{k'}\Delta^-)c\theta'\right\}s\theta' \\
&+&\left\{
\left[\varepsilon_{k'}(\Delta^++2m\varepsilon_{k})s^2\theta'+(m\Delta^++2\varepsilon_{k}\varepsilon_{k'}^2)c^2\theta'\right]
\varepsilon_{k}^-c\theta-2k\varepsilon_{k}\varepsilon_{k'}C\theta'\right\} s\theta c\varphi' \\
&+&\left\{\left[\varepsilon_{k}(\Delta^++2m\varepsilon_{k'})s^2\theta
+( m\Delta^+ + 2\varepsilon_{k}^2\varepsilon_{k'}) (1+c^2\theta)\right]\varepsilon_{k'}^-c\theta'
-4k'\varepsilon_{k}\varepsilon_{k'}C\theta\right\} s\theta' c^2\varphi'\cr
{\sqrt2\kappa_{24}\over \alpha\pi}&=&-\Delta^-\left\{(\varepsilon_{k}-c^2\theta \varepsilon_{k}^-)S\theta'+mS\theta c\varphi'-
(\varepsilon_{k}^+ - \varepsilon_{k}^-c^2\theta ) S\theta'c^2\varphi'  \right\} \\
{\sqrt{2}\kappa_{31}\over\alpha\pi}&=&
\left\{\left[\varepsilon_{k}(\Delta^++2m\varepsilon_{k'})-
\varepsilon_{k}^-(\varepsilon_{k}-\varepsilon_{k'})^2c^2\theta\right]\varepsilon_{k'}^-c\theta'
-2k'\varepsilon_{k}\varepsilon_{k'}C\theta\right\} s\theta' c\varphi' \\
&-&\left\{(b^2_--m\varepsilon_{k}\Delta^-)c\theta+2kk'\varepsilon_{k}\varepsilon_{k'}c\theta'\right\}s\theta
+\varepsilon_{k}^-\varepsilon_{k'}^-(\varepsilon_{k}-\varepsilon_{k'})^2s\theta c\theta s^2\theta' c^2\varphi' \\
{\sqrt{2}\kappa_{32}\over\alpha\pi}&=&
\left\{2k\varepsilon_{k}\varepsilon_{k'}C\theta'+(b_-^2-m\varepsilon_{k}\Delta^-)c\theta\right\}s\theta\\
&+&\left\{\left[\varepsilon_{k}(\Delta^++2m\varepsilon_{k'})s^2\theta+(m\Delta^++2\varepsilon_{k}^2\varepsilon_{k'})c^2\theta\right]
\varepsilon_{k'}^-c\theta'-2k'\varepsilon_{k}\varepsilon_{k'}C\theta\right\}s\theta'c\varphi' \\
&+&\left\{\left[\varepsilon_{k'} (\Delta^++2m \varepsilon_{k}) s^2\theta'
+(m\Delta^+ + 2\varepsilon_{k}\varepsilon_{k'}^2)(1+c^2\theta')\right]\varepsilon_{k}^-  c\theta -4kk'\varepsilon_{k} \varepsilon_{k'} c\theta'\right\} s\theta c^2\varphi' \\
{\kappa_{33}\over\alpha\pi}&=&\left\{
-\varepsilon_{k'}\varepsilon_{k'}\left[(\Delta^+-2m^2)s^2\theta s^2\theta'+2C\theta C\theta'\right]
-b_-^2 c^2\theta c^2\theta'
+m\Delta^-\left[\varepsilon_{k'}s^2\theta'c^2\theta-\varepsilon_{k}s^2\theta c^2\theta'\right]
 \right\}c\varphi'\\
&+&\left[2kk'\varepsilon_{k}\varepsilon_{k'}+\varepsilon_{k}^-\varepsilon_{k'}^-
(\varepsilon_{k}-\varepsilon_{k'})^2c\theta c\theta'\right]s\theta s\theta'c^2\varphi'\\
{\kappa_{34}\over\alpha\pi}&=&-\Delta^-\varepsilon_{k}^- s\theta S\theta'c\theta s^2\varphi'\\
{\sqrt{2}\kappa_{41}\over\alpha\pi}&=&-\Delta^-\left\{
-S\theta(\varepsilon_{k'}-\varepsilon_{k'}^-c^2\theta')+mS\theta'c\varphi'+S\theta s^2\theta'\varepsilon_{k'}^-c^2\varphi'\right\} \\
{\sqrt{2}\kappa_{42}\over\alpha\pi}&=&\Delta^-  \left\{(\varepsilon_{k'} - \varepsilon_{k'}^-c^2\theta')S\theta
+ mS\theta'c\varphi' -( \varepsilon_{k'}^+  - \varepsilon_{k'}^-c^2\theta') S\theta c^2\varphi'   \right\} \\
{\kappa_{43}\over\alpha\pi}&=&\Delta^-\varepsilon_{k'}^- s\theta' S\theta c\theta's^2\varphi'\\
{\kappa_{44}\over\alpha\pi}&=&(-b_-^2-2\varepsilon_{k}\varepsilon_{k'}C\theta C\theta')c\varphi'-\Delta^+S\theta S\theta' c^2\varphi'\\
\end{eqnarray*}

\subsection{Pseudo-Vector}

Pseudovector kernels will be given as a sum of the pseudoscalar ones
plus a term $\delta_{ij}$ which depends on variables $t,t'$ defined in (\ref{ttp}) and vanishes on energy shell ($t=t'=0$).
\[\kappa_{ij} = \kappa^{ps}_{ij} +\delta_{ij}\]
The following expressions for $\delta_{ij}$ are valid only for
$x-x'>0$ -- with $x,x'$ defined by (\ref{x})-- and because of that,  coefficients
(\ref{cde}) are not symmetric in the exchange $(i,k,\theta)\leftrightarrow(i',k',\theta')$.
For $x-x'<0$, the corresponding expressions are obtained by replacing $t\to -t'$, $t' \to -t$
and their symmetry properties restaured.

\bigskip
$J=0^+$:
\begin{eqnarray*}
{\delta_{11}\over\alpha\pi}&=& m^2\left\{m^2tt'+(t-t')\Delta^--(t+t')
(\varepsilon_{k'}C\theta'-\varepsilon_{k}C\theta)\right\}-m^2tt'S\theta S\theta'c\varphi'\nonumber\\
{\delta_{12}\over\alpha\pi}&=&m\left\{m^2tt'+\varepsilon_{k}^2(t-t') + \varepsilon_{k}(t+t')C\theta\right\} S\theta'
+m\left\{m^2tt'-\varepsilon_{k'}^2(t-t')-\varepsilon_{k'}C\theta'(t+t')\right\} S\theta c\varphi'\nonumber\\
{\delta_{21}\over\alpha\pi}&=& m\left\{m^2tt'-\varepsilon_{k'}^2(t-t')-\varepsilon_{k'}(t+t')C\theta'\right\}S\theta
+m\left\{m^2tt'+\varepsilon_{k}^2(t-t')+\varepsilon_{k}(t+t')C\theta\right\}S\theta'c\varphi'\label{eq1_pv}\\
{\delta_{22}\over\alpha\pi}&=&m^2tt' S\theta S\theta'
-m^2\left[m^2tt'+(t-t')\Delta^-  -(t+t')(\varepsilon_{k'}C\theta'-\varepsilon_{k}C\theta)\right]c\varphi'\nonumber
\end{eqnarray*}

\subsection{Vector}

Vector kernels are written in the form
\[ \kappa_{ij}= 2 m^2 tt'{m^2\over\mu^2} \; \upsilon_{ij} + \chi_{ij}  \]
in which $\chi_{ij}$ correspond to the $\mu=0$ case. The
$\upsilon_{ij}$ contribution, due to  $\mu$-dependent term
in the vector propagator, appears as being of shell corrections.
Positronium kernels are simply given by
$\kappa^{(PS)}_{ij}=-\chi_{ij}$.

$J=0^+$:
\begin{eqnarray}
-{\kappa_{11}\over2\alpha\pi}&=&2m^2tt'\frac{m^2}{\mu^2}(m^2+S\theta S\theta'c\varphi')
+ (b_-^2-2\epsilon_{k'}^2\epsilon_{k}^2)\nonumber\\
-{\kappa_{12}\over2\alpha\pi}&=&2m^3tt'\frac{m^2}{\mu^2}(S\theta'-S\theta c\varphi')+ m\Delta^- S\theta'\nonumber\\
-{\kappa_{21}\over2\alpha\pi}&=&2m^3tt'\frac{m^2}{\mu^2}(S\theta-S\theta' c\varphi')- m\Delta^- S\theta \label{vkern}\\
-{\kappa_{22}\over2\alpha\pi}&=&2m^2tt'\frac{m^2}{\mu^2}(S\theta S\theta'+m^2 c\varphi')-
(\Delta^+S\theta S\theta'
+2\epsilon_{k}\epsilon_{k'}(\epsilon_{k}\epsilon_{k'}+C\theta C\theta') c\varphi')\nonumber
\end{eqnarray}

$J=1^+,a=0$:
\begin{eqnarray}
{\kappa_{11}\over2\alpha\pi}&=&-2m^2tt'\frac{m^2}{\mu^2}\left[m^2c\theta c\theta'+\epsilon_{k}\epsilon_{k'}s\theta s\theta' c\varphi' \right]
+\left[m^2\Delta^+c\theta c\theta'+ 2\epsilon_{k}\epsilon_{k'}(kk'+2m^2 s\theta s\theta'c\varphi')\right] \nonumber\\
{\kappa_{12}\over2\alpha\pi}&=&2m^3tt'\frac{m^2}{\mu^2}\left[\epsilon_{k'}c\theta s\theta'-\epsilon_{k}s\theta c\theta' c\varphi' \right]
-m\epsilon_{k'}\left[\Delta^+c\theta s\theta'
- 2\epsilon_{k}\epsilon_{k'} s\theta c\theta'  c\varphi'   \right]\nonumber\\
{\kappa_{21}\over2\alpha\pi}&=&2m^3tt'\frac{m^2}{\mu^2}\left[\epsilon_{k}s\theta c\theta'-\epsilon_{k'}c\theta s\theta' c\varphi' \right]
-m\epsilon_{k}\left[\Delta^+s\theta c\theta'
- 2\epsilon_{k}\epsilon_{k'} c\theta s\theta'  c\varphi'   \right]\label{kappa_J1a0_V}\\
{\kappa_{22}\over2\alpha\pi} &=&-2m^2tt'\frac{m^2}{\mu^2}\left[\epsilon_{k}\epsilon_{k'}s\theta s\theta'+m^2c\theta c\theta' c\varphi' \right]
+\left[\epsilon_{k}\epsilon_{k'}\Delta^+s\theta s\theta'
+2\epsilon_{k}\epsilon_{k'}(kk'+\epsilon_{k}\epsilon_{k'}c\theta c\theta') c\varphi' \right]\nonumber
\end{eqnarray}

$J=1^+,a=1:$

\bigskip
The $\mu$-independent kernels $\chi_{ij}$ are given by:

\begin{eqnarray*}
{\chi_{11}\over\alpha\pi}&=&
2\varepsilon_{k}\varepsilon_{k'} \left[2C\theta C\theta'+
(m\Delta^+ +\varepsilon_{k'} \varepsilon_{k}^-c^2\theta +\varepsilon_{k} \varepsilon_{k'}^-c^2\theta')\right]
+[4kk'\varepsilon_{k}\varepsilon_{k'}+\Delta^+(kk'+\varepsilon_{k}^- \varepsilon_{k'}^-)]s\theta s\theta' c\varphi'\\
&+& 2\varepsilon_{k}\varepsilon_{k'}\varepsilon_{k}^-\varepsilon_{k'}^- s^2\theta s^2\theta' c^2\varphi'
\\
{\chi_{12}\over\alpha\pi}&=&\left[kk'(4\varepsilon_{k}\varepsilon_{k'}-\Delta^+)
+\varepsilon_k^-\varepsilon_{k'}^- \Delta^+ c\theta c\theta'\right]s\theta s\theta' c\varphi'\\
&-&2\varepsilon_{k}\varepsilon_{k'}\varepsilon_{k}^-(2\varepsilon_{k'}-\varepsilon_{k'}^-s^2\theta')s^2\theta c^2\varphi'
+2\varepsilon_{k}\varepsilon_{k'}\left[
\varepsilon_{k}\varepsilon_{k'}(c^2\theta'-c^2\theta)+m( \varepsilon_{k}s^2\theta'-\varepsilon_{k'}s^2\theta)\right]\\
{\chi_{13}\over\sqrt{2}\alpha\pi}&=&-2\varepsilon_{k}\varepsilon_{k'}
(kk'c\theta+\varepsilon_{k}\varepsilon_{k'}^-c\theta') s\theta'
-\left\{\varepsilon_k^-\Delta^+(m+\varepsilon_{k'}^-s^2\theta') c\theta-4kk'\varepsilon_{k}\varepsilon_{k'}c\theta'\right\}s\theta c\varphi'
+2\varepsilon_{k}\varepsilon_{k'}\varepsilon_k^-\varepsilon_{k'}^-s^2\theta s\theta'c\theta'c^2\varphi'\\
{\chi_{14}\over\sqrt{2}\alpha\pi}&=&m\Delta^-S\theta c\varphi'\\
{\chi_{21}\over\alpha\pi}&=&\left[kk'(4\varepsilon_{k}\varepsilon_{k'}-\Delta^+)+\varepsilon_k^-\varepsilon_{k'}^-
\Delta^+c\theta c\theta'\right]s\theta s\theta' c\varphi' \\
&-&2 \varepsilon_{k}\varepsilon_{k'}\varepsilon_{k'}^-(2\varepsilon_{k}-\varepsilon_{k}^-s^2\theta) s^2\theta'c^2\varphi'
-2\varepsilon_{k}\varepsilon_{k'}\left[\varepsilon_{k}\varepsilon_{k'}(c^2\theta'-c^2\theta)
+m(\varepsilon_{k}s^2\theta'-\varepsilon_{k'}s^2\theta)\right]\\
{\chi_{22}\over\alpha\pi}&=&-2\varepsilon_{k}\varepsilon_{k'}\left\{ m(\varepsilon_{k}+\varepsilon_{k'})
+ \varepsilon_{k}\varepsilon_{k'}^-c^2\theta' + \varepsilon_{k'}\varepsilon_{k}^-c^2\theta +2C\theta C\theta'\right\}
+\left\{kk'(4\varepsilon_{k}\varepsilon_{k'}+\Delta^+)  + \varepsilon_{k}^-\varepsilon_{k'}^- \Delta^+ c\theta c\theta' \right\}s\theta s\theta'c\varphi'\\
&+&2\varepsilon_{k}\varepsilon_{k'}
\left\{(\varepsilon_{k'}^++\varepsilon_{k'}^-c^2\theta')(\varepsilon_{k}^++\varepsilon_{k}^-c^2\theta)+4C\theta C\theta' \right\}c^2\varphi' \\
{\chi_{23}\over\sqrt{2}\alpha\pi}&=&
2\varepsilon_{k}\varepsilon_{k'}(kk'c\theta+\varepsilon_{k'}^-\varepsilon_k c\theta')s\theta'
-2\varepsilon_{k}\varepsilon_{k'}\left\{2kk'c\theta+(\varepsilon_k^++\varepsilon_k^-c^2\theta)
\varepsilon_{k'}^-  c\theta'\right\}s\theta' c^2\varphi'\\
&+&\left\{4kk'\varepsilon_{k}\varepsilon_{k'}c\theta'
-\varepsilon_k^-\Delta^+(m+\varepsilon_{k'}^-s^2\theta')c\theta \right\} s\theta c\varphi' \\
{\chi_{24}\over\sqrt{2}\alpha\pi}&=&-m\Delta^-S\theta c\varphi' \\
{\chi_{31}\over\sqrt{2}\alpha \pi}&=&-2\varepsilon_k\varepsilon_{k'}(kk'c\theta'+\varepsilon_{k'}\varepsilon_{k}^-c\theta) s\theta
-\left\{\varepsilon_{k'}^-\Delta^+(m+\varepsilon_{k}^-s^2\theta)c\theta'-4kk'\varepsilon_k\varepsilon_{k'} c\theta\right\}s\theta'c\varphi'
+2\varepsilon_k\varepsilon_{k'}\varepsilon_{k}^-\varepsilon_{k'}^- c\theta s\theta s^2\theta'c^2\varphi'\\
{\chi_{32}\over\sqrt{2}\alpha\pi}
&=&2\varepsilon_{k}\varepsilon_{k'}(kk'c\theta'+\varepsilon_{k}^-\varepsilon_{k'} c\theta)s\theta
-2\varepsilon_{k}\varepsilon_{k'}\left\{2kk'c\theta'+(\varepsilon_{k'}^++\varepsilon_{k'}^-c^2\theta')
\varepsilon_{k}^-  c\theta\right\}s\theta c^2\varphi'\\
&+&\left\{4kk'\varepsilon_{k}\varepsilon_{k'}c\theta
-\varepsilon_{k'}^-\Delta^+(m+\varepsilon_{k}^-s^2\theta)c\theta' \right\}s\theta' c\varphi'  \\
{\chi_{33}\over\alpha\pi}&=&
4\varepsilon_{k}\varepsilon_{k'}(kk'+\varepsilon_{k}^-\varepsilon_{k'}^-)s\theta s\theta' c^2\varphi'
+2\left\{4\varepsilon_{k}\varepsilon_{k'} C\theta C\theta'
+\Delta^+(\varepsilon_{k}\varepsilon_{k'} +  \varepsilon_{k}^-\varepsilon_{k'}^-c^2\theta c^2\theta'
-\varepsilon_{k'}\varepsilon_{k}^-c^2\theta -\varepsilon_{k}\varepsilon_{k'}^-c^2\theta') \right\}c\varphi' \\
{\chi_{34}\over\alpha\pi} &=&0 \\
{\chi_{41}\over\sqrt{2}\alpha\pi}&=&- m\Delta^- S\theta' c\varphi'\\
{\chi_{42}\over\alpha\pi}        &=&  m\Delta^- S\theta' c\varphi'\\
{\chi_{43}\over\alpha\pi}&=&0\\
{\chi_{44}\over\alpha\pi}&=&2 (2\varepsilon_{k}^2\varepsilon_{k'}^2 - b_-^2) c\varphi'
\end{eqnarray*}

and the $\upsilon_{ij}$  contribution reads:
\begin{eqnarray*}
{\upsilon_{11}\over\alpha\pi}&=&
m[\varepsilon_{k}^- c^2\theta+\varepsilon_{k'}^- c^2\theta'-(\varepsilon_{k}+\varepsilon_{k'})]
- [k k'+ \varepsilon_k^-\varepsilon_{k'}^- c\theta c\theta']s\theta s\theta' c\varphi'
- \varepsilon_k^-\varepsilon_{k'}^-s^2\theta s^2\theta'c^2\varphi'   \\
{\upsilon_{12}\over\alpha\pi}&=&
m[(\varepsilon_{k}-\varepsilon_{k'})-\varepsilon_{k}^-c^2\theta+\varepsilon_{k'}^-c^2\theta']
+[kk'-\varepsilon_k^-\varepsilon_{k'}^- c\theta c\theta']s\theta s\theta'c\varphi'
-\varepsilon_k^-(\varepsilon_{k'}^+-\varepsilon_{k'}^-c^2\theta')s^2\theta c^2\varphi'\\
{\upsilon_{13}\over\sqrt{2}\alpha\pi}&=&-m\varepsilon_{k'}^- s\theta'c\theta'
+\varepsilon_k^-(m+\varepsilon_{k'}^-s^2\theta')s\theta c\theta c\varphi'
-\varepsilon_k^-\varepsilon_{k'}^- s^2\theta s\theta'c\theta'c^2\varphi'\\
{\upsilon_{14}\over\sqrt{2}\alpha\pi}&=&(m+\varepsilon_{k}^- s^2\theta)S\theta'
-mS\theta c\varphi' -\varepsilon_{k}^-s^2\theta S\theta'c^2\varphi' \\
{\upsilon_{21}\over\alpha\pi}&=&-m[(\varepsilon_{k}-\varepsilon_{k'})+\varepsilon_{k'}^-c^2\theta'-\varepsilon_{k}^-c^2\theta]
+[k k'-\varepsilon_k^- \varepsilon_{k'}^- c\theta c\theta']s\theta s\theta'c\varphi'
-\varepsilon_{k'}^- (\varepsilon_{k}^+- \varepsilon_{k'}^-c^2\theta)s^2\theta'c^2\varphi'\\
{\upsilon_{22}\over\alpha\pi}&=&
m[(\varepsilon_{k}+\varepsilon_{k'})-\varepsilon_{k}^-c^2\theta-\varepsilon_{k'}^-c^2\theta']
-(kk'+\varepsilon_k^- \varepsilon_{k'}^- c\theta c\theta') s\theta s\theta'c\varphi'
- (2m+\varepsilon_{k}^- s^2\theta)(2m+\varepsilon_{k'}^-s^2\theta')c^2\varphi'\\
{\upsilon_{23}\over\sqrt{2} \alpha \pi}
&=&m\varepsilon_{k'}^-s\theta'c\theta'
+\varepsilon_k^- s\theta c\theta(m+\varepsilon_{k'}^-s^2\theta')c\varphi'
-\varepsilon_{k'}^-(\varepsilon_{k}^+ -\varepsilon_{k}^-c^2\theta ) s\theta'c\theta'c^2\varphi'\\
{\upsilon_{24}\over \sqrt{2} \alpha \pi}&=&(m+\varepsilon_{k}^-s^2\theta)S\theta'
+ mS\theta c\varphi' - (2m+\varepsilon_{k}^- s^2\theta) S\theta' c^2\varphi'  \\
{\upsilon_{31}\over\sqrt{2}\alpha\pi} &=&-m\varepsilon_k^- s\theta c\theta
+ \varepsilon_{k'}^- (m+\varepsilon_{k}^-s^2\theta ) s\theta'c\theta'c\varphi'
-\varepsilon_k^-\varepsilon_{k'}^- s^2\theta' s\theta c\theta c^2\varphi'\\
{\upsilon_{32}\over\sqrt{2}\alpha\pi} &=&m \varepsilon_k^- s\theta c\theta
+\varepsilon_{k'}^- s\theta' c\theta'(m+\varepsilon_{k}^-s^2\theta)c\varphi'
-\varepsilon_{k}^-(\varepsilon_{k'}^+ -\varepsilon_{k'}^-c^2\theta') s\theta c\theta c^2\varphi'\\
{\upsilon_{33}\over \alpha\pi} &=&-2(m+\varepsilon_{k'}^-s^2\theta')(m+\varepsilon_{k}^-s^2\theta)c\varphi'
-2\varepsilon_{k}^- \varepsilon_{k'}^- c\theta c\theta's\theta s\theta' c^2\varphi'\\
{\upsilon_{34}\over\alpha\pi}&=&2\varepsilon_k^- s\theta c\theta  S\theta's^2\varphi'\\
{\upsilon_{41}\over\sqrt{2}\alpha\pi}&=&(m+\varepsilon_{k'}^- s^2\theta')S\theta
-mS\theta'c\varphi' -\varepsilon_{k'}^-S\theta s^2\theta'c^2\varphi' \\
{\upsilon_{42}\over\alpha\pi}&=&(m+\varepsilon_{k'}^-s^2\theta')S\theta
+ mS\theta' c\varphi' -(2m+\varepsilon_{k'}^- s^2\theta') S\theta   c^2\varphi' \\
{\upsilon_{43}\over\alpha\pi}&=&2\varepsilon_{k'}^-s\theta'c\theta'S\theta s^2\varphi'\\
{\upsilon_{44}\over\alpha\pi}&=&-2(m^2+S\theta S\theta'c\varphi')c\varphi'\\
\end{eqnarray*}

\section {Relations between the components of $J=1$ state}\label{app3}

The wave function of the $J=1$ state is represented in two forms:
in the form (\ref{nz2}) with the components $\varphi_i$ and in the form
(\ref{nz8}) with the components $f_i$.
The formulas expressing the components $\varphi_i$ in terms of the $f_i$,
in approximation $M\approx 2m$, are
given in Appendix C from \cite{CDKM_PR_98}. Here we give these
relations for arbitrary $M$. Note that $\varphi_3$ and $\varphi_6$ only
differ relative to \cite{CDKM_PR_98}. We denote below $z=\cos\theta$.
\begin{eqnarray}
\varphi_1 &=& \frac{m^2(2\varepsilon_k + m)}{4\varepsilon_k k^2}f_2
+\frac{m^2}{4\varepsilon_k(\varepsilon_k + m)}(\sqrt{2}f_1 - f_3 + zf_4
- \sqrt{3}zf_6)\ ,
\nonumber\\
\varphi_2 &=& \frac{m}{4\varepsilon_k}(\sqrt{2}f_1 - f_2 -f_3 -2zf_4)\ ,
\nonumber\\
\varphi_3 &=& -\frac{\sqrt{2}(2\varepsilon_k-M)^2 k}
{16\varepsilon_k^2(\varepsilon_k +m)}zf_1
-\frac{(2\varepsilon_k + m)(2\varepsilon_k-M)^2}
{16\varepsilon_k^2 k}zf_2
+\frac{(4\varepsilon_k^2 + 8M\varepsilon_k + M^2)k}
{16\varepsilon_k^2(\varepsilon_k + m)}zf_3
\nonumber \\
&&+\frac{3M}{4k}\left(1-\frac{z^2(2\varepsilon_k-M)^2 k^2}
{12M\varepsilon_k^2(\varepsilon_k + m)}\right)f_4
+\frac{\sqrt{3}M}{4k}\left(1 +
\frac{z^2(2\varepsilon_k-M)^2 k^2}
{4M\varepsilon_k^2(\varepsilon_k + m)}\right)f_6\ ,
\nonumber\\
\varphi_4 &=& -\frac{3m}{2k}f_4 + \frac{\sqrt{3}m}{2k}f_6\ ,
\nonumber\\
\varphi_5
&=&\frac{1}{2}\sqrt{\frac{3}{2}}\frac{m^2}{k\varepsilon_k}f_5\ ,
\nonumber\\
\varphi_6 &=&
\frac{(2\varepsilon_k-M)^2}{8m\varepsilon_k}(\sqrt{2}f_1-f_2
+zf_4-\sqrt{3}zf_6)
-\frac{(4\varepsilon_k^2+8M\varepsilon_k+M^2)}{8m\varepsilon_k}f_3\ .
\label{ba5}
\end{eqnarray}

The state with $J=1,a=1$ is determined by eq. (\ref{nz2_1}) as
a decomposition in four orthogonal spin structures $S^{(1)}_{i\mu}$.
These four structures are expressed by eq. (\ref{nzf2}) in terms of six
structures $S_{j\mu}$, defined in (\ref{eq12d}),  with the coefficients
$h_{ij}$ given below. These coefficient are found as follows.
We
substitute the formulas (\ref{eq4c}) into  (\ref{ba5}), then  eqs. (\ref{ba5})
-- into (\ref{nz2}). In this way, the way function
$\phi^{(1)}_{\mu}$ is expressed in terms of the four functions $g^{(1)}_i$,
i.e., obtains the form (\ref{nz2_1}).
The
coefficients at the front of $g^{(1)}_i$  are the structures $S^{(1)}_{i\mu}$.
Collecting these coefficients, we find $S^{(1)}_{i\mu}$ in
terms of  six structures $S_{j\mu}$, in the form of eq. (\ref{nzf2})  with the
following coefficients $h_{ij}$:
\begin{eqnarray}\label{hij}
&&h_{11}=\frac{\sqrt{3}m^2}{4\varepsilon_k(\varepsilon_k+m)},\quad
h_{12}=\frac{\sqrt{3}m}{4\varepsilon_k},\quad
h_{13}=-\frac{\sqrt{3}(\varepsilon_k-m)(4\varepsilon_k^2+M^2)z}
{16 \varepsilon_k^2 k},
\nonumber\\
&&h_{14}=h_{15}=0,\quad
h_{16}=\frac{\sqrt{3}(4\varepsilon_k^2+M^2)}
{8\varepsilon_k m},
\nonumber\\
&&h_{21}=\frac{\sqrt{3}m^2[\varepsilon_k(1-z^2)+m(1+z^2)]}
{4\varepsilon_k k^2(1-z^2)},\quad
h_{22}=-\frac{\sqrt{3}m}{4\varepsilon_k},
\nonumber\\
&&h_{23}=-\frac{\sqrt{3}(4\varepsilon_k^2+M^2)
[\varepsilon_k(1-z^2)+m(1+z^2)]z}
{16\varepsilon_k^2 k(1-z^2)},\quad
h_{24}=\frac{\sqrt{3}mz}{k(1-z^2)},
\nonumber\\
&&h_{25}=0,\quad
h_{26}=-\frac{\sqrt{3}(4\varepsilon_k^2+M^2)(1+z^2)}
{8\varepsilon_k m (1-z^2)}
\nonumber\\
&&h_{31}=\frac{\sqrt{3}m^2 z}
{2\varepsilon_k(\varepsilon_k+m)\sqrt{2(1-z^2)}}, \quad
h_{32}=0,\quad
h_{33}=-\frac{\sqrt{3}(\varepsilon_k-m)(4\varepsilon_k^2+M^2)z^2}
{8\varepsilon_k^2 k \sqrt{2(1-z^2)}},
\nonumber\\
&&h_{34}=-\frac{\sqrt{3}m}{k\sqrt{2(1-z^2)}}, \quad
h_{35}=0,\quad
h_{36}=\frac{\sqrt{3}(4\varepsilon_k^2+M^2)z}
{4\varepsilon_k m \sqrt{2(1-z^2)}},
\nonumber\\
&&h_{41}=h_{42}=h_{43}=h_{44}=0,\quad
h_{45}=\frac{\sqrt{3}m^2}{2\varepsilon_k k \sqrt{2(1-z^2)}},\quad h_{46}=0.
\end{eqnarray}

\end{document}